\documentclass[%
 preprint,
showpacs,
amsmath,amssymb,
aps,
pra,
floatfix,
]{revtex4-1}
\usepackage{color}
\usepackage{amsmath,amssymb,graphicx}
\usepackage{upgreek}
\usepackage{graphicx}
\usepackage{dcolumn}
\usepackage{bm}
\usepackage{lipsum} 
\usepackage{float}
\usepackage{color}
\usepackage{ulem}

\begin{document}

\newcommand{\degree}{^\circ}
\newcommand{\unit}[1]{\,\mathrm{#1}}
\newcommand{\Figureref}[2][]{Figure~\ref{#2}#1}
\newcommand{\Eqnref}[1]{Equation~(\ref{#1})}
\newcommand{\bra}[1]{\left<\mathrm{#1}\right|}
\newcommand{\ket}[1]{\left|\mathrm{#1}\right>}

\newcommand{\comment}[1]{}

\title{Collective state synthesis in an optical cavity using Rydberg atom blockade}

\author{Santosh Kumar}
  \author{Jiteng Sheng}
      \author{Jonathon A. Sedlacek}
 \author{Haoquan Fan}
\author{James P. Shaffer}
\email{Corresponding author: shaffer@nhn.ou.edu}
\affiliation{Homer L. Dodge Department of Physics and Astronomy, The University of Oklahoma, 440 W. Brooks St. Norman, OK 73019, USA}

\date{\today}

\begin{abstract}
We investigate the coherent manipulation of interacting Rydberg atoms placed inside a high-finesse optical cavity for the deterministic preparation of strongly coupled light-matter systems. We consider a four-level diamond scheme with one common Rydberg level for $N$ interacting atoms. One side of the diamond is used to excite the atoms into a collective `superatom' Rydberg state using either $\pi$-pulses or stimulated Raman adiabatic passage (STIRAP) pulses. The upper transition on the other side of the diamond is used to transfer the collective state to one that is coupled to a field mode of an optical cavity. Due to the strong interaction between the atoms in the Rydberg level, the Rydberg blockade mechanism plays a key role in the deterministic quantum state synthesis of the atoms in the cavity. We use numerical simulation to show that non-classical states of light can be generated and that the state that is coupled to the cavity field is a collective one. We also investigate how different decay mechanisms affect this interacting many-body system. We also analyze our system in the case of two Rydberg excitations within the blockade volume. The simulations are carried out with parameters corresponding to realizable high-finesse optical cavities and alkali atoms like rubidium. 
\begin{description}
\item[PACS numbers] {32.80.Ee, 42.50.Ex, 37.30.+i, 42.50.†}
\end{description}
\end{abstract}

\maketitle

\section{Introduction}
We investigate using Rydberg atom interactions to deterministically synthesize collective quantum states that can be strongly coupled to an optical cavity for applications in quantum information science, quantum optics and the production of quantum light fields. The method enables a single or several collective excitations to be prepared in a high-finesse optical cavity. The number of excitations is precisely controlled so that photon number states can be produced using the coupled atoms-cavity system. The system we describe in this paper differs from typical Jaynes-Cummings (JC) and Tavis-Cummings (TC) physics in that strong atom-atom interactions play a key role in the initial state preparation, which is the focus of this paper.

Rydberg atoms have many exotic properties. For example, Rydberg atoms are large, interact strongly with one another, have relatively long lifetimes, and can be manipulated with external electric and magnetic fields \cite{Gallagher94}. Particularly important to the work presented in this paper, the interactions between ultracold Rydberg atoms can be larger than typical frequency stabilized laser spectral bandwidths and kinetic energies at distances of over $10\,\mu$m \cite{Schwettmann2006,AAMOP2014}. The interactions can be manipulated easily with electric fields, relative to valence states \cite{Cabral2011}. There is a broad range of behavior that can be induced when the multi-level character of the interactions is taken into account.

Rydberg atom blockade is a result of the strong interactions between Rydberg atoms. In Rydberg atom blockade, the long range interactions between Rydberg atoms suppress multiple excitations in a volume in which the interactions are large enough to shift the excitation laser out of resonance with the Rydberg atom transition frequency \cite{PRL04_Gould,PRL07_Pfau}. It has been realized over the last 15 years that Rydberg atom blockade is useful for a host of applications, particularly in quantum optics and quantum information science. Rydberg atom blockade can be used to control quantum dynamics and prepare interesting quantum states \cite{PRL01_Lukin,NatPhys09_Saffman,PRL14_Walker}. Rabi oscillations have been observed in effective two-level ultracold Rydberg atoms \cite{PRL08_Weidemuller,NPJ08_Weidemuller,PRL08_Saffman,PRL14_Browaeys,NatPhy09_Grangier,NatPhys12_Kuzmich} demonstrating long coherence times and collective behavior. Collective Rabi oscillations on Rydberg transitions have even been observed in a thermal  rubidium (Rb) vapor on a time scale below $1\,$ns \cite{PRL11_Pfau}.

Atoms contained in cavities, both optical and microwave, have been important systems for investigating the quantum-classical boundary in physics. The ability to entangle atoms with light presented by cavity quantum electrodynamics has also served as a model system for applications in quantum information science \cite{Nat08_Kimble, RMP01_Haroche}. A cavity quantum electrodynamical system with a single excitation or controlled number of excitations is useful both for fundamental studies \cite{Stenholm1973} and applications that use quantum behavior, particularly those involving quantum entanglement \cite{PRL00_Zheng,PRL_11_Rempe,NJP_Reiter12}. Entangled states can be created between distant atoms in one cavity or separate cavities \cite{PRA99_Zoller, PRA08_Morigi, PRA12_Kumar, Nat12_Rempe, PRL_13_Blatt}. Trapping of a single atom in a high-finesse optical cavity has been achieved \cite{PRA99_Kimble,PRL03_Kimble}. However, it is technically challenging to couple the light in the cavity to a single atom because of the small absorption cross-section \cite{PRL99_Rempe,Science00_Kimble,PRL07_Chapman,PRL11_Mikio}. The problem of small absorption cross-sections can be handled by using an atomic ensemble, where light more strongly interacts with collective atomic states.

In this paper, we theoretically show that one can excite collective $N$ atom superposition states using Rydberg atom interactions that can be strongly coupled to a high-finesse cavity. We observe $\sqrt N$ enhancement of the Rabi oscillations within the cavity showing that the collective state can be realized \cite{PRL09_Wallraff,PRL10_Adams} and analyze the decay mechanisms that affect this interacting many-body system. Collectively enhanced cooperativity factors help to improve the efficiency of photon generation out of the cavity. The efficiency can be improved by increasing the number of atoms. We also show that by chirping the laser excitation pulses in time, the single atom Rydberg blockade effect can be suppressed and we can create two Rydberg excitations within the blockade radius that are coupled to the cavity.

We use a four-level diamond type atomic scheme as shown in Fig.~\ref{fig:1}(b). One side of the diamond is used to collectively excite the atoms to the Rydberg level using a pair of $\pi$-pulses or a pair of counter-intuitive stimulated Raman adiabatic passage (STIRAP) pulses with Rabi frequencies $P_1(t)$ and $P_2(t)$. By using a laser pulse, $\Omega(t)$, and the single atom cavity coupling strength, $g$, the state of the collective atoms-cavity system can be controlled. Rydberg atom blockade plays an important role in this process, since we are interested in cases where a single or controlled finite number of coherently shared excitation(s) are produced in a collection of $5 P_{3/2}$ state Rb atoms. The $N$ interacting atoms-cavity system is within the reach of experimental realization \cite{PRA98_Kimble,PRL13_Rempe}. The system can be used to prepare single photons using a single  excitation and more complex quantum light fields by preparing multiple excitations in the cavity.

Recently, a JC model in the optical domain has been proposed for investigating Rydberg blockaded atomic ensembles in a cavity. The transmission through the cavity was studied numerically by Monte-Carlo simulation \cite{PRA10_Molmer}. In contrast to this work and others, our studies resemble a pumping scheme for inverting the population of a laser, except the upper state of the lasing transition is prepared in a collective state. The pumping scheme is similar to a Raman or superradiant laser and work done on low atom number masers and micromasers. However, for these types of lasers and masers, the atoms are coherent but non-interacting \cite{EPL92_Zoller,PRL93_Haake,PRA96_Haake}. Our approach is useful for single photon generation \cite{PRA13_Rempe,PRA13_Pfau}, as well as the generation of other types of quantum light fields, and coherent optical manipulation for quantum information processing \cite{Nat08_Girvin,PRL08_Molmer,RMP10_Molmer,PRA12_Kumar,Nat12_Rempe,Thompson06}. The collective atoms-cavity system can also be important for generating unique quantum states such as superposition states of the atomic ensemble \cite{PRL03_Polzik,RMP10_Polzik}. The approach does not require phase matching and it may be easier to use this system than those based on phase matching to create more elaborate quantum light fields \cite{PRA08_Walker,PRA_12_Pfau}. Creating arbitrary quantum light fields remains an important challenge for quantum optics, so it is useful to explore a wide range of different strategies.
The method that we explore in this paper may have technological and experimental advantages because the collective state is prepared on a set of transitions which have significantly different excitation wavelengths, $> 15\,$nm for all transitions in Rb, than the cavity mode to which it is coupled. The cavity output can be filtered and detected with low background.

\section{Model}

\begin{figure}[htbp]
\includegraphics[width=6in]{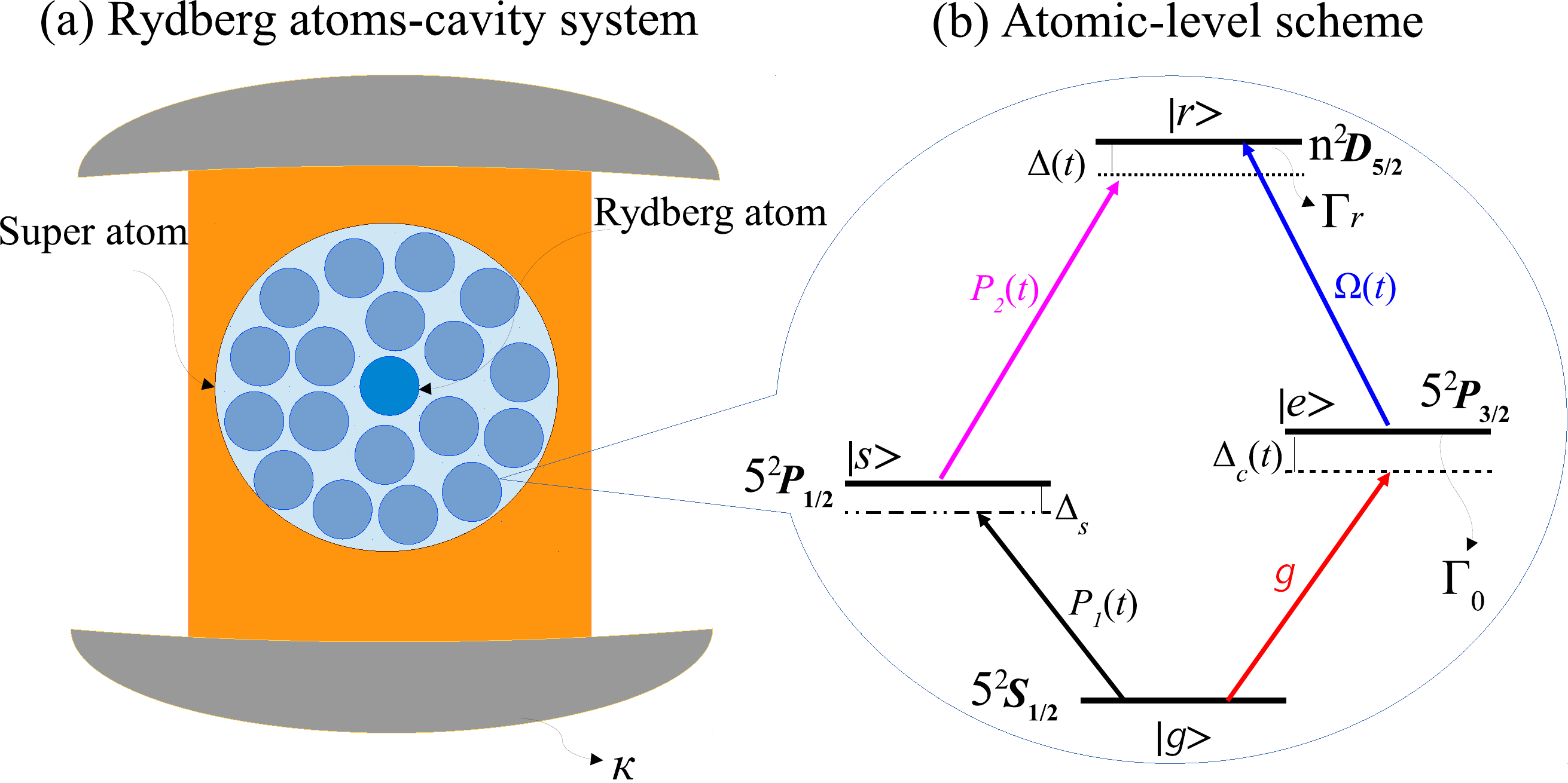}
\caption[short caption]{ (a) A Rydberg atom collective state in an optical cavity indicating Rydberg blockade. (b) Level scheme of a four-level atomic system driven by three laser pulses on transitions $\vert g \rangle \leftrightarrow \vert s \rangle, \vert s \rangle \leftrightarrow \vert r \rangle$ and $\vert r \rangle \leftrightarrow \vert e \rangle $ with Rabi frequencies $P_1(t), P_2(t)$ and $\Omega(t)$, respectively. The transition  $\vert g \rangle \leftrightarrow \vert e \rangle$ is coupled to an optical cavity with coupling strength $g$. $ \Gamma_r$ is the decay rate of the Rydberg state, $ \Gamma_{\perp} = \Gamma_0/2$ is the transverse spontaneous decay rate of the $\vert e \rangle$ state and $\kappa$ is the cavity decay rate. Energy levels are shown for Rb.
\label{fig:1}}
\end{figure}

Fig.~\ref{fig:1} shows a conceptual drawing of the experimental setup. The setup consists of a collection of atoms trapped in a high-finesse optical cavity. The atomic energy level scheme is a four-level diamond scheme with atomic states denoted by $\vert g \rangle$, $\vert e \rangle$, $\vert s \rangle$ and $\vert r \rangle$. $\hbar \omega_i$ is the energy of the $i^{th}$ atomic level, therefore the transition frequencies $\omega_{ij} = \omega_j - \omega_i$ where $i$ and $j$ are labels corresponding to the respective states $\vert g \rangle$, $\vert e \rangle$, $\vert s \rangle$ and $\vert r \rangle$. The Rabi frequencies are taken to be time dependent since we are concerned with applying pulses for the preparation of a collective state that we can couple to the cavity as well as with controlling the photons emitted from the cavity. The decay processes that we consider explicitly in our model are also shown. We envision the atoms to be held in the cavity with a far off resonance dipole trap (FORT) or a magnetic trap. The dipole trapping wavelength can ideally be set to a magic wavelength so that the frequency of the $\vert g \rangle \leftrightarrow \vert e \rangle$ transition is fixed in the cavity. We discuss example design parameters in this section as well as the following 2. Additional decay processes and the consequences of the main approximations used are also discussed later in the paper.

One side of the diamond is used to collectively excite the atoms to a Rydberg level using a pair of $\pi$-pulses or counter-intuitive STIRAP pulses with Rabi frequencies $P_1(t)$ and $P_2(t)$. $P_1(t)$ drives the $\vert g \rangle \leftrightarrow \vert s \rangle$ transition, frequency $\omega_{gs}$, while $P_2(t)$ drives the $\vert s \rangle \leftrightarrow \vert r \rangle$ transition, frequency $\omega_{sr}$. If the interaction between the Rydberg atoms is strong enough so that the $\vert g \rangle \leftrightarrow \vert r \rangle$ transition is shifted out of resonance after the first Rydberg excitation occurs, further population of $\vert r \rangle$ is blocked. 

The other side of the diamond system is used to produce a collective state that is close to resonance with a field mode of a high-finesse optical cavity. The optical cavity mode is coupled to the $\vert g \rangle \leftrightarrow \vert e \rangle$ transition since the resonant frequency of the cavity, $\omega_{c}$, is assumed to be close to the atomic transition frequency, $\omega_{ge}$. In fact, $\omega_{c} = \omega_{ge}$ for most of the calculations that follow. The control laser is treated classically and has a frequency $\omega_{l}$ that couples to the $\vert e \rangle \leftrightarrow \vert r \rangle$ transition at frequency $\omega_{er}$ with a Rabi frequency $\Omega(t)$. The cavity field is treated as a quantum field. The single atom states of the coupled atoms-cavity field system are products of $\vert e\rangle$ or $\vert g\rangle$, each coupled to a ladder of photon number states.

The density matrix equations are not easily solvable for an arbitrary number of atoms because the number of states increases dramatically with $N$. One way to reduce the basis set size in a many-body system is to use a mean-field approximation. For an ultracold Rydberg gas this means each highly excited atom shifts the Rydberg levels of surrounding atoms out of resonance with the laser field \cite{PRA07_Rost}. In this approach, atomic coherence is damped out and the quantum master equations reduce to many-body rate equations. This method is useful for understanding excitation suppression due to Rydberg atom blockade in cold gases \cite{PRL04_Weidemuller}. A second way to practically address the increasing basis set size is to utilize the idea of local blockade and truncate the basis set when the number of excitations in the system exceeds some limit \cite{PRA09_Cote,PRA05_Robicheaux}. The second method is appropriate when the system dynamics are constrained to a subspace of the complete basis. Rydberg atom dipole blockade naturally justifies this approach. In the spirit of the latter approach, we carry-out our calculations in a truncated Hilbert space, similar to that used by Saffman and Walker \cite{PRA02_Saffman}.


To prepare the initial collective excitation of the atoms, $P_{1}(t)$ and $P_{2}(t)$ are detuned from the atomic transition as shown in Fig.~\ref{fig:1}(b). $\Delta_s$ is the detuning from the intermediate level $\vert s \rangle$, assumed to be fixed in time, and $\Delta(t)$ is the detuning from the Rydberg level $\vert r \rangle$. The one photon detuning $\Delta_s$ is taken to be larger than either of the excitation Rabi frequencies for excitation by $\pi$-pulses, $\Delta_s \equiv \omega_s-\omega_g -\omega_{p_1}> P_1(t), P_2(t)$ and the detuning satisfies the two photon condition,
\begin{equation}
\omega_{p_{2}} + \omega_{p_{1}} = \omega_{r} - \omega_{g},
\end{equation}
where $\omega_{p_{1},p_{2}}$ are the laser frequencies associated with $P_1(t)$ and $P_2(t)$, respectively. Under these circumstances, the residence time in level $\vert s \rangle$ is small and $\vert s \rangle$ can be adiabatically eliminated. For STIRAP pulses, the one photon detuning $\Delta_s$ has a very small effect and the required conditions are a 2-photon resonance between initial and final states and adiabatic evolution of the pulses. The adiabatic approximation means that the Rydberg state excitation can be replaced by a single photon excitation in this setup. However, most ultracold Rydberg work uses a 2-photon scheme like this one and a 2-photon excitation from the ground state of an alkali atom allows access to $n$S states, which generally have simpler interactions, or $n$D states, which have more sub-states to use. 

The effective Hamiltonian for the excitation side of the diamond system can be written in a collective state basis as,
\begin{eqnarray}
H_{1} = \frac{\hbar}{2} \left[ - 2(\Delta(t)+ \Delta_{s}) \vert R \rangle \langle R \vert  -4 (\Delta(t)+ \Delta_{s}) \vert RR \rangle \langle RR \vert + \{S_1(t) \vert G \rangle \langle R \vert  + S_2(t)  \vert R \rangle \langle RR \vert  \right. \nonumber \\ \left. + \mathrm{H.c.} \} \right],~~~~~
\end{eqnarray}
where
\begin{equation}
S_1(t)= \frac{\sqrt{N}S(t)}{2},~   S_2(t)= \frac{\sqrt{N-1}S(t)}{\sqrt{2}(2 +\Delta(t)/\Delta_{s})} ~\mathrm{and}~ S(t) = P_{1}(t) P_{2}(t)/\Delta_s,
\end{equation}
\begin{equation}
\vert G \rangle   \equiv \vert g \rangle = \vert g_{1}...g_{i}...g_{N} \rangle,
\end{equation}
\begin{equation}
\vert  R \rangle   \equiv  1/\sqrt{N} \sum_{i=1}^{N}\vert r_{i} \rangle = 1/\sqrt{N} \sum_{i=1}^{N} \vert g_{1}...r_{i}...g_{N} \rangle,
\end{equation}
\begin{eqnarray}
\vert  RR \rangle  & \equiv &  1/\sqrt{N_e} \sum_{i \neq j = 1}^{N} \vert r_{i}r_{j} \rangle = 1/\sqrt{N_e} \sum_{i \neq j = 1}^{N} \vert g_{1}...r_{i}...r_{j}...g_{N} \rangle,
\end{eqnarray}
and
\begin{eqnarray}
N_e = \frac{N\,!}{(N-2)\,! \,2\,!}.
\end{eqnarray}
The Hamiltonian for the emission side of the diamond system for $N$ interacting atoms with truncated Hilbert space bounded by the two Rydberg excitations is
\begin{eqnarray}
H_{2} = \hbar/2  \left[ \omega_{c} \hat{C}^{\dagger} \hat{C}  - 2\Delta_{c} \left(  \vert \mathrm{E,0} \rangle \langle \mathrm{E,0} \vert + \vert  \mathrm{E,1} \rangle \langle \mathrm{E,1} \vert + 2\vert  \mathrm{EE,0} \rangle \langle \mathrm{EE,0} \vert +  \vert  \mathrm{ER,0} \rangle \langle \mathrm{ER,0} \vert   \right) \right. \nonumber \\ \left. + \left\{\Omega(t) \left( \vert  \mathrm{R,0} \rangle \langle \mathrm{E,0} \vert + \vert  \mathrm{R,1} \rangle \langle \mathrm{E,1} \vert + \sqrt{2}\vert  \mathrm{EE,0} \rangle \langle \mathrm{ER,0} \vert + \sqrt{2}\vert  \mathrm{ER,0} \rangle \langle \mathrm{RR,0} \vert  \right) \right. \right. \nonumber \\ \left. \left. +g \left( \sqrt{N}  \vert  \mathrm{G,1} \rangle \langle \mathrm{E,0} \vert +  \sqrt{N} \vert  \mathrm{G,2} \rangle \langle \mathrm{E,1} \vert + \sqrt{(N-1)}\vert  \mathrm{R,1} \rangle \langle \mathrm{ER,0} \vert  \right.\right.\right. \nonumber \\ \left.\left.\left. + \sqrt{2(N-1)} \vert \mathrm{E,1} \rangle \langle \mathrm{EE,0} \vert \right) + \mathrm{H.c.} \right\} \right],~~~~
\end{eqnarray}
where $\Omega(t)$ is the Rabi frequency of the control laser, $\hat{C}^{\dagger} (\hat{C})$ is the creation (annihilation) operator for the near resonant cavity mode, $\Delta_{c}$ is the cavity detuning, $g$ is the coupling strength of one atom to the cavity and
\begin{eqnarray}
\ket  E   \equiv  1/\sqrt{N} \sum_{i=1}^{N}\vert e_{i} \rangle =  1/\sqrt{N} \sum_{i=1}^{N} \vert g_{1}...e_{i}...g_{N} \rangle,
\end{eqnarray}
\begin{eqnarray}
\vert  \mathrm{EE} \rangle  & \equiv & 1/\sqrt{N_e}  \sum_{i \neq j = 1}^{N} \vert e_{i}e_{j} \rangle,
\end{eqnarray}
and
\begin{eqnarray}
\vert  \mathrm{ER} \rangle  & \equiv &   1/\sqrt{2N_e}  \sum_{i \neq j = 1}^{N} \vert e_{i}r_{j} \rangle.
\end{eqnarray}
On the emission side of the system, the first element of the ket represents the collective atomic state. The second number of the ket represents the Fock, or photon-number state, of the near resonance mode of the cavity. $\vert  \mathrm{E} \rangle$ and $\vert  \mathrm{EE} \rangle$ are the collective states we are interested in preparing in the cavity \cite{OptCom00_Fleischhauer}. Although much of the interesting physics in this system is due to the cavity quantum electrodynamical system, it is important to note that we are also addressing the preparation of the atomic collective state and the effect of several different decay mechanisms. As a result, both $H_1$ and $H_2$ are important. Stark shifts caused by the excitation lasers can be compensated with detunings.

In our calculations, we assume that the interaction volume is less than the blockade volume. This allows for the modeling of the blockade effect through a straightforward detuning of states with more than a single Rydberg excitation. The level shift between a pair of atoms labeled $i$ and $j$ each in a Rydberg state can be estimated using
\begin{equation}
\Delta_{R} = \frac{C_{p}}{\hbar |l_{i} - l_{j}|^p},\label{interaction}
\end{equation}
where two types of simplified interatomic interactions can be considered: (a) dipole-dipole interaction where $p = 3$, and (b) van der Waals interaction where $p = 6$. $C_p$ are the dispersion coefficients, and $l_{i}$ and $l_{j}$ are the positions of the $i^{th}$ and $j^{th}$ atoms, respectively. Here, we are primarily interested in repulsive potentials due to the atom-atom interaction. More realistic Rydberg atom interaction potentials have a complicated dependence on internuclear separation, including many avoided crossings that can be manipulated with electric and magnetic fields \cite{Schwettmann2006,AAMOP2014,Cabral2011,NatPhy09_Richard,PRA07_Richard}. We use Eq.~\ref{interaction} to estimate the magnitude of the interaction whose exact value and radial dependence is not critical provided it is large enough for Rydberg blockade for the present discussion. The Hamiltonian for atom-atom interaction can be written as
\begin{eqnarray}
H_{a,a} = \hbar/2 \sum_{i<j} \vert r_i \rangle \langle r_i \vert \Delta_{R}  \vert r_j \rangle \langle r_j \vert.
\end{eqnarray}

The condition that all atoms in the interaction volume lie within the blockade volume translates to
\begin{equation}
\Delta_{R} >>\mathrm{max} \left[ \frac{\vert P_1(t) \vert^2 + \vert P_2(t) \vert^2}{\sqrt{2\vert P_1(t) \vert^2 + \Gamma_s^2/4}}\right], \label{blockadecond}
\end{equation}
where $\Gamma_s$ is the spontaneous decay rate of the intermediate state $\vert s \rangle$. The situation where all atoms are within a blockade volume can be realized by using sufficiently narrow bandwidth lasers to form an interaction volume with Rydberg atom interactions strong enough to satisfy Eq.~\ref{blockadecond}. In this work, the atoms are also assumed to be cold enough that they can be approximated as stationary, or frozen in space. These ideas are extendable to countable numbers of excitations by increasing the interaction volume or decreasing the blockade volume. We also consider the case where we want to compensate the shift due to the one Rydberg excitation blockade, $\Delta_R$. This shift can be compensated by changing the time dependent detuning $\Delta(t)$ of the excitation fields, so that we can excite two Rydberg atoms in the blockade volume. More elaborate states can be created by, for example, applying more pulses to create several collective excitation states. Note that for 2 excitations, rather flat avoided crossings that occur in Rydberg atom interactions can be used to create 2 excitation states with minimized spatial correlation \cite{Cabral2011,NatPhy09_Richard,PRA07_Richard}.


The total Hamiltonian of the $N$ atom system coupled to the cavity is
\begin{equation}
H_{T} =H_{1} +H_{2} +H_{a,a}.
\end{equation}
The density matrix operator $\hat{\rho}$ of the $N$ interacting atoms and all the fields obeys Liouville's equation
\begin{eqnarray}
\frac{\partial \hat{\rho}}{\partial t} = -\frac{i}{\hbar}[H_{T},\hat{\rho}] + \mathcal{L}\hat{\rho},
 \label{evolution}
\end{eqnarray}
where the decays are due to spontaneous emission from $\vert e \rangle $ and $\vert r \rangle$ and loss from the cavity. These effects are included in the second term of the equation using the Lindblad operator $\mathcal{L}$. $\mathcal{L} \hat{\rho}$ can be written as
\begin{eqnarray}
\mathcal{L}\hat{\rho} =  \mathcal{L}_{r} \hat{\rho} +\mathcal{L}_{\perp} \hat{\rho} + \mathcal{L}_{\kappa} \hat{\rho} ,  ~~~
\end{eqnarray}
with
\begin{eqnarray}
\mathcal{L}_{r} \hat{\rho} &=& \Gamma_r \big(2L_{i}\hat{\rho}L^{ \dagger}_r - [ L_r L^{\dagger}_r, \hat{\rho} ]  \big) ,~~
\end{eqnarray}
\begin{eqnarray}
\mathcal{L}_{\perp} \hat{\rho} &=& \Gamma_{\perp} \big(2L_{\perp}\hat{\rho}L^{ \dagger}_{\perp} - [ L_{\perp} L^{\dagger}_{\perp}, \hat{\rho} ]  \big) ,~~
\end{eqnarray}
\begin{eqnarray}
\mathcal{L}_{\kappa} \hat{\rho} &=& \kappa \big(2L_{\kappa}\hat{\rho}L^{ \dagger}_{\kappa} - [ L_{\kappa} L^{\dagger}_{\kappa}, \hat{\rho} ]  \big) ,~~
\end{eqnarray}
\begin{eqnarray}
L^{ \dagger}_{r} =  \vert R,0 \rangle \langle E,0 \vert  + \vert R,1 \rangle \langle E,1 \vert  + \sqrt{2} \vert RR,0 \rangle \langle ER,0 \vert + \sqrt{2}\vert ER,0 \rangle \langle EE,0 \vert  \nonumber \\+ \sqrt{2(N-1)} \vert RR,0 \rangle \langle R,0 \vert ,
\end{eqnarray}
\begin{eqnarray}
L^{ \dagger}_{\perp} = \sqrt{N(N-1)} \vert ER,0 \rangle \langle L,0 \vert  +\sqrt{N(N-1)/2} \vert EE,0 \rangle \langle L,0 \vert  +\sqrt{N} \vert E,1 \rangle \langle L,0 \vert  \nonumber \\+ \sqrt{N} \vert E,0 \rangle \langle L,0 \vert ,
\end{eqnarray}
and
\begin{eqnarray}
L^{ \dagger}_{\kappa} &=&  \vert G, 2 \rangle \langle G, 1 \vert + \vert G, 1 \rangle \langle G, 0 \vert +  \vert E, 1 \rangle \langle E, 0 \vert +  \vert R, 1 \rangle \langle R, 0 \vert,
\end{eqnarray}
where $\Gamma_{r} $ is the Rydberg state decay rate, $\Gamma_{\perp}$ is the transverse spontaneous emission rate into the free space modes and $\kappa$ is the cavity decay rate. A dummy state $\vert L,0\rangle$ is used to model decay into non-cavity modes and loss from the collective state manifold due to dephasing. Here the `0' includes all other modes of the light field to simplify the notation. In particular, we use the dummy state to model the effect of additional dephasing mechanisms described in the Discussion section and to estimate their effect. 

We define the parameter $\Gamma_0 = 2 \pi \times 6 \,$MHz, the free space spontaneous decay rate corresponding to $^{87}$Rb, as a scale for the decay rates used for the calculations that follow. The total Rydberg state decay rate is $\Gamma_r \sim  2\pi \times 1.4$ kHz for $n = 90$. The black body radiation decay rate of the Rydberg state is very small, $\Gamma_{bb}\sim 2 \pi \times 0.2 $ kHz for $n = 90$, compared to the total decay rate. Consequently, we ignore black body decay which would cause atoms to leave our model system \cite{PRA09_Beterov}. At large principal quantum numbers, $n$, where the interactions are strongest, which are most appropriate for the scheme described in this paper, the neglect of black body decay is best justified. $\Gamma_r$ is small but we include it for completeness in the Liouville equations. 

The time variation of the electric field of the laser pulses is of the form,
\begin{equation}
 E(t) = A_0~\mathrm{sech}  [\frac{t-t_0}{\tau}].
\end{equation}
$A_0$ is the amplitude, $t_0$ is the time when the pulse reaches its maximum value and $\tau$ is the pulse width. $ 2 \pi \times \tau^{-1}$ is taken to be smaller than the shift due to the interaction of Rydberg atoms $\Delta_R$.

\section{Cavity and Interaction Parameter Estimates}

The key parameter for determining the emission rate into the cavity for a single atom is $g^2/(\kappa \Gamma_{\perp})$, which is the ratio of the rate of emission into the cavity mode to the loss due to spontaneous emission into other electromagnetic field modes. This ratio is an expression for the Purcell factor, $F_p = g^2/(\kappa \Gamma_{\perp})$, described in more detail in the next section, and is related to the single atom cooperativity factor, $C = F_p/2$. The cavity decay rate $\kappa$ is determined by the mirror reflectivities. The single atom coupling strength $g$ is determined by the atomic transition dipole moment, $\mu$, for the transition near the cavity resonance and the mode volume of the cavity, $V[L_c, R_c, \lambda]$. Realistic cavity parameters limit $F_p$. For collective excitations of $N$ atoms, the effective Purcell factor and collectively enhanced cooperativity increase, since $\mu$ increases like $\sqrt{N}$.

For our numerical calculations, we use parameters that have been achieved in recent experimental setups \cite{NatPhys09_Saffman,PRL14_Walker,NatPhys12_Kuzmich,PRL13_Rempe} and choose $\,^{87}$Rb to investigate the general system. To reach the strong coupling regime, we can achieve a single atom cavity coupling strength
\begin{equation}
g = \mu \sqrt{\frac{\omega_c}{2\hbar \epsilon_0 V[L_c, R_c, \lambda]}}
\end{equation}
of up to $2\pi \times 50\,$MHz for realistic cavity parameters. The cavity mode volume is
\begin{eqnarray}
V[L_c,R_c,\lambda] = \frac{\pi}{4} w_0[L_c,R_c,\lambda]^2 L_c,  ~~~
\end{eqnarray}
and the beam waist is
\begin{eqnarray}
w_0[L_c,R_c,\lambda] =\sqrt{\frac{\lambda}{\pi} \sqrt{\frac{L_c}{2}\left(R_c-\frac{L_c}{2}\right)}}.
\end{eqnarray}
$R_c$ is the radius of curvature of the mirrors, $L_c$ is the cavity length and $\lambda$ is the wavelength of the cavity mode. We concentrate on numerical results corresponding to conditions near the strong-coupling regime, $g >> \kappa, \Gamma_{\perp}, \Gamma_r$ as well as the leaky cavity regime, $\kappa >> g^2/\kappa >> \Gamma_{\perp}, \Gamma_r$. Choosing realistic parameters, $g$ and $\kappa$, does not allow a clean separation of the strong-coupling and leaky cavity regimes. Relevant cavity parameters are length $L_c \approx 50$ to $300\,\mu$m and beam waist $w_0 \approx 15$ to $22~\mu $m. A mirror radius of curvature of around $R_c = 25\,$mm is suitable for the cavity.

The blockade radius depends on the states of the Rydberg atoms. The blockade radius for principal quantum number $n=90$, corresponding to the $\sim 2 \pi \times 30\,$MHz excitation Rabi frequency, is $\sim 9\,\mu$m. At $n=90$, the shifts due to the interaction of Rydberg atoms $\Delta_R \approx 2 \pi \times 220$ MHz at $6.3\,\mu$m \cite{Schwettmann2006,AAMOP2014}. For densities of $\sim 10^{12}$ cm$^{-3}$, there are $\sim 3 \times 10^{3}$ atoms in the blockade volume with $\sqrt{N}$ shot to shot atom number fluctuations of $\sim 50$ atoms. The possible number of atoms in the blockade volume far exceeds the atom numbers considered in this paper. Small atom numbers, as studied here, can be generated by decreasing the trap density.

\section{The Purcell Regime}

The rate of spontaneous emission into the cavity mode can be enhanced by the interaction of the atoms with a matched resonant cavity, commonly called the Purcell effect. The Purcell regime occurs when $\kappa > g > \Gamma_{\perp}$. The enhancement, $F_p$, written in terms of the cavity mode volume and mirror reflectivities, $R_1$ and $R_2$, is \cite{QO_Fabre10},
\begin{eqnarray}
F_{p}  =   \frac{3\lambda^2L_c}{2\pi^2}\frac{F(R_1,R_2)}{V[L_c,R_c,\lambda]}, ~~~
\end{eqnarray}
where the cavity finesse is \begin{eqnarray}
F(R_1,R_2) \approx \frac{2\pi}{(1-R_1)+(1-R_2)}.  ~~~
\end{eqnarray}
The decay rate into the cavity mode due to the Purcell effect is then
\begin{eqnarray}
\Gamma_{p}  =   F_p \Gamma_{0}.  ~~~
\end{eqnarray}
As a realizable example, we take $R_1 = 0.999985$ and $R_2 = 0.99985$ with radius of curvature $R_c = 25$ mm. We can obtain a finesse $F \approx 3.8 \times 10^4$ at $780\,$nm with these reflectivities. For Purcell factor $F_p  \approx 10$, the cavity length is $L_c \approx 50$ $\mu$m, which still allows the positioning of a 30 $\mu$m FORT inside the cavity. For this example, $g \approx 2\pi \times 50 \,$MHz and $\kappa \approx 2\pi \times 72\,$MHz.

In the Purcell regime, the coupling to the cavity, accounted for by $g$, can be replaced with an enhanced spontaneous emission rate into the cavity mode. We explore this regime by replacing the direct coupling of the atoms to the cavity with an incoherent decay term into one mode of the electromagnetic field
\begin{eqnarray}
L^{ \dagger}_{p} = \sqrt{N} \vert E,0 \rangle \langle G,1 \vert  +\sqrt{N} \vert E,1 \rangle \langle G,2 \vert  + \sqrt{N-1}\vert ER,0 \rangle \langle R,1 \vert  \nonumber \\+\sqrt{2(N-1)} \vert EE,0 \rangle \langle E,1 \vert
\end{eqnarray}
with
\begin{eqnarray}
\mathcal{L}_{p} \hat{\rho} &=& \Gamma_{p} \big(2L_{p}\hat{\rho}L^{ \dagger}_{p} - [ L_{p} L^{\dagger}_{p}, \hat{\rho} ]  \big).~~
\end{eqnarray}
Effectively, the cavity is explicitly removed from the problem and the picture becomes one where an atom, or collective excitation, in free space preferentially emits into a specific mode of the electromagnetic field. The Purcell picture provides physical insight into preparing and detecting a countable number of excitations created in a cavity. 

\section{Verification of the Model}

First, for low $N$ and a single Rydberg excitation as a target, we consider the possibility of 2 Rydberg excitations occurring in our system. In these calculations, we show that, for $\Delta_R >>0$ and $\Delta = 0$, there is insignificant leakage out of the Hilbert space bounded by a single Rydberg excitation. For calculations with $N \geq3$, we use these results to justify the truncation of the Hilbert space at the 2 Rydberg excitation level. These results are also supported by numerical results of many other works done on Rydberg blockade. In these studies it is demonstrated that a large probability ($> 0.98$) for single Rydberg excitation of a superatom can be achieved while the probability of multiple excitations is negligible \cite{RMP10_Molmer,PRA11_Bergamini,PRA13_Molmer}.

\begin{figure}[htbp]
\includegraphics[height=150mm,width=160mm]{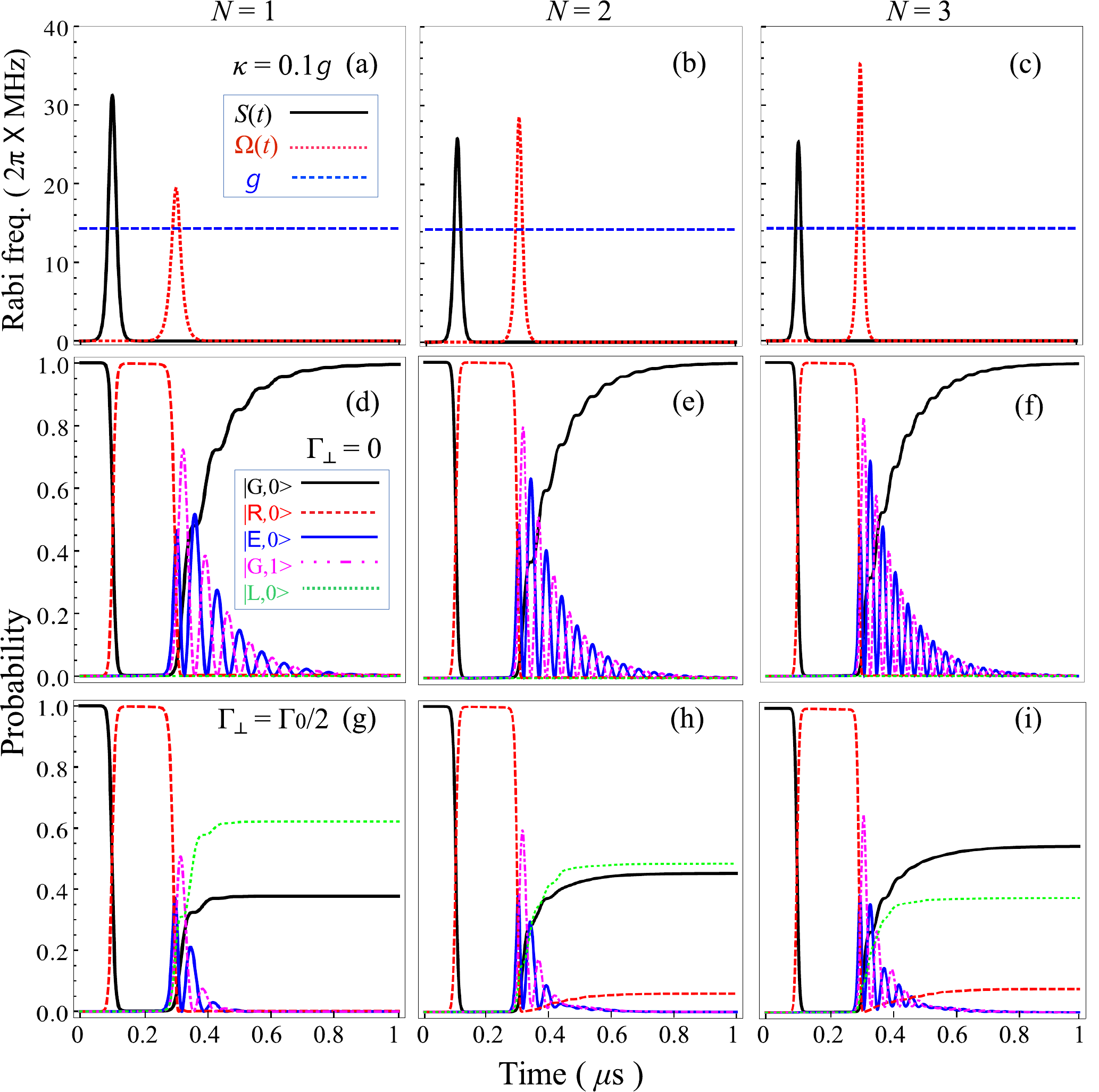}
\caption[short caption]{(a-c) The time variations of the Rabi frequencies of the laser pulses applied to a four-level system for $N$ interacting atoms. The effective Rabi frequency $S(t)$ is represented by the black solid curve. The red dotted curve shows the Rabi frequency of the control laser pulse $\Omega(t)$ and the blue dashed curve shows the cavity coupling strength $g$. Panels (d-i) show the corresponding probabilities for population of states $\vert G,0 \rangle$ (black solid), $\vert R,0 \rangle$ (red dashed), $\vert E,0 \rangle$ (blue solid), $\vert G,1 \rangle$ (magenta dot-dashed) and $\vert L,0 \rangle$ (green dotted). Parameters for the calculation are $\Gamma_r = 2 \pi \times 1.4 \,$kHz, $\Gamma_0 = 2\pi \times 6\,$MHz, $\kappa =2\pi \times 1.4\,$MHz $= 0.1 g$, $\Delta_c = 0$, $\Delta_R = 2\pi \times 220\,$MHz, $\Delta(t) = -2\pi \times 110\,$MHz and $\Delta_s = 2\pi \times 110\,$MHz.
\label{fig:2}}
\end{figure}

In Fig.~\ref{fig:2}, we show the results for $N = 1-3$ atoms with one Rydberg excitation. The model is efficient for investigating the case of $N > 3$, but here we focus our attention on a few interacting atoms within the cavity. The time dependence and peak amplitudes of the Rabi frequencies are shown in Fig.~\ref{fig:2}(a-c), while the probabilities for various states are shown in Fig.~\ref{fig:2}(d-i). For $N=1$ and $N=2$, the results are exact within the bounds of the model. To evaluate the solution for $N\geq3$, we ignore the higher order excitation states. The population in those states are $<$ 0.0015 justifying the truncation at the 2 excitation level for the work addressed in this paper. 

To further verify the model, we numerically investigate the preparation of a single collective excitation and the observation of a $\sqrt{N}$ enhancement of the Rabi oscillations of the superatom in the cavity. Collective states coupled to the cavity will exhibit a $\sqrt{N}$ enhancement in their effective atoms-cavity coupling strength, $\sqrt{N} g$. 

For $N = 1$, we used two $\pi$ pulses, $P_{1}(t)$ and $P_{2}(t)$, applied at the same time to prepare an atom in the Rydberg state. Subsequently, a control laser pulse with Rabi frequency $\Omega(t)$ is turned on to transfer the collective state to one that is near to resonance with the cavity. The population oscillates within the cavity between $\vert E,0 \rangle$ $\leftrightarrow$ $\vert G,1 \rangle$. Due to the cavity decay rate, $\kappa = 0.1 g$, these oscillations decay with time  \cite{Grynberg82}. For $N=2$ and $N=3$, two counter-intuitive STIRAP pulses \cite{PRA11_Bergamini,PRA13_Molmer} with pulse areas of $\pi$ were used to create a collective state with a single Rydberg excitation. We also increased the amplitude of the control laser pulse to optimize the transfer probability. In Fig.~\ref{fig:2}(e) and Fig.~\ref{fig:2}(f), the frequency of the Rabi oscillations are enhanced by $\sqrt{N}$ in the cavity. The decay rate and detuning of the cavity were the same as in Fig.~\ref{fig:2}(d). We also obtained the same results as shown in Fig.~\ref{fig:2}(e) and Fig.~\ref{fig:2}(f) using two superposed $\pi$ pulses (i.e. without using a STIRAP sequence). The plots in Fig.~\ref{fig:2} demonstrate that only a single excitation is prepared in the cavity and it collectively interacts with the cavity field. This result shows our numerical approach is consistent with expectations.

\section{Numerical Results}

In Fig.~\ref{fig:2}(g-i), we can see that the probability of populating the cavity ground state, $\vert G,0 \rangle$, is increased with increasing number of atoms. For $N>1$, due to imperfect $\pi$-pulses, the population returning to the Rydberg state increases compared to Fig.~\ref{fig:2}(d-f) where there is no decay \cite{PRL08_Molmer}. These results suggest that the probability of populating the cavity ground state depends on the number of atoms and the spontaneous emission rate out of the cavity. As we increase the coupling to the cavity mode, it, not surprisingly, makes the photon production from the cavity more efficient. This is the result of increasing the effective (collectively enhanced) cooperativity.

\begin{figure}[htbp]
\includegraphics[width=5.6in]{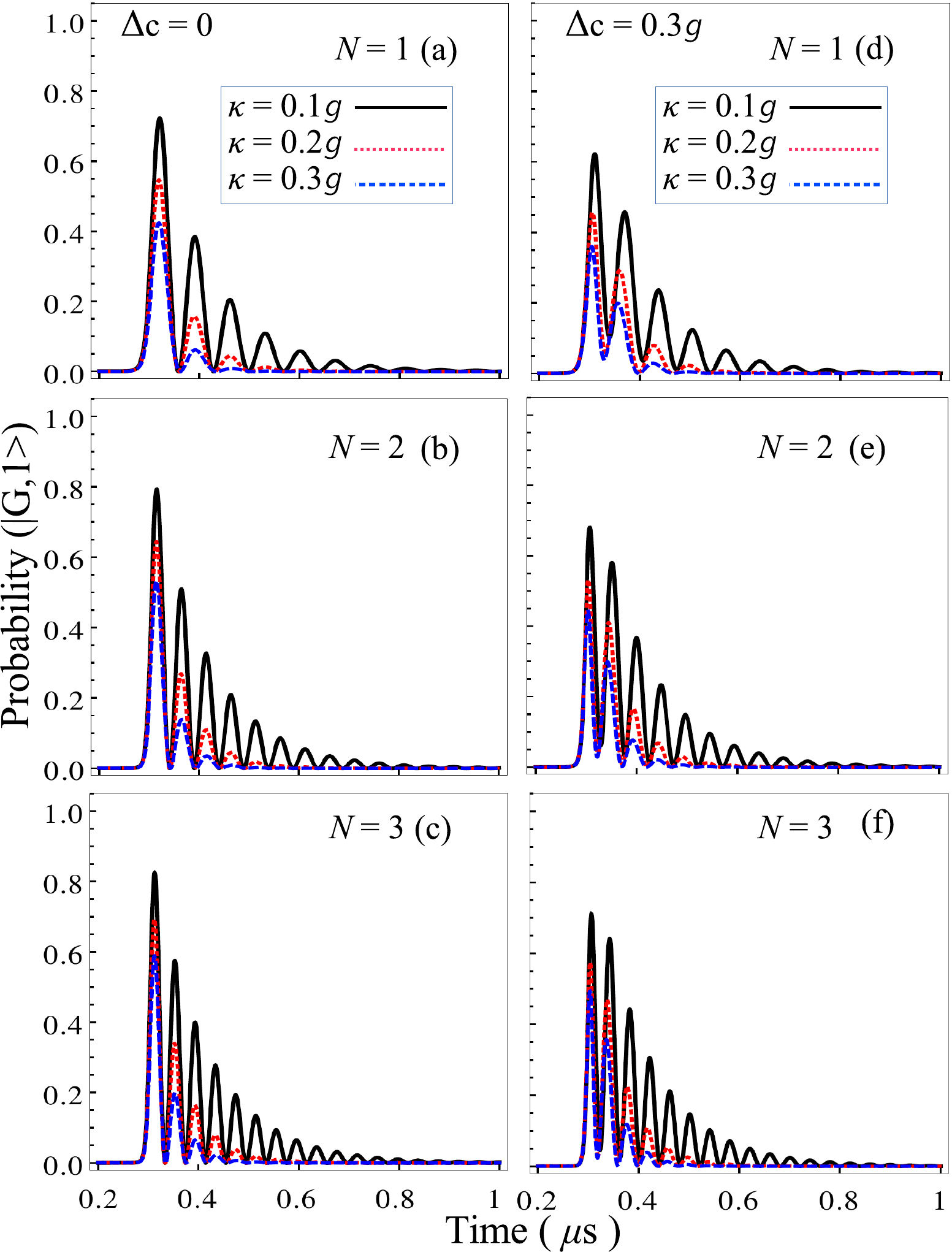}
\caption[short caption]{Time variation of Rabi oscillations for $N$ interacting atoms in the cavity for cavity detuning (a-c) $\Delta_c = 0$ and (d-f) $\Delta_{c} = 2\pi \times$ 4.3 MHz $ = 0.3 g$. Three different cavity decay rates are shown. The black solid curve corresponds to $\kappa = 2\pi \times$ 1.4 MHz $=  0.1 g$, the red dotted curve corresponds to $\kappa = 2\pi \times$ 2.85 MHz $= 0.2 g$, the blue dashed curve corresponds to $\kappa = 2\pi \times$ 4.3 MHz $= 0.3 g$, and $\Gamma_{\perp} = 0$. Other parameters are the same as in Fig.~\ref{fig:2}.
\label{fig:3}}
\end{figure}

Fig.~\ref{fig:3} shows the time variation of the Rabi oscillations for $N$ atoms in a resonant cavity, with $\Delta_c = 0$ [Fig.~\ref{fig:3}(a-c)], and $\Delta_c = 0.3 g$ [Fig.~\ref{fig:3}(d-f)], for different cavity decay rates $\kappa = 0.1, 0.2, 0.3 g$. As the cavity decay rate is increased, the Rabi oscillations are more quickly damped out \cite{Petrosyan06}. However, increasing the number of atoms enhances the frequency of the oscillations by the $\sqrt{N}$ factor as shown in Fig.~\ref{fig:3}(a-c). These oscillation frequencies are further enhanced by detuning the cavity \cite{PRA86_Agarwal} at the expense of a decreasing probability amplitude for photon production as shown in Fig.~\ref{fig:3}(d-f). These results are consistent with the JC model.

\begin{figure}[htbp]
\includegraphics[width=3in]{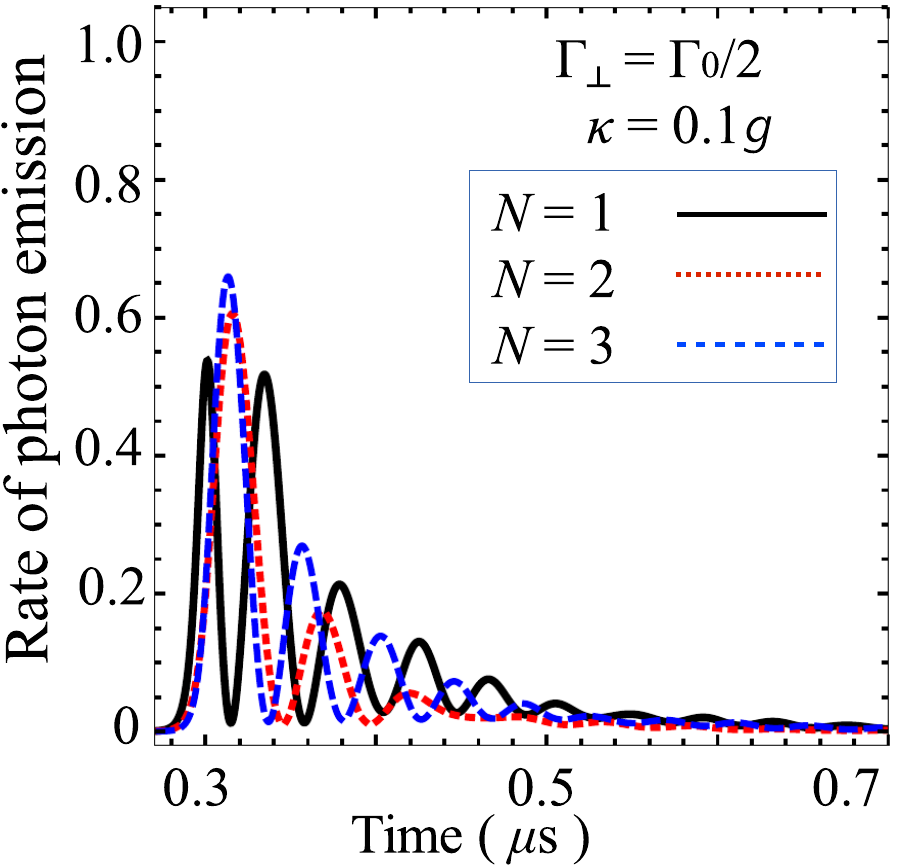}
\caption[short caption]{Rate of single photon emission efficiency for $N = 1-3$ atoms with one Rydberg excitation vs time. The black solid curve represents the photon emission with $N=1$, the red dotted curve corresponds to $N=2$ and the blue dashed curve corresponds to $N=3$. The cavity decay rate is $\kappa = 2\pi \times$ 1.4 MHz $=  0.1 g$ and $\Gamma_{\perp} =2\pi \times$ 3 MHz $= \Gamma_0/2$. Other parameters are the same as in Fig.~\ref{fig:2}. \label{fig:4}}
\end{figure}

The rate of single photon emission for a given number of atoms versus time is shown in Fig.~\ref{fig:4} with cavity decay rate $\kappa = 0.1 g$ and $\Gamma_{\perp}= \Gamma_0/2$. In this case, we observe oscillations in the rate of photon emission that change with $N$. The most interesting thing here is that the rate of photon emission is more strongly peaked at early times by increasing $N$. These results demonstrate that preparing the collective excitation helps to have the photon emitted at a specific time, due to an increase in the effective cooperativity from the collective state.

\begin{figure}[htbp]
\includegraphics[width=5.6in]{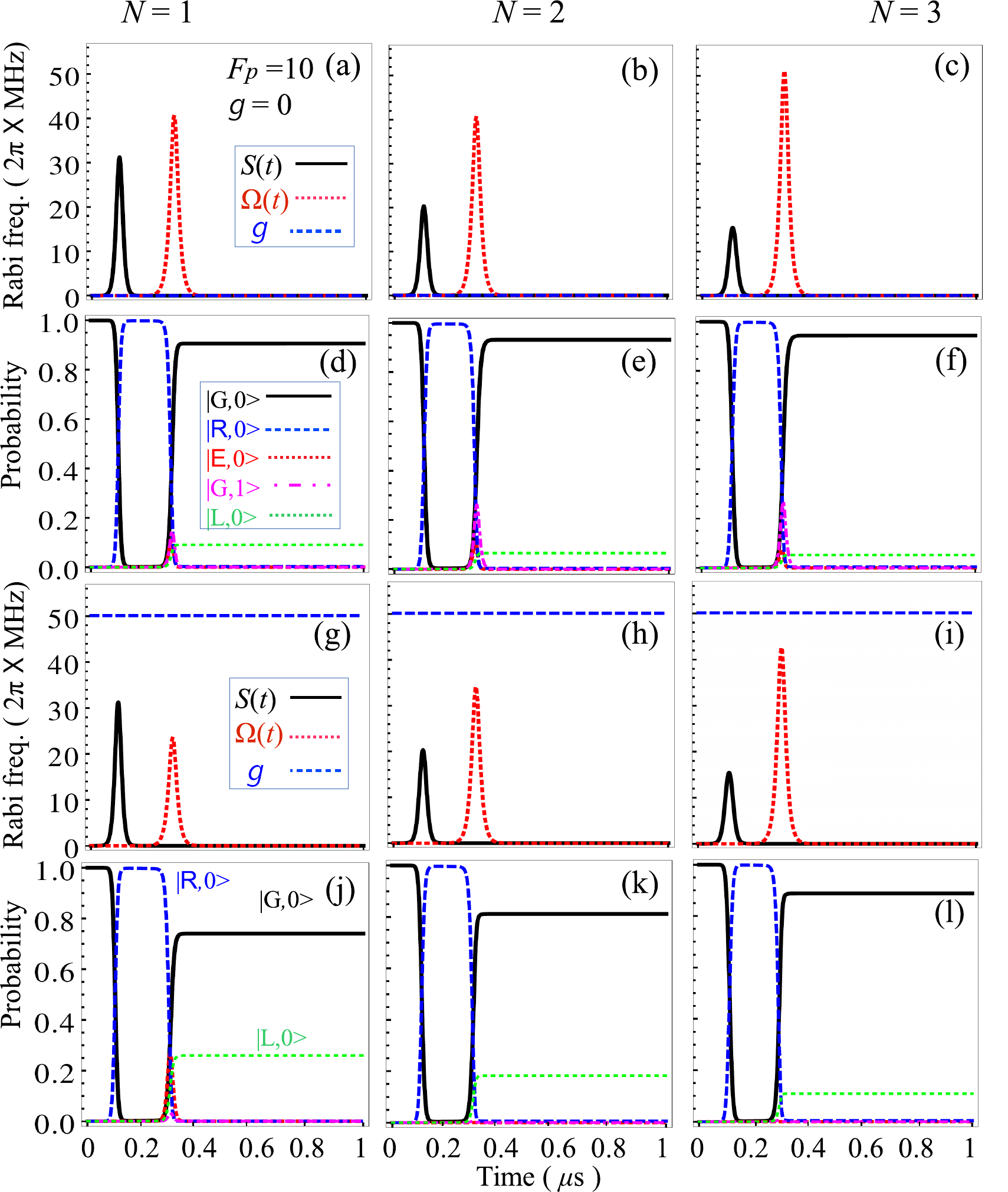}
\caption[short caption]{Panels (a-f) show the results with an incoherent decay in the Purcell regime. Panels (g-l) show the results with an enhanced spontaneous emission rate into the cavity mode in the weak cavity coupling regime. (a-c, g-i) The time variations of the Rabi frequencies of the laser pulses applied to a four-level system with $N$ interacting atoms. The effective Rabi frequency $S(t)$ is represented by the black solid curve. The red dotted curve shows the Rabi frequency of the control laser pulse $\Omega(t)$ and the blue dashed curve shows the cavity coupling strength $g$. Panels (d-f, j-l) show the corresponding probabilities for the population of states $\vert G,0 \rangle$ (black solid), $\vert E,0 \rangle$ (red dot), $\vert R,0 \rangle$ (blue dashed), $\vert G,1 \rangle$ (magenta dot-dashed) and $\vert L,0 \rangle$ (green dotted). The cavity decay rate is $\kappa = 2\pi \times$ 72.6 MHz. Other parameters are the same as in Fig.~\ref{fig:2}. 
\label{fig:5}}
\end{figure}

In Fig.~\ref{fig:5}, we show the results for $N = 1-3$ atoms for the full model and with an incoherent decay based on an enhanced spontaneous emission into the cavity mode as discussed in section IV. The time dependence and amplitudes of the Rabi frequencies for the incoherent decay associated with the Purcell effect are shown in Fig.~\ref{fig:5}(a-c), while the probabilities for various states are shown in Fig.~\ref{fig:5}(d-f). Rabi frequencies for the full model are shown in Fig.~\ref{fig:5}(g-i) with the associated probabilities for each state in Fig.~\ref{fig:5}(j-l). In both cases, we observe that the photon emission efficiency is enhanced with number of atoms because the cooperativity is effectively increasing due to the collectively enhanced $\mu$. The efficiency, for a given number of atoms, for an incoherent decay for the Purcell effect based calculation is greater than in the full model because there is significant transient population of the cavity excited state in the full calculation. Population of $\vert E,0 \rangle$ can spontaneously decay into a free space electromagnetic field mode, degrading the probability that the photon is emitted into the cavity mode.

\begin{figure}[htbp]
\setlength{\linewidth}{\textwidth}
\setlength{\hsize}{\textwidth}
\centering
\includegraphics[width=6.4in]{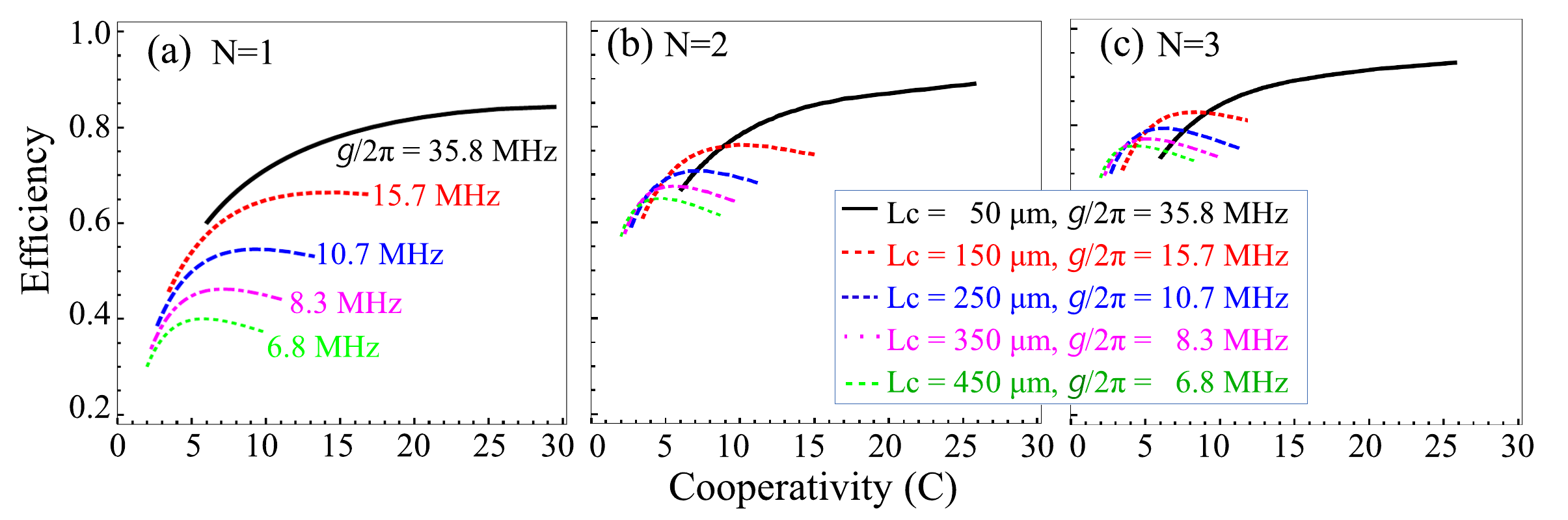}
\caption[short caption]{Photon emission efficiency as a function of the cooperativity, C, for $N$ interacting atoms with different cavity length, $L_c$. The mirror radius of curvature $R_c$ = 10 cm. $\kappa$ is decreasing to increase C for fixed $g$. Other parameters are the same as in Fig.~\ref{fig:2}.
 \label{fig:6}}
\end{figure}

In Fig.~\ref{fig:6}, the photon emission efficiency is plotted as a function of the cooperativity, C, for $N$ interacting atoms with different cavity length $L_c$. As we decrease $L_c$, the photon emission efficiency can be improved because $L_c$ increases the single atom $g$. The efficiency of photon emission also depends on the control laser pulse which transfers the Rydberg population into the cavity. For large cavity lengths, we can see that the photon emission efficiency first increases with C and then decreases. Although C is increasing, $\kappa$ is decreasing. The efficiency decreases because $\kappa$ becomes smaller compared to $\Gamma_{\perp}$. As we increase the number of atoms, the photon emission efficiency can be further improved by increasing the effective cooperativity but only when cavity loss beats out spontaneous emission. 

\begin{figure}[htbp]
\includegraphics[width=5.6in]{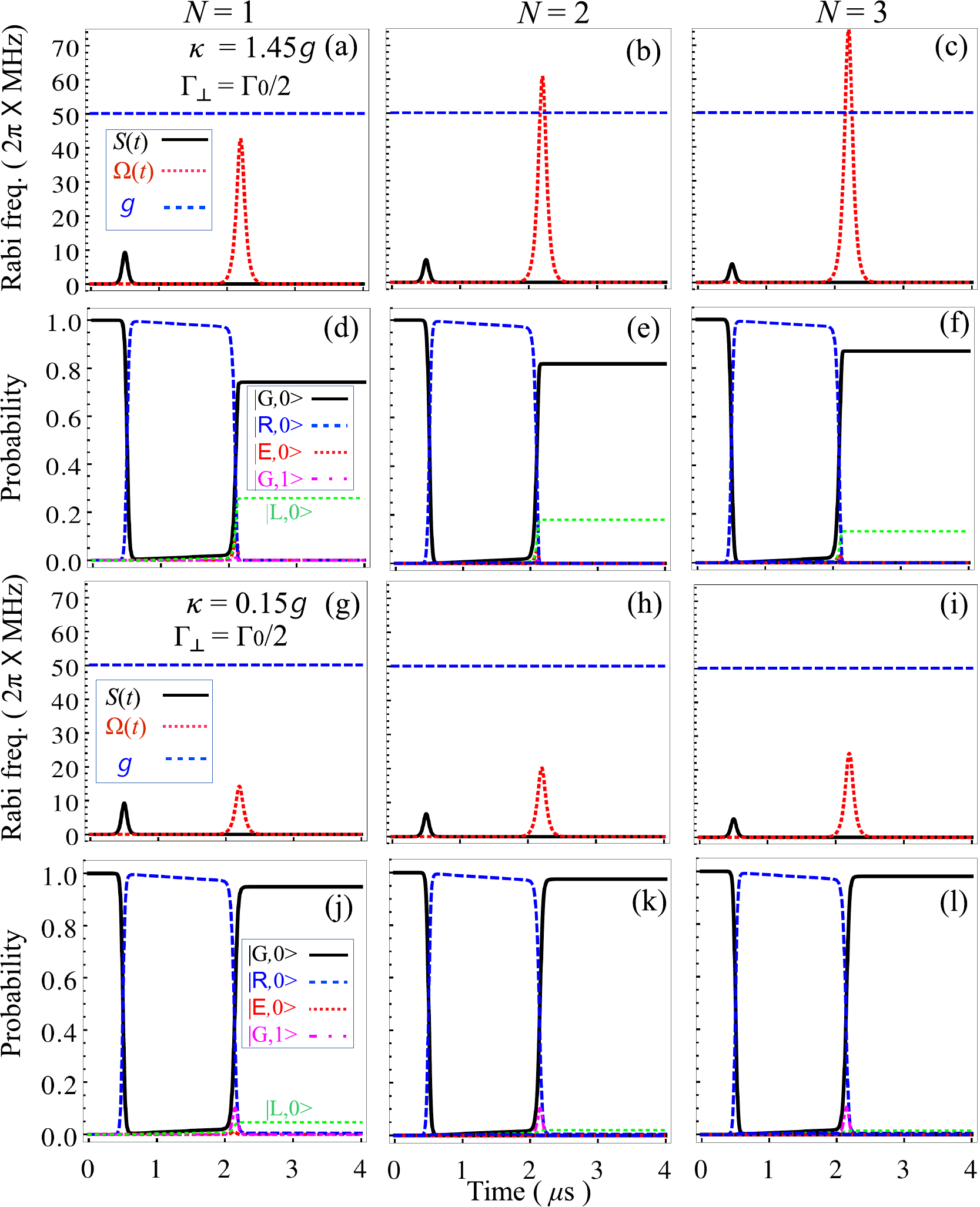}
\caption[short caption]{The time variations of the Rabi frequencies of the laser pulses applied to the four-level system for $N$ interacting atoms with (a-c)  $\kappa  = 2\pi \times$ 72.6 MHz $= 1.45 g$, and (g-i) $\kappa  = 2\pi \times$ 8 MHz $=  0.15 g$. The effective Rabi frequency $S(t)$ is represented by the black solid curve. The red dotted curve shows the Rabi frequency of the control laser pulse $\Omega(t)$ and the blue dashed curve shows the cavity coupling strength $g$. Panels (d-f, j-l) show the corresponding probabilities for the population of states $\vert G,0 \rangle$ (black solid), $\vert E,0 \rangle$ (red dotted), $\vert R,0 \rangle$ (blue dashed), $\vert G,1 \rangle$ (magenta dot-dashed) and $\vert L,0 \rangle$ (green dotted) with (d-f)  $\kappa = 1.45 g$, and (j-l) $\kappa = 0.15 g$. Other parameters are the same as in Fig.~\ref{fig:2}.
\label{fig:7}}
\end{figure}

In Fig.~\ref{fig:7}, we show the results for $N = 1-3$ atoms in the weak ($\kappa > g>\Gamma_{\perp}$) and strong ($ g>\kappa >\Gamma_{\perp}$) coupling regime. The time dependence and peak amplitudes of the Rabi frequencies for the weak coupling regime are shown in Fig.~\ref{fig:7}(a-c) and for the strong coupling regime in Fig.~\ref{fig:7}(g-i). The probabilities for various states are shown in Fig.~\ref{fig:7}(d-f) and Fig.~\ref{fig:7}(j-l). In Fig.~\ref{fig:7}(d-f), the Rabi oscillations between $\vert E,0 \rangle$ and $\vert G,1 \rangle$ cannot be observed due to the comparatively large cavity decay rate $\kappa$. $\vert G,1 \rangle$ decays and produces a photon that escapes the cavity. If we increase the value of the single atom cooperativity, for a given number of atoms, the population of the $\vert L,0 \rangle$ state decreases and the population of the cavity ground state is enhanced. If we increase the number of atoms, we observe that the ground state population is also increased. Thus, the increase in effective cooperativity resulting from an increase in the number of atoms plays a constructive role in enhancing the efficiency of photon production.
In Fig.~\ref{fig:7}(j-l), the Rabi oscillations between $\vert E,0 \rangle$ and $\vert G,1 \rangle$ cannot be observed because there is a coherent population transfer from the Rydberg state $\vert R,0 \rangle$ to the cavity ground state $\vert G, 1 \rangle$. The transient population in $\vert E,0 \rangle$ is approximately zero. This reduces the loss due to spontaneous emission into the dummy state $\vert L,0 \rangle$ and improves the photon emission efficiency into the cavity mode \cite{Law97,Khun10}.

\begin{figure}[htbp]
\includegraphics[width=6in]{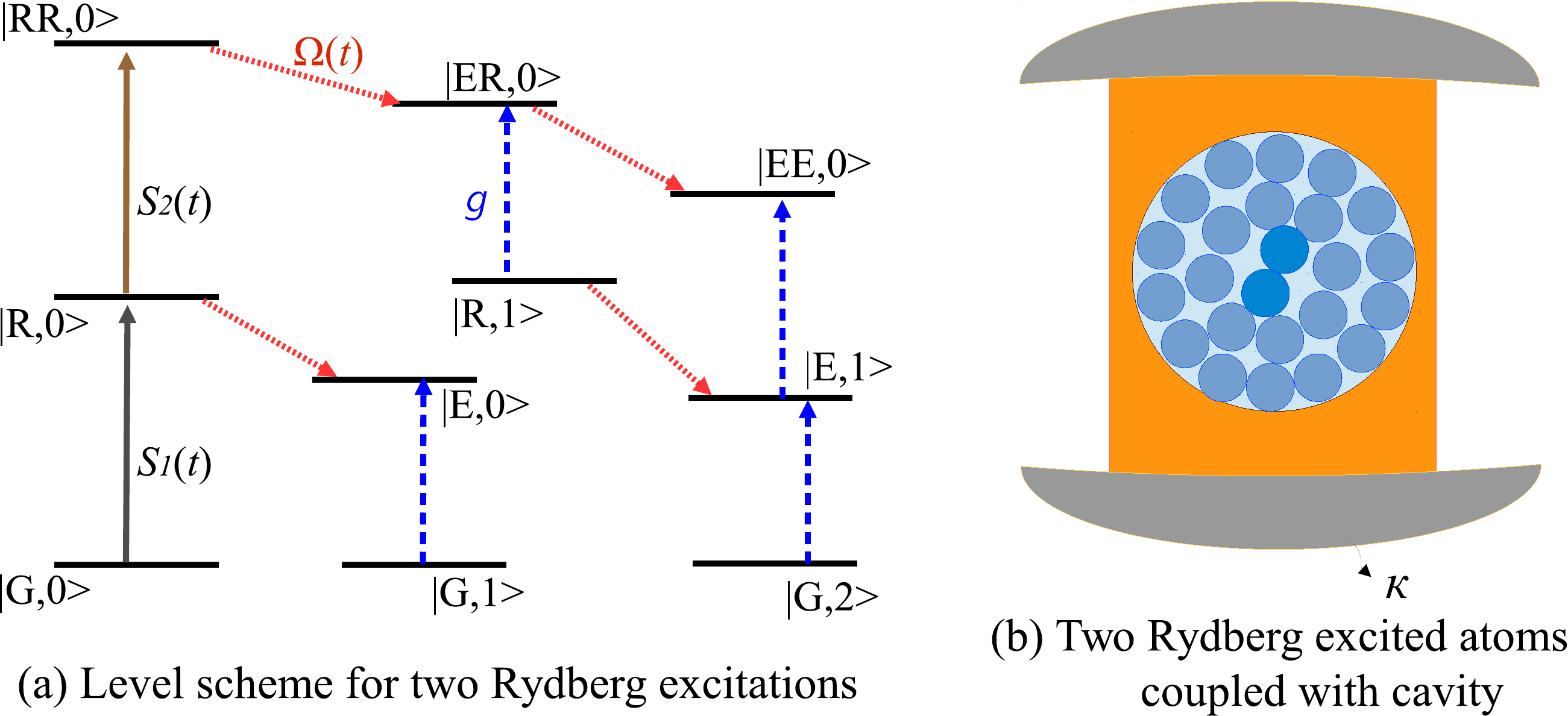}
\caption[short caption]{(a) Level scheme for two Rydberg excitations with $N$ interacting atoms coupled to a cavity with coupling strength $g$ [blue dashed line]. $S_1(t)$ [black solid line] and $S_2(t)$ [brown solid line] are the effective Rabi frequencies of the laser pulses for one and two Rydberg excitations, respectively. $\Omega(t)$ [red dotted line] is the Rabi frequency of the control laser pulse. (b) Pictorial view of two Rydberg excited atoms coupled with the cavity. The excitation is effectively a macrodimer state.\label{fig:8}}
\end{figure}

We can also investigate the more complicated situation where two Rydberg atoms are excited within the blockade radius. The level scheme for this case is shown in Fig.~\ref{fig:8}. Using a sequence of pulses with effective Rabi frequencies $S_1(t)$ and $S_2(t)$ while changing the one photon Rydberg level detuning $\Delta(t)$ to compensate the blockade shift $\Delta_R$, we can excite two Rydberg atoms into the state $\vert RR,0 \rangle$ \cite{PRL12_Cross,PRA_Rolston10}. Then we have two channels: (1) $\vert RR,0 \rangle \rightarrow \vert ER,0 \rangle \rightarrow \vert EE,0 \rangle \rightarrow \vert E,1 \rangle \rightarrow \vert G,2 \rangle$ and (2) $\vert RR,0 \rangle \rightarrow \vert ER,0 \rangle \rightarrow \vert R,1 \rangle \rightarrow \vert E,1 \rangle \rightarrow \vert G,2 \rangle$ by which $ \vert G,2 \rangle$ can be produced. By applying a control laser pulse with Rabi frequency $\Omega(t)$ and the cavity coupling strength $g$, we can observe emission of two photons from different atomic dipoles out of the cavity mode. By using Eq.~\ref{evolution}, we can calculate the photon correlations.

\begin{figure}[htbp]
\setlength{\linewidth}{\textwidth}
\setlength{\hsize}{\textwidth}
\centering
\includegraphics[width=6in]{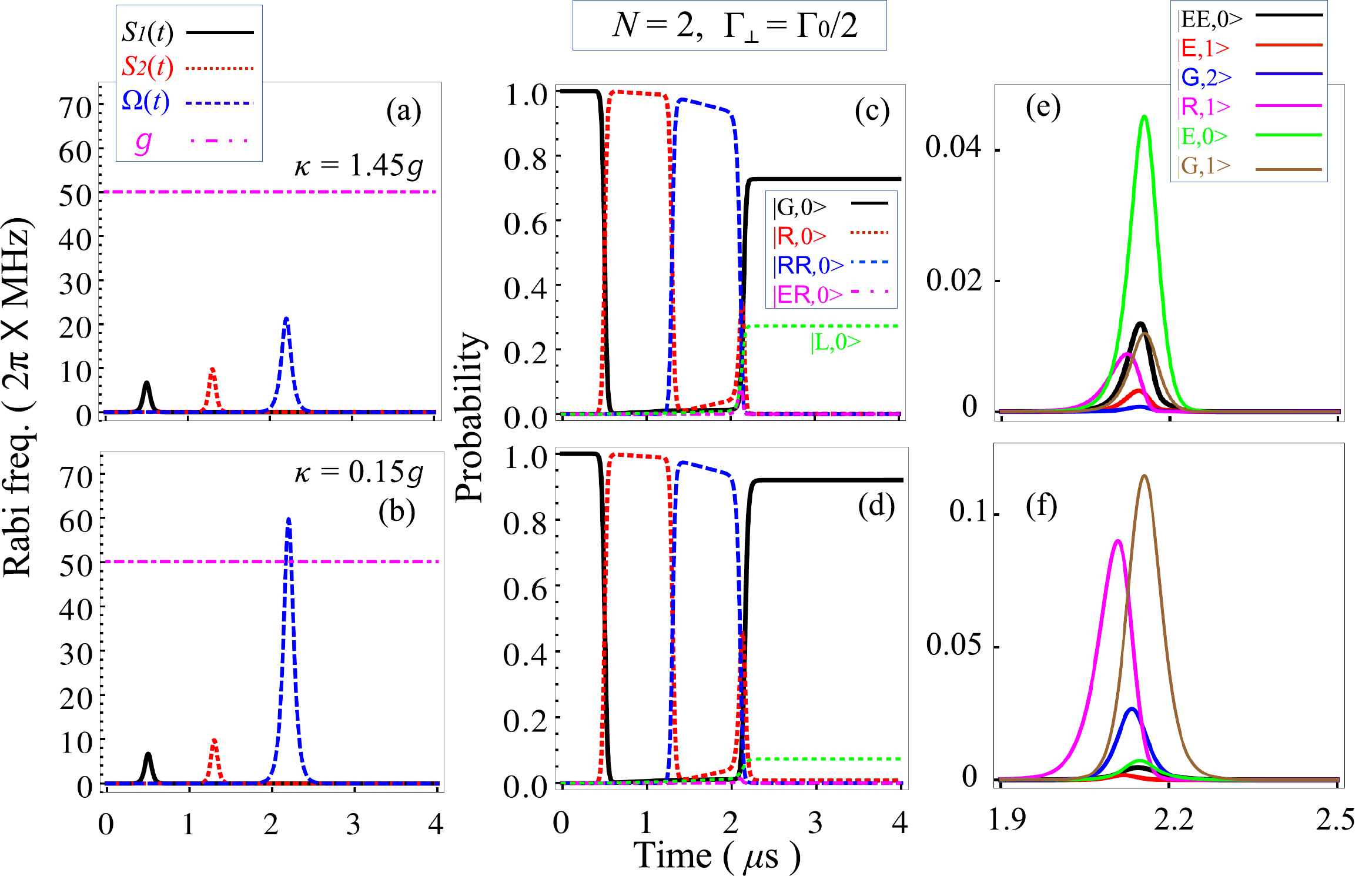}
\caption[short caption]{Coherent manipulation of two interacting Rydberg atoms with two excitations. (a, b) The time variations of the Rabi frequencies of the laser pulses applied to the four-level system for $N = 2$ interacting atoms with (a)  $\kappa = 2\pi \times$ 72.6 MHz $= 1.45 g$, and (b) $\kappa = 2\pi \times$ 8 MHz $= 0.15 g$. Rabi frequencies $S_{1}(t)$ and $S_{2}(t)$ are represented by the black solid and red dotted curves, respectively, and generate the first and second Rydberg excitations, respectively. The blue dashed curve shows $\Omega(t)$ and the magenta dot-dashed curve shows $g$. (c, d) Probability of finding atoms in $\vert G,0 \rangle$ (black solid), $\vert R,0 \rangle$ (red dotted), $\vert RR,0 \rangle$ (blue dashed), $\vert ER,0 \rangle$ (magenta dot-dashed) and $\vert L,0 \rangle$ (green dotted). (e, f) Probability of finding atoms in $\vert EE,0 \rangle$ (black line), $\vert E,1 \rangle$ (red line), $\vert G,2 \rangle$ (blue line), $\vert R,1\rangle$ (magenta line), $\vert E,0 \rangle$ (green line), $\vert G,1 \rangle$ (brown line). Parameters for the calculation are $\Gamma_r = 2 \pi \times 1.4 \,$kHz, $\Gamma_{\perp} = 2\pi \times 3$ MHz $ = \Gamma_0/2$, $\Delta_c = 0$, $\Delta_R = 2\pi \times 220\,$MHz, $\Delta_s = 2\pi \times 110\,$MHz and $\Delta(t)$ changes with time from $-2\pi \times 110$ to $55$ MHz.
 \label{fig:9}}
\end{figure}

In Fig.~\ref{fig:9}(a, b), we show the Rabi frequencies versus time for the system of $N = 2$ interacting atoms. Rabi frequency $S_{1}(t)$ drives the first excitation while $S_{2}(t)$ drives the second excitation. The cavity decay rates are $\kappa = 1.45 g$ and $\kappa = 0.15 g$ in Fig.~\ref{fig:9}(a, b) respectively. In Fig.~\ref{fig:9}(c, d), we can see that a 1 Rydberg excitation state is generated by using an effective Rabi pulse $S_1(t)$. Subsequently, by applying another effective Rabi pulse $S_2(t)$, we can generate a 2 Rydberg excitation state. This state is similar to a Rydberg atom macrodimer \cite{NatPhy09_Richard}. A single off-resonant excitation of the macrodimer can also be used, equivalent to a photoassociation of the molecule \cite{SchwettJMODOPT}. By using control laser pulse $\Omega(t)$ with cavity coupling strength $g$, we observe that more transient population builds up in $\vert E,0 \rangle$ and $\vert EE,0 \rangle$ compared to $\vert R,1 \rangle$ and $\vert G,1 \rangle$. This situation increases the spontaneous emission from $\vert E,0 \rangle$ and $\vert EE,0 \rangle$ to the dummy state $\vert L,0\rangle$. The photon emission efficiency in the weak cavity coupling regime is reduced due to the spontaneous emission. If we increase the cooperativity the photon emission efficiency can be improved. Fig.~\ref{fig:9}(d, f) corresponds to the strong coupling regime. In this case, we can populate the 2 Rydberg excitation state and observe photons emitted from the cavity. We see that the photons are quickly emitted from the cavity and photon emission is improved because the transient population of $\vert E,0 \rangle$ and $\vert EE,0 \rangle$ is negligible compare to $\vert R,1 \rangle$ and $\vert G,1 \rangle$. The population is coherently transferred from the Rydberg state to the cavity ground state, similar to the single excitation case.

\begin{figure}[htbp]
\setlength{\linewidth}{\textwidth}
\setlength{\hsize}{\textwidth}
\centering
\includegraphics[width=6in]{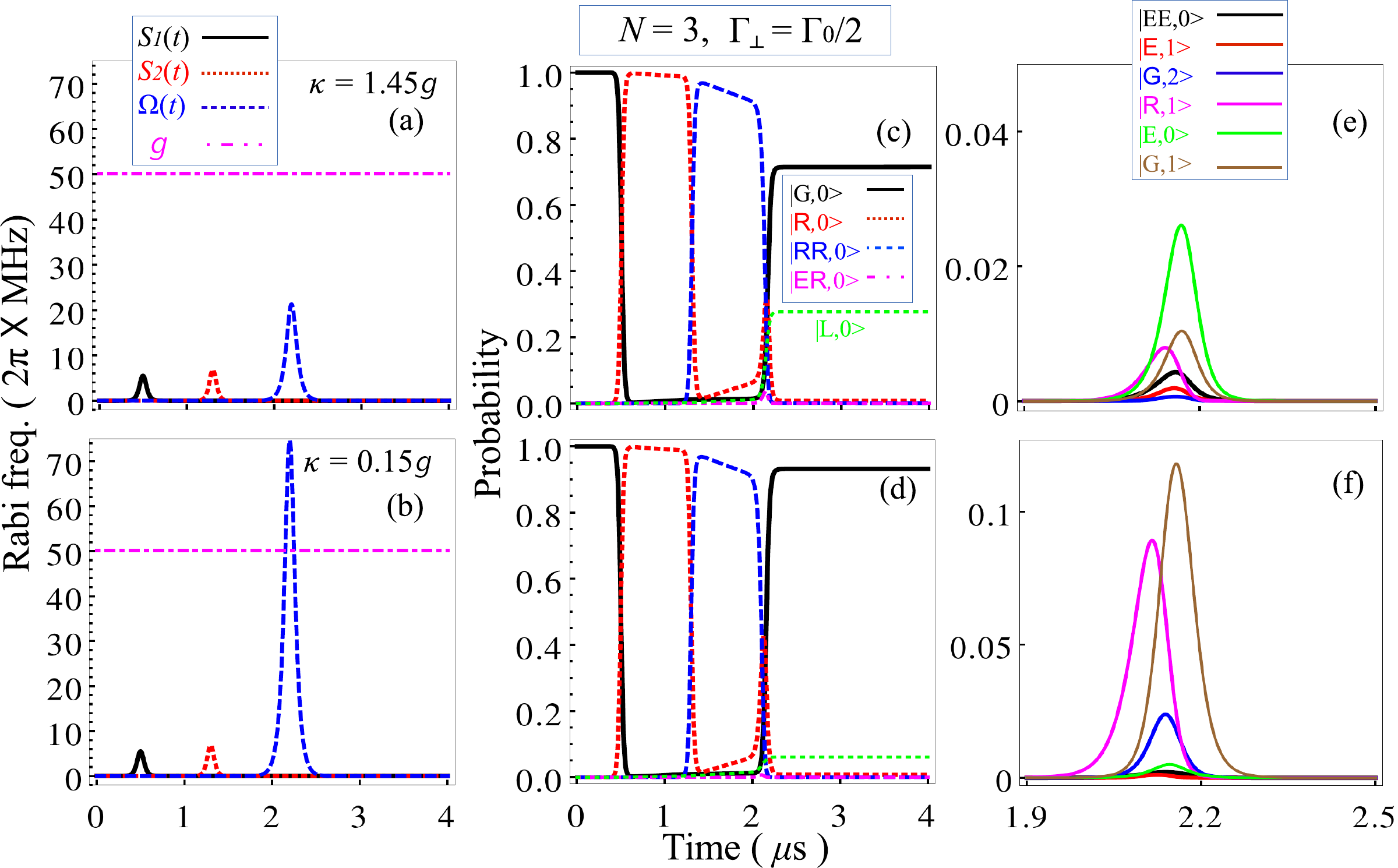}
\caption[short caption]{Coherent manipulation of three interacting Rydberg atoms with two excitations. (a, b) The time variations of the Rabi frequencies of the laser pulses applied to the four-level system for $N = 3$ interacting atoms with (a)  $\kappa = 2\pi \times$ 72.6 MHz $= 1.45 g$, and (b) $\kappa = 2\pi \times$ 8 MHz $= 0.15 g$. Rabi frequencies $S_{1}(t)$ and $S_{2}(t)$ are represented by the black solid and red dotted curves, respectively, and generate the first and second Rydberg excitations, respectively. The blue dashed curve shows $\Omega(t)$ and the magenta dot-dashed curve shows $g$. (c,d) Probability of finding atoms in $\vert G,0 \rangle$ (black solid), $\vert R,0 \rangle$ (red dotted), $\vert RR,0 \rangle$ (blue dashed), $\vert ER,0 \rangle$ (magenta dot-dashed) and $\vert L,0 \rangle$ (green dotted). (e,f) Probability of finding atoms in $\vert EE,0 \rangle$ (black line), $\vert E,1 \rangle$ (red line), $\vert G,2 \rangle$ (blue line), $\vert R,1\rangle$ (magenta line), $\vert E,0 \rangle$ (green line), $\vert G,1 \rangle$ (brown line). Other parameters are the same as in Fig.~\ref{fig:9}.
 \label{fig:10}}
\end{figure}

In Fig.~\ref{fig:10}, for $N=3$ interacting atoms, the detunings and blockade shift are the same as in Fig.~\ref{fig:9}. Similar to Fig.~\ref{fig:9}, the photons are quickly emitted from the cavity and some population decays into the dummy state $\vert L,0 \rangle$. The photon emission probability is larger in this case because the effective cooperativity is larger. Although small, as we increase the number of atoms the photon emission probability is enhanced, due to the effective increase in $g$.

\begin{figure}[htbp]
\setlength{\linewidth}{\textwidth}
\setlength{\hsize}{\textwidth}
\centering
\includegraphics[width=4.8in]{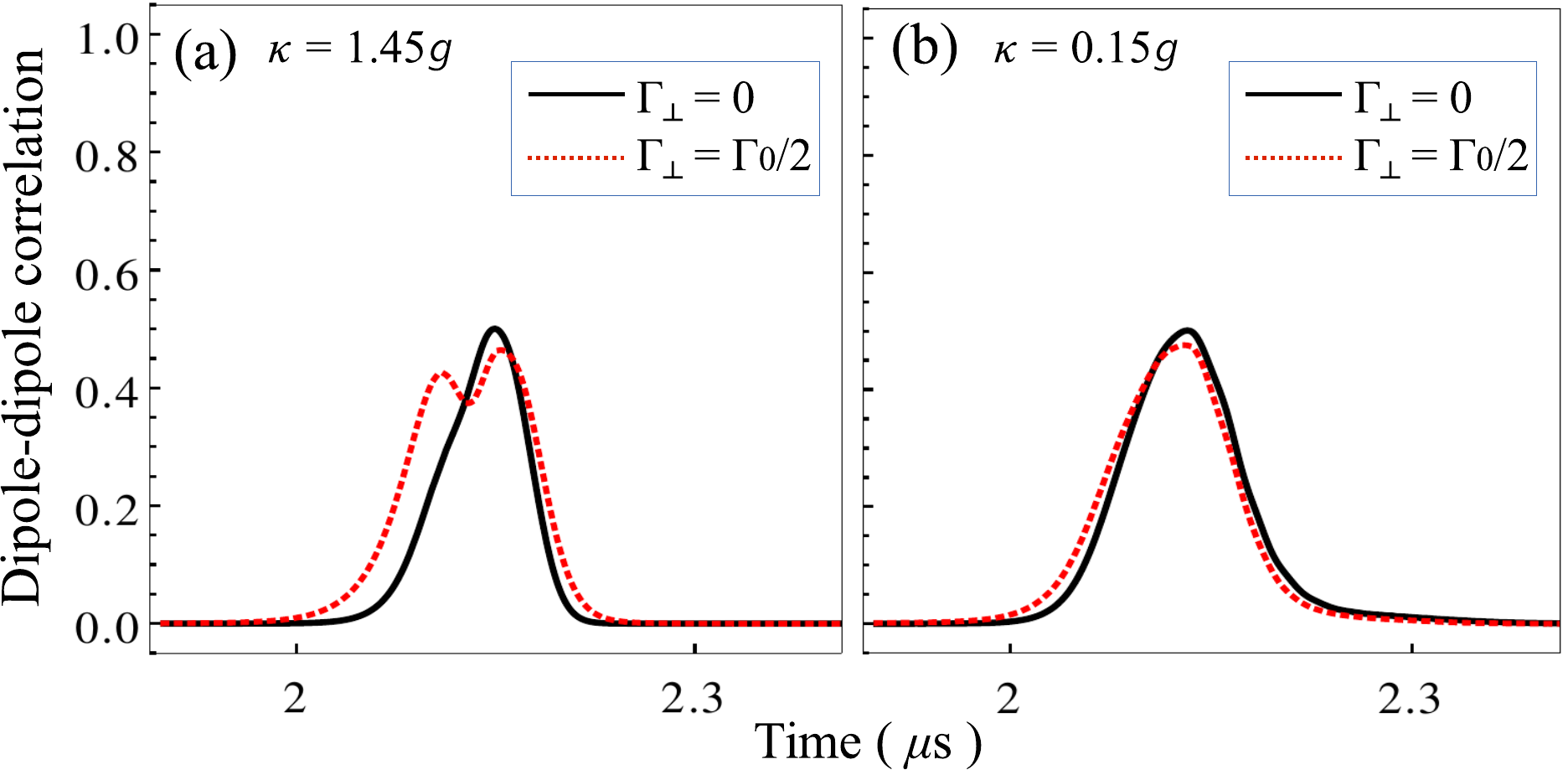}
\caption[short caption]{Expectation value of the dipole-dipole correlation versus time for the case of 2 interacting atoms with (a)  $\kappa = 2\pi \times$ 72.6 MHz $= 1.45 g$, and (b) $\kappa = 2\pi \times$ 8 MHz $= 0.15 g$. Other parameters are the same as in Fig.~\ref{fig:9}.
 \label{fig:11}}
\end{figure}

The expectation value of the dipole-dipole correlation for the case of two Rydberg excited atoms is shown in Fig.~\ref{fig:11}.
This quantity is given by
\begin{equation}
\langle \mu_{ij} \rangle =  \langle \Psi \vert D^{+}_i D^{-}_j \vert \Psi \rangle
\end{equation}
and measures the cross-correlation existing between the dipoles belonging to the two atoms \cite{Grynberg82}. $D^{\pm}_i$ are the raising and lowering operators for the cavity transition which can be defined as $D^{+}_i \vert g \rangle_i = \vert e \rangle_i$ and $D^{-}_i \vert e \rangle_i = \vert g \rangle_i$, where subscript $i$ represents the $i^{th}$ atom. In Fig.~\ref{fig:11}(a,b), the normalized dipole-dipole correlation function as a function of time is shown. We see that the dipoles are correlated. The laser pulses and cavity coupling time dependencies are the same as shown in Fig.~\ref{fig:9}(a,b). In Fig.~\ref{fig:11}(a), with spontaneous emission, the prominent dip observed is due to the transient population in the cavity excited state that results from spontaneous emission outside the cavity. 

\begin{figure}[htbp]
\setlength{\linewidth}{\textwidth}
\setlength{\hsize}{\textwidth}
\centering
\includegraphics[width=4.6in]{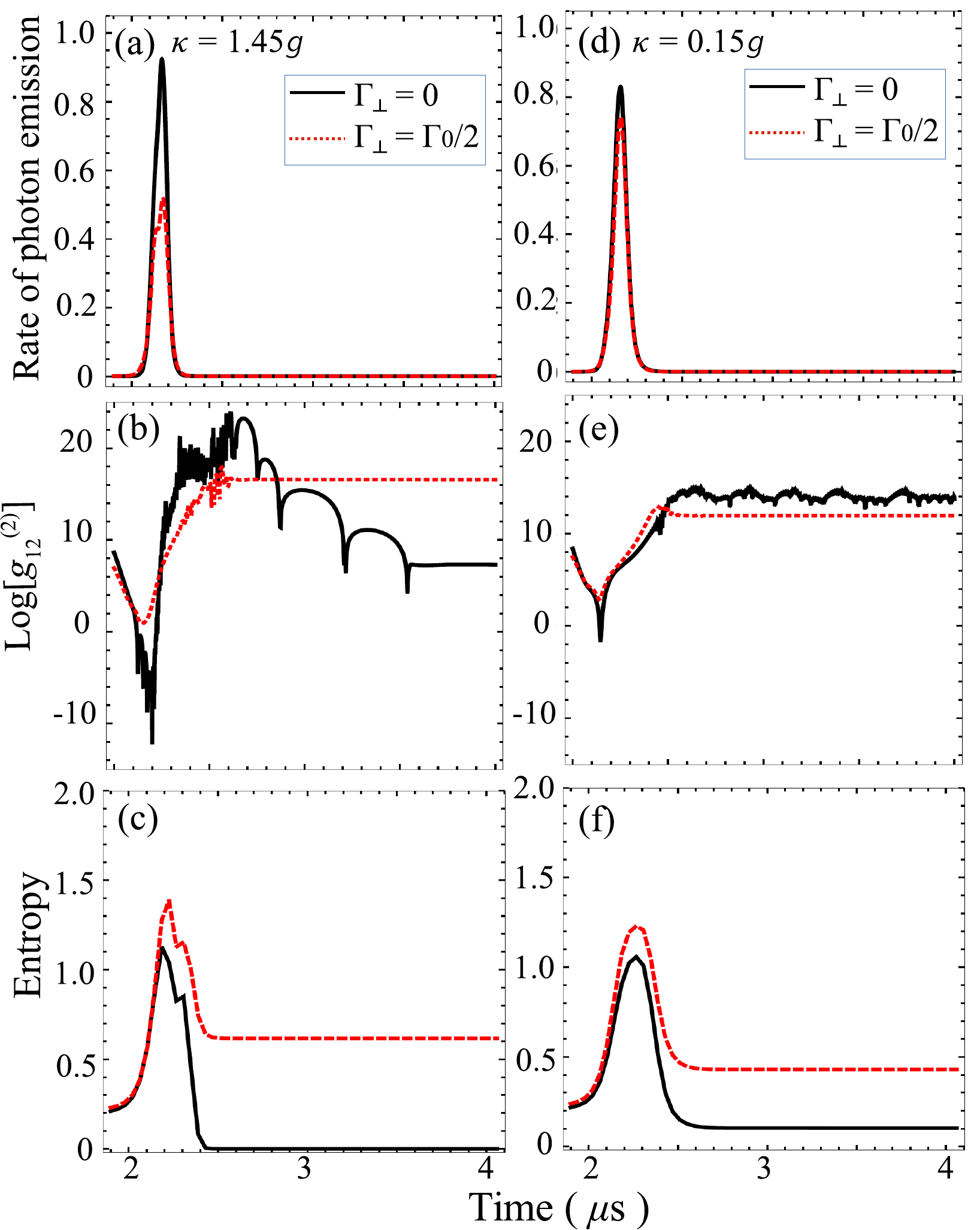}
\caption[short caption]{(a,b) The rate of photon emission efficiency for 2 interacting atoms with two Rydberg excitations. (c,d) The logarithm of the second order correlation function Log[g$^{(2)}_{12}$] of photons emitted by two different atoms with two Rydberg excitations versus time. (e,f) Variation of total entropy of the system with time. (a,c,e) $\kappa = 2\pi \times$ 72.6 MHz $= 1.45 g$ and (b,d,f) $\kappa = 2\pi \times$ 8 MHz $= 0.15 g$. Other parameters are the same as in Fig.~\ref{fig:9}.
 \label{fig:12}}
\end{figure}

Fig.~\ref{fig:12}(a,b) shows the rate of photon emission efficiency for two interacting Rydberg atoms with two excitations for cavity decay rates $\kappa = 1.45 g$ and $\kappa = 0.15 g$ respectively. In Fig.~\ref{fig:12}(a,b), the weak cavity coupling case has larger photon emission amplitude than the strong coupling regime. With the addition of spontaneous emission, the rate of photon emission decreases significantly in the weak cavity coupling case, again due to the transient population in the cavity excited state. 

The second order correlation function of the photons emitted by the two different atoms is given by
\begin{equation}
 \mathrm{g}^{(2)}_{ij}  =  \frac{\langle \Psi \vert D^{+}_i D^{-}_iD^{+}_j D^{-}_j \vert \Psi\rangle}{\langle \Psi \vert D^{+}_i D^{-}_i \vert \Psi\rangle\langle \Psi \vert D^{+}_j D^{-}_j \vert \Psi\rangle},
\end{equation}
where subscripts $i, j$ represent the $i^{th}$ and $j^{th}$ atoms respectively. If Log[g$^{(2)}_{ij}] > 0$, radiated photons tend to be bunched while 
if Log[g$^{(2)}_{ij}] < 0$, the photons are anti-bunched. This function is plotted in Fig.~\ref{fig:12}(a,b) with and without spontaneous emission as a function of time for $\kappa = 1.45 g$ and $\kappa = 0.15 g$. In these cases, we observe that Log[g$^{(2)}_{ij}$] is smaller than zero for short times when spontaneous emission is neglected. This indicates that the photons are anti-bunched at short times when the photons are preferentially emitted. The value of Log[g$^{(2)}_{ij}$] is greater than zero for longer times, indicating the emission of photons is bunched after the initial photon pulse. If we consider the spontaneous emission into the dummy state $\vert L, 0 \rangle$ then Log[g$^{(2)}_{ij}$] is always greater than zero. Log[g$^{(2)}_{ij}] > 0$ implies the emission of photons is always bunched for the more realistic case. The additional dephasing destroys the interferences of the 2 oscillating dipoles in this case.

To define the entanglement and purity of the quantum systems, there are several measures. One of them is the von Neumann entropy of the quantum system. This is a very sensitive measure of the purity of the quantum system and it determines the degree of correlation between the collective excitations and photon states \cite{Phoenix1988,PRA91_Knight,JPB04_Johnson}. The von Neumann entropy of the atoms-cavity system can be defined as
\begin{eqnarray}
S (\rho) = -\sum_{j} \eta_j \mathrm{ln}  \eta_j,
\label{entropy}
\end{eqnarray}
with density operator
\begin{eqnarray}
\rho = -\sum_{j} \eta_j \vert j\rangle \langle j \vert,
\label{entropy1}
\end{eqnarray}
where the $\eta_j$'s are the diagonalized eigenvalues of the system and $j$ is the number of states in the Hilbert space. Since our system is multi-partite, and there is no general measure to study the entanglement for multi-partite system, we approximate our system as a bipartite system of atoms and photons. Then, the sum of the entropies of the 2 sub-systems atoms, $S (\rho_a)$, and photons, $S (\rho_p)$, should be larger than the total entropy of the system, $S (\rho_a) + S (\rho_p) \geq S (\rho)$. The bound for  the entropy for these two interacting sub-systems can be defined by the Araki and Lieb \cite{Araki1970} triangle inequality,
 \begin{eqnarray}
| S (\rho_a) - S (\rho_p) | \leq S (\rho) \leq  | S (\rho_a) + S (\rho_p) |.
\label{entropybound}
\end{eqnarray}

The total entropy of the system is plotted in Fig.~\ref{fig:12} taking into account and also neglecting spontaneous emission as a function of time. Fig.~\ref{fig:12}(e,f) shows the variation of total entropy with cavity decay rates $\kappa = 1.45 g$ and $\kappa = 0.15 g$. As photons are emitted from the cavity, the entropy of the system first increases but then decreases due to the loss of the photons. In Fig.~\ref{fig:12}(e,f), the entropy of the system is larger for the realistic case where spontaneous emission out of the cavity is considered. In both cases, after the complete emission of the photons, the entropy of the photonic sub-system has reached zero with time and system is in a pure entangled atomic state \citep{PRL_90_Julio,PRA99_Knight}.

\section{Discussion}

In the literature, there are several different ways to generate single photons \cite{PRA13_Rempe,PRA13_Pfau,Nat00_Moerner,AIP11_Migdall, Nat14_Ohshima}. One way is to use four-wave mixing \cite{PRA08_Walker}. The approach we have investigated in this work is superficially similar to the four-wave mixing method because of the atomic level scheme. One fundamental difference is that the method described here does not require phase matching. In the four-wave mixing method, the phase matching condition needs to be fulfilled to observe the photon which will be emitted in a diffraction-limited solid angle. The approach described in this paper uses an alternative way to generate single photons by preparing a single collective state using Rydberg atom blockade in a cavity. The lasers are time dependent in such a way that the preparation, and control are not present at the same time. Technologically, Rydberg blockade allows one to deterministically control the number of excitations in the cavity and the cross-section for coupling to the cavity is enhanced by the collective nature of the initial state. This addresses some of the problems inherent in trapping single atoms in an optical cavity to generate single photons. The diamond configuration enables one to isolate the quantum state preparation from the single photon output. The output is in a single mode determined by the properties of the optical cavity.

A similar atomic level scheme has been studied for collective excitation for Raman and superradiant lasers \cite{EPL92_Zoller,PRL93_Haake,PRA96_Haake}. In these systems, $N$ atoms are excited collectively without considering atom-atom interactions. Our model shares the structure of the excitation scheme, but Rydberg blockade prepares an $N$ atom entangled state with a controlled number of excitations or controlled number of singly excited superatoms. In a more general case, spatial correlations can be written into the sample. This is quite different from these laser schemes as our approach produces a state similar to a micromaser but with deterministic control over the number of excitations in the cavity. The system we described can be useful for studying  superradiant behavior because a collective state can be deterministically prepared as the initial state. The cavity coupling can be used to analyze the collective quantum state by detecting the emitted radiation \cite{Grynberg82}.

Most of the models discussed previously for Rydberg atoms in a cavity use ground state atoms as an initial state. The JC model for the atoms initially in the ground state in an optical cavity with a single Rydberg excitation as an essential piece of the dynamics has been proposed in Ref. \cite{PRA10_Molmer}. This model is different from our approach. In our model, we first excite a single collective state by using the Rydberg blockade mechanism and then by manipulating a control laser field, generate photons from the cavity. In addition, the preparation lasers are all far detuned from the cavity field frequency. Now, if we consider the Rydberg excited state as an initial state of the collective atoms-cavity system, then the photon emission efficiency can be increased due to the collectively enhanced cooperativity and can be further improved once the strong coupling regime is achieved due to a negligible transient population in the cavity excited state \cite{Grynberg82}.

We have neglected several dephasing processes in our analysis because we anticipate these effects are small, or can be made small with a judicious choice of parameters. The effects that are of most concern are the spatial dependence of the atoms-cavity coupling, dependence of the Rydberg atom interactions on internuclear separation, motion of the atoms, and interaction between Rydberg excitations when there are 2 excitations simultaneously in the cavity. Taking these effects into account with the straightforward Liouville equations we have used in our calculations is involved, so we estimate their effect on our current results. Light pressure forces on the atoms in the cavity are negligible because we have assumed the cavity is on resonance with the $5^2S_{1/2} \leftrightarrow5^2P_{3/2}$ transition for times after the atoms are excited to the $5^2P_{3/2}$ state, the exception being Fig.~\ref{fig:3} where we show a detuned cavity to demonstrate that our solution is producing expected results. There is no asymmetry in the absorption of the photons in the cavity under these conditions and the photon recoil is small. The recoil temperature of $\,^{87}$Rb is $362\,$nK. Additionally, for a large fraction of the time the atoms are interacting with the cavity, there are no photons inside it.

We have run our calculations with additional dephasing rates by coupling the singly and doubly excited Rydberg states to the dummy state. The results of these calculations with additional dephasing provide an estimate of the effects of dephasing on the collective excitations. In these calculations, we used the same rates for each transition. Both processes were modeled with the Lindblad operator.  We examined additional dephasing rates up to $50\,$kHz. The results for the photon production probability are reduced by around $\sim 4\%$ for $50\,$kHz of dephasing on each transition for the excitation timescales used in this work in the 2 excitation case. The reduction in the overall photon production scales linearly as these dephasing rates are increased.

Practically, ultracold samples at temperatures $\sim100\,$nK can be generated and used for an experiment to reduce the decoherence resulting from thermal motion, although photon recoil can limit the temperature to $362\,$nK. At temperatures of $\sim 1\,\mu$K, the atoms only move $\sim 100\,$nm in $10\,\mu$s. Because this is a small fraction of a blockade radius, free evolution of the atoms can be neglected at $\sim 1\,\mu$K. The corresponding Doppler shifts of the atoms at $\sim 1\,\mu$K are $\sim 2 \pi \times 10\,$kHz. Doppler shifts on this order can lead to reductions in the photon production rate of $\sim 1\%$. The laser spectral bandwidths and the inverse of the times over which the collective states freely evolve are broader than the Doppler widths of the atoms that make up the collective excitations. In our case, for very low, but experimentally achievable temperatures, the frozen gas approximation is valid to around $1\%$ in the calculated photon production probability.

In general, multiple excitation states within the blockade volume and states consisting of more than 1 superatom will have spatial correlations \cite{PRL_11_Raithel,Nat12_Schausz,PRL14_Arimondo,Urvoy15,PRL_12_Schempp} because the interaction between the Rydberg atoms depends on their internuclear separation and the orientation of the internuclear axis relative to any fields. However, as mentioned earlier in the paper, 2 Rydberg atoms can be excited around an avoided crossing where there exists a broad region where the energy has weak dependence on the internuclear separation. These types of potentials can be observed when the multi-level nature of the Rydberg atom interactions and external magnetic and electric fields are considered. In fact, the flat regions of the Rydberg potentials, so called `stationary regions', are regions where Rydberg atom pair excitations occur most readily. The acceleration of the atoms after excitation is determined by the gradient of the interaction potential and has been observed experimentally \cite{Schwettmann2006,NatPhy09_Richard,PRA07_Richard}. The acceleration on these types of interaction potentials can be negligible as demonstrated in these works - in these experiments, we had to wait $\sim 200 \,\mu$s to observe a difference between a thermally expanding cloud of cold atoms at $80\,\mu$K and the Rydberg atom collision partners \cite{PRA07_Richard}. Scaling this time to a $1\,\mu$K temperature gives a time of $\sim 20\,\mu$s. Similarly, if we assume the atoms constantly accelerate to the final velocity measured in \cite{NatPhy09_Richard,PRA07_Richard}, $17\,$cm$\,$s$^{-1}$, in $\sim 100\,\mu$s, it takes $\sim 5\,\mu$s to reach a thermal velocity that corresponds to $1\,\mu$K. The time over which the 2 Rydberg excitations exist in our case is only $\sim 1\,\mu$s, so this effect can be negligible with an optimized choice of interacting Rydberg state, or use of electric and magnetic fields to engineer the interaction potentials \cite{NatPhy09_Richard,PRL10_Phol1,Petrosyan_PRL1132014}. Recall that the acceleration will cease when the 2 Rydberg excitation state is projected onto the non-interacting atom collective state. For interaction potentials where there is significant radial dependence, the efficiency of the 2 excitation dynamics will be reduced and there will be spatial correlations in the state that is excited. Changes in the efficiency can depend strongly on the Rydberg atom interaction potentials.

For the calculations in this paper on 2 excitation states, the collective excitations are not independent of one another but are analogous to a doubly excited Rydberg molecule or macrodimer. In this case, the dephasing rates of these entities should be approximately equal to those occurring in a 2 independent excitation case where 2 superatoms are present, provided there are no additional open molecular decay channels. In many cases, the decay time for the macrodimer is determined by the radiative decay of the Rydberg atoms that make up the molecule \cite{SchwettJMODOPT,Petrosyan_PRL1132014}. Ground state-Rydberg atom molecules are not important because the internuclear spacings are greater than the size of the Rydberg orbital \cite{Nat09_Bendkowsky,Booth2015}.

The cavity coupling strength also has spatial dependence which must be considered. The size of the collective excitation is very large compared to the wavelength of the cavity transition.  The relative sizes can be quantified by comparing the blockade radius for $n=90$ to the wavelength corresponding to the cavity resonance, $9\,\mu\mathrm{m}\gg 780\,$nm. The cavity coupling strength is effectively averaged over the spatial dependence of the cavity field mode, reducing the average cavity coupling strength by $\sqrt{2}$ in this simple argument. A rough estimate places half the atoms at nodes and the other half at antinodes. Using this approximation to estimate the spread of atomic energy shifts for the atom-cavity coupling, yields a spread of atom energies of $\sim \sqrt{N} g$. It is interesting to explore the dephasing caused by the interaction of a collective excitation that is spatially larger than the modulation in the cavity coupling in more detail, but such an investigation is outside the scope of the work here. Numerically, this can be approached using quantum Monte Carlo methods which are computationally intensive but possible \cite{PRA10_Molmer}. For a collective state spread over a large number cavity resonant wavelengths, the overall effect is primarily to reduce the number of atoms participating in the atoms-cavity dynamics since atoms in the antinodes of the cavity are affected by the cavity coupling while those at nodes essentially behave as if in free-space. The atoms located at antinodes dominate the dynamics because they are coupled more strongly to the cavity. The spatial dependence will lead to fluctuations of the atoms-cavity coupling parameters that will decrease as the number of atoms in the collective state increases. Although our calculations are for low atom number collective states, the results scale to higher atom number collective states, at least with respect to the spatial dependence of the cavity mode.

The discussion of the dephasing mechanisms here points to a common factor that is important for a broad range of Rydberg atom quantum optics experiments. The product of the time the atoms exist in a particular collective state and the dephasing rate due to atomic motion, including heating mechanisms that arise in an experimental scheme, needs to be small. As a consequence, our method requires cold samples, $\sim 1\,\mu$K, and fast excitation dynamics, $<1\,\mu$s. In addition, to create multiple excitation states, we require potentials that are engineered to reduce the acceleration of atom pairs after excitation and provide a broad range of internuclear separations over which atom pairs can be excited, if spatial correlations are undesirable. Speeding up the excitation processes used in our scheme can compensate for some of the losses due to the additional dephasing mechanisms if, as in the case investigated, the Rydberg atom interaction strength is large enough to accommodate an increase in the Rabi frequencies of the laser pulses.

The Purcell factors and cooperativities we have used can be justified on the grounds that some experiments have achieved cavities with sufficient finesse and mode volumes to at least approach, or realize, the parameter values ($F_p$, $g$, $\kappa$) used here \cite{Heinzen87,Reichel10,Yalla14}. Lower cooperativity factors, of course, will reduce the efficiency of the approach that we have presented, however, a smaller cooperativity factor does not preclude the qualitative observation of the proposed scheme to deterministically synthesizing quantum states in a resonant, high-finesse cavity. Realizing this goal is interesting and can motivate improvements in the cavity design.

To summarize, we showed that Rydberg blockade allows one to deterministically prepare a fixed number of collective excitations inside a cavity using our model. This is impossible for the case of direct excitation of atoms inside a cavity, since, for noninteracting atoms, the excitation is probabilistic. In the case of 2 excitation states generated by the blockade mechanism, we can deterministically observe interesting dynamics and analyze the correlations between two, or more, atoms and the photons emitted from the cavity. The 2 excitation states we examined in this paper can be understood as macrodimers excited on flat relatively featureless interatomic potentials that are subsequently coherently transferred to noninteracting, collective excitation states. Both the excitation and transfer processes produce collective states as the excitations are shared by the atoms in the sample. Using macrodimer states that are not featureless can be used to introduce spatial correlations into the collective excitations. A detailed treatment of the spatial correlations introduced by different interatomic potentials lies outside the scope of this paper, but is a future goal. The engineering of spatial correlations by using Rydberg atom interactions in a collective ensemble is interesting and could be useful for experiments in quantum optics, for example in studying superradiance. In this paper, we showed that the collective state dynamics can be controlled deterministically by demonstrating results for collective dynamics that are similar to the TC model for two excited atoms. We have observed anti-bunched photon emission when we neglect spontaneous emission out of the cavity. With spontaneous emission the emitted light is bunched in the weak ($\kappa > g>\Gamma_{\perp}$) and strong ($ g>\kappa >\Gamma_{\perp}$) coupling regime. Similarly, the entropy of the system decreases with time, while the second order correlation function of the photons emitted by the 2 collective excitations is enhanced. The efficiency of photon production is enhanced by increasing the effective cooperativity by increasing $g$ through the cavity parameters or the number of atoms, as would be expected.

\section{Conclusion}

We have investigated a four-level system of $N$ entangled atoms placed inside a high-finesse cavity. The Rydberg blockade mechanism plays an important role in preparing the atoms in a collective state. We have used numerical simulations to show that one can deterministically excite collective $N$ atom superposition states using Rydberg blockade and laser pulses. Rydberg blockade allows one to deterministically prepare a fixed number of collective excitations inside a cavity. This is impossible for the case of direct excitation of atoms inside a cavity since that process is probabilistic. In the case of 2 collective excitation states generated by the blockade mechanism, we can deterministically observe bunched photon emission from the cavity similar to the case of 2 excited atoms initially prepared in a cavity. The scheme is useful for the generation of single photons as well as other types of quantum light fields. It could also be useful for coherent optical manipulation in quantum information processing and the study of interesting quantum optical phenomena such as superradiance \cite{PhyRep93_Gross}. The rates used for the calculations are realistic and experimentally achievable. The most important dephasing mechanisms were included or addressed. This system merits further theoretical and experimental study because it demonstrates flexibility for deterministically synthesizing collective quantum states that can be analyzed by interrogating the photons leaving the cavity.
\acknowledgements
This work was supported by AFOSR (FA9550-12-1-0282, FA9550-15-1-0381) and NSF (PHY-1104424).


\begin{thebibliography}{93}%
\makeatletter
\providecommand \@ifxundefined [1]{%
 \@ifx{#1\undefined}
}%
\providecommand \@ifnum [1]{%
 \ifnum #1\expandafter \@firstoftwo
 \else \expandafter \@secondoftwo
 \fi
}%
\providecommand \@ifx [1]{%
 \ifx #1\expandafter \@firstoftwo
 \else \expandafter \@secondoftwo
 \fi
}%
\providecommand \natexlab [1]{#1}%
\providecommand \enquote  [1]{``#1''}%
\providecommand \bibnamefont  [1]{#1}%
\providecommand \bibfnamefont [1]{#1}%
\providecommand \citenamefont [1]{#1}%
\providecommand \href@noop [0]{\@secondoftwo}%
\providecommand \href [0]{\begingroup \@sanitize@url \@href}%
\providecommand \@href[1]{\@@startlink{#1}\@@href}%
\providecommand \@@href[1]{\endgroup#1\@@endlink}%
\providecommand \@sanitize@url [0]{\catcode `\\12\catcode `\$12\catcode
  `\&12\catcode `\#12\catcode `\^12\catcode `\_12\catcode `\%12\relax}%
\providecommand \@@startlink[1]{}%
\providecommand \@@endlink[0]{}%
\providecommand \url  [0]{\begingroup\@sanitize@url \@url }%
\providecommand \@url [1]{\endgroup\@href {#1}{\urlprefix }}%
\providecommand \urlprefix  [0]{URL }%
\providecommand \Eprint [0]{\href }%
\providecommand \doibase [0]{http://dx.doi.org/}%
\providecommand \selectlanguage [0]{\@gobble}%
\providecommand \bibinfo  [0]{\@secondoftwo}%
\providecommand \bibfield  [0]{\@secondoftwo}%
\providecommand \translation [1]{[#1]}%
\providecommand \BibitemOpen [0]{}%
\providecommand \bibitemStop [0]{}%
\providecommand \bibitemNoStop [0]{.\EOS\space}%
\providecommand \EOS [0]{\spacefactor3000\relax}%
\providecommand \BibitemShut  [1]{\csname bibitem#1\endcsname}%
\let\auto@bib@innerbib\@empty
\bibitem [{\citenamefont {Gallagher}(1994)}]{Gallagher94}%
  \BibitemOpen
  \bibfield  {author} {\bibinfo {author} {\bibfnamefont {T.}~\bibnamefont
  {Gallagher}},\ }\href {http://dx.doi.org/10.1017/CBO9780511524530} {\emph
  {\bibinfo {title} {Rydberg Atoms}}}\ (\bibinfo  {publisher} {Cambridge
  University Press, New York},\ \bibinfo {year} {1994})\BibitemShut {NoStop}%
\bibitem [{\citenamefont {Schwettmann}\ \emph {et~al.}(2006)\citenamefont
  {Schwettmann}, \citenamefont {Crawford}, \citenamefont {Overstreet},\ and\
  \citenamefont {Shaffer}}]{Schwettmann2006}%
  \BibitemOpen
  \bibfield  {author} {\bibinfo {author} {\bibfnamefont {A.}~\bibnamefont
  {Schwettmann}}, \bibinfo {author} {\bibfnamefont {J.}~\bibnamefont
  {Crawford}}, \bibinfo {author} {\bibfnamefont {K.~R.}\ \bibnamefont
  {Overstreet}}, \ and\ \bibinfo {author} {\bibfnamefont {J.~P.}\ \bibnamefont
  {Shaffer}},\ }\href {\doibase 10.1103/PhysRevA.74.020701} {\bibfield
  {journal} {\bibinfo  {journal} {Phys. Rev. A}\ }\textbf {\bibinfo {volume}
  {74}},\ \bibinfo {pages} {020701(R)} (\bibinfo {year} {2006})}\BibitemShut
  {NoStop}%
\bibitem [{\citenamefont {Marcassa}\ and\ \citenamefont
  {Shaffer}(2014)}]{AAMOP2014}%
  \BibitemOpen
  \bibfield  {author} {\bibinfo {author} {\bibfnamefont {L.}~\bibnamefont
  {Marcassa}}\ and\ \bibinfo {author} {\bibfnamefont {J.~P.}\ \bibnamefont
  {Shaffer}},\ }\href@noop {} {\bibfield  {journal} {\bibinfo  {journal} {Adv.
  At. Mol. Phys.}\ }\textbf {\bibinfo {volume} {63}},\ \bibinfo {pages} {47}
  (\bibinfo {year} {2014})}\BibitemShut {NoStop}%
\bibitem [{\citenamefont {Cabral}\ \emph {et~al.}(2011)\citenamefont {Cabral},
  \citenamefont {Kondo}, \citenamefont {Gonçalves}, \citenamefont
  {Nascimento}, \citenamefont {Marcassa}, \citenamefont {Booth}, \citenamefont
  {Tallant}, \citenamefont {Schwettmann}, \citenamefont {Overstreet},
  \citenamefont {Sedlacek},\ and\ \citenamefont {Shaffer}}]{Cabral2011}%
  \BibitemOpen
  \bibfield  {author} {\bibinfo {author} {\bibfnamefont {J.~S.}\ \bibnamefont
  {Cabral}}, \bibinfo {author} {\bibfnamefont {J.~M.}\ \bibnamefont {Kondo}},
  \bibinfo {author} {\bibfnamefont {L.~F.}\ \bibnamefont {Gonçalves}},
  \bibinfo {author} {\bibfnamefont {V.~A.}\ \bibnamefont {Nascimento}},
  \bibinfo {author} {\bibfnamefont {L.~G.}\ \bibnamefont {Marcassa}}, \bibinfo
  {author} {\bibfnamefont {D.}~\bibnamefont {Booth}}, \bibinfo {author}
  {\bibfnamefont {J.}~\bibnamefont {Tallant}}, \bibinfo {author} {\bibfnamefont
  {A.}~\bibnamefont {Schwettmann}}, \bibinfo {author} {\bibfnamefont {K.~R.}\
  \bibnamefont {Overstreet}}, \bibinfo {author} {\bibfnamefont {J.~A.}\
  \bibnamefont {Sedlacek}}, \ and\ \bibinfo {author} {\bibfnamefont {J.~P.}\
  \bibnamefont {Shaffer}},\ }\href
  {http://stacks.iop.org/0953-4075/44/i=18/a=184007} {\bibfield  {journal}
  {\bibinfo  {journal} {J. Phys. B: At. Mol. Opt. Phys.}\ }\textbf {\bibinfo
  {volume} {44}},\ \bibinfo {pages} {184007} (\bibinfo {year}
  {2011})}\BibitemShut {NoStop}%
\bibitem [{\citenamefont {Tong}\ \emph {et~al.}(2004)\citenamefont {Tong},
  \citenamefont {Farooqi}, \citenamefont {Stanojevic}, \citenamefont
  {Krishnan}, \citenamefont {Zhang}, \citenamefont {C\^ot\'e}, \citenamefont
  {Eyler},\ and\ \citenamefont {Gould}}]{PRL04_Gould}%
  \BibitemOpen
  \bibfield  {author} {\bibinfo {author} {\bibfnamefont {D.}~\bibnamefont
  {Tong}}, \bibinfo {author} {\bibfnamefont {S.~M.}\ \bibnamefont {Farooqi}},
  \bibinfo {author} {\bibfnamefont {J.}~\bibnamefont {Stanojevic}}, \bibinfo
  {author} {\bibfnamefont {S.}~\bibnamefont {Krishnan}}, \bibinfo {author}
  {\bibfnamefont {Y.~P.}\ \bibnamefont {Zhang}}, \bibinfo {author}
  {\bibfnamefont {R.}~\bibnamefont {C\^ot\'e}}, \bibinfo {author}
  {\bibfnamefont {E.~E.}\ \bibnamefont {Eyler}}, \ and\ \bibinfo {author}
  {\bibfnamefont {P.~L.}\ \bibnamefont {Gould}},\ }\href {\doibase
  10.1103/PhysRevLett.93.063001} {\bibfield  {journal} {\bibinfo  {journal}
  {Phys. Rev. Lett.}\ }\textbf {\bibinfo {volume} {93}},\ \bibinfo {pages}
  {063001} (\bibinfo {year} {2004})}\BibitemShut {NoStop}%
\bibitem [{\citenamefont {Heidemann}\ \emph {et~al.}(2007)\citenamefont
  {Heidemann}, \citenamefont {Raitzsch}, \citenamefont {Bendkowsky},
  \citenamefont {Butscher}, \citenamefont {L\"ow}, \citenamefont {Santos},\
  and\ \citenamefont {Pfau}}]{PRL07_Pfau}%
  \BibitemOpen
  \bibfield  {author} {\bibinfo {author} {\bibfnamefont {R.}~\bibnamefont
  {Heidemann}}, \bibinfo {author} {\bibfnamefont {U.}~\bibnamefont {Raitzsch}},
  \bibinfo {author} {\bibfnamefont {V.}~\bibnamefont {Bendkowsky}}, \bibinfo
  {author} {\bibfnamefont {B.}~\bibnamefont {Butscher}}, \bibinfo {author}
  {\bibfnamefont {R.}~\bibnamefont {L\"ow}}, \bibinfo {author} {\bibfnamefont
  {L.}~\bibnamefont {Santos}}, \ and\ \bibinfo {author} {\bibfnamefont
  {T.}~\bibnamefont {Pfau}},\ }\href {\doibase 10.1103/PhysRevLett.99.163601}
  {\bibfield  {journal} {\bibinfo  {journal} {Phys. Rev. Lett.}\ }\textbf
  {\bibinfo {volume} {99}},\ \bibinfo {pages} {163601} (\bibinfo {year}
  {2007})}\BibitemShut {NoStop}%
\bibitem [{\citenamefont {Lukin}\ \emph {et~al.}(2001)\citenamefont {Lukin},
  \citenamefont {Fleischhauer}, \citenamefont {C\^ot\'e}, \citenamefont {Duan},
  \citenamefont {Jaksch}, \citenamefont {Cirac},\ and\ \citenamefont
  {Zoller}}]{PRL01_Lukin}%
  \BibitemOpen
  \bibfield  {author} {\bibinfo {author} {\bibfnamefont {M.~D.}\ \bibnamefont
  {Lukin}}, \bibinfo {author} {\bibfnamefont {M.}~\bibnamefont {Fleischhauer}},
  \bibinfo {author} {\bibfnamefont {R.}~\bibnamefont {C\^ot\'e}}, \bibinfo
  {author} {\bibfnamefont {L.~M.}\ \bibnamefont {Duan}}, \bibinfo {author}
  {\bibfnamefont {D.}~\bibnamefont {Jaksch}}, \bibinfo {author} {\bibfnamefont
  {J.~I.}\ \bibnamefont {Cirac}}, \ and\ \bibinfo {author} {\bibfnamefont
  {P.}~\bibnamefont {Zoller}},\ }\href {\doibase 10.1103/PhysRevLett.87.037901}
  {\bibfield  {journal} {\bibinfo  {journal} {Phys. Rev. Lett.}\ }\textbf
  {\bibinfo {volume} {87}},\ \bibinfo {pages} {037901} (\bibinfo {year}
  {2001})}\BibitemShut {NoStop}%
\bibitem [{\citenamefont {Dudin}\ \emph {et~al.}(2009)\citenamefont {Dudin},
  \citenamefont {Li}, \citenamefont {Bariani},\ and\ \citenamefont
  {Kuzmich}}]{NatPhys09_Saffman}%
  \BibitemOpen
  \bibfield  {author} {\bibinfo {author} {\bibfnamefont {Y.~O.}\ \bibnamefont
  {Dudin}}, \bibinfo {author} {\bibfnamefont {L.}~\bibnamefont {Li}}, \bibinfo
  {author} {\bibfnamefont {F.}~\bibnamefont {Bariani}}, \ and\ \bibinfo
  {author} {\bibfnamefont {A.}~\bibnamefont {Kuzmich}},\ }\href {\doibase
  10.1038/nphys1178} {\bibfield  {journal} {\bibinfo  {journal} {Nat. Phys.}\
  }\textbf {\bibinfo {volume} {5}},\ \bibinfo {pages} {110} (\bibinfo {year}
  {2009})}\BibitemShut {NoStop}%
\bibitem [{\citenamefont {Ebert}\ \emph {et~al.}(2014)\citenamefont {Ebert},
  \citenamefont {Gill}, \citenamefont {Gibbons}, \citenamefont {Zhang},
  \citenamefont {Saffman},\ and\ \citenamefont {Walker}}]{PRL14_Walker}%
  \BibitemOpen
  \bibfield  {author} {\bibinfo {author} {\bibfnamefont {M.}~\bibnamefont
  {Ebert}}, \bibinfo {author} {\bibfnamefont {A.}~\bibnamefont {Gill}},
  \bibinfo {author} {\bibfnamefont {M.}~\bibnamefont {Gibbons}}, \bibinfo
  {author} {\bibfnamefont {X.}~\bibnamefont {Zhang}}, \bibinfo {author}
  {\bibfnamefont {M.}~\bibnamefont {Saffman}}, \ and\ \bibinfo {author}
  {\bibfnamefont {T.~G.}\ \bibnamefont {Walker}},\ }\href {\doibase
  10.1103/PhysRevLett.112.043602} {\bibfield  {journal} {\bibinfo  {journal}
  {Phys. Rev. Lett.}\ }\textbf {\bibinfo {volume} {112}},\ \bibinfo {pages}
  {043602} (\bibinfo {year} {2014})}\BibitemShut {NoStop}%
\bibitem [{\citenamefont {Reetz-Lamour}\ \emph {et~al.}(2008)\citenamefont
  {Reetz-Lamour}, \citenamefont {Amthor}, \citenamefont {Deiglmayr},\ and\
  \citenamefont {Weidem\"uller}}]{PRL08_Weidemuller}%
  \BibitemOpen
  \bibfield  {author} {\bibinfo {author} {\bibfnamefont {M.}~\bibnamefont
  {Reetz-Lamour}}, \bibinfo {author} {\bibfnamefont {T.}~\bibnamefont
  {Amthor}}, \bibinfo {author} {\bibfnamefont {J.}~\bibnamefont {Deiglmayr}}, \
  and\ \bibinfo {author} {\bibfnamefont {M.}~\bibnamefont {Weidem\"uller}},\
  }\href {\doibase 10.1103/PhysRevLett.100.253001} {\bibfield  {journal}
  {\bibinfo  {journal} {Phys. Rev. Lett.}\ }\textbf {\bibinfo {volume} {100}},\
  \bibinfo {pages} {253001} (\bibinfo {year} {2008})}\BibitemShut {NoStop}%
\bibitem [{\citenamefont {Reetz-lamour}\ \emph {et~al.}(2008)\citenamefont
  {Reetz-lamour}, \citenamefont {Deiglmayr}, \citenamefont {Amthor},\ and\
  \citenamefont {Weidem\"uller}}]{NPJ08_Weidemuller}%
  \BibitemOpen
  \bibfield  {author} {\bibinfo {author} {\bibfnamefont {M.}~\bibnamefont
  {Reetz-lamour}}, \bibinfo {author} {\bibfnamefont {J.}~\bibnamefont
  {Deiglmayr}}, \bibinfo {author} {\bibfnamefont {T.}~\bibnamefont {Amthor}}, \
  and\ \bibinfo {author} {\bibfnamefont {M.}~\bibnamefont {Weidem\"uller}},\
  }\href {\doibase 10.1088/1367-2630/10/4/045026} {\bibfield  {journal}
  {\bibinfo  {journal} {New J. Phys.}\ }\textbf {\bibinfo {volume} {10}},\
  \bibinfo {pages} {045026} (\bibinfo {year} {2008})}\BibitemShut {NoStop}%
\bibitem [{\citenamefont {Johnson}\ \emph {et~al.}(2008)\citenamefont
  {Johnson}, \citenamefont {Urban}, \citenamefont {Henage}, \citenamefont
  {Isenhower}, \citenamefont {Yavuz}, \citenamefont {Walker},\ and\
  \citenamefont {Saffman}}]{PRL08_Saffman}%
  \BibitemOpen
  \bibfield  {author} {\bibinfo {author} {\bibfnamefont {T.~A.}\ \bibnamefont
  {Johnson}}, \bibinfo {author} {\bibfnamefont {E.}~\bibnamefont {Urban}},
  \bibinfo {author} {\bibfnamefont {T.}~\bibnamefont {Henage}}, \bibinfo
  {author} {\bibfnamefont {L.}~\bibnamefont {Isenhower}}, \bibinfo {author}
  {\bibfnamefont {D.~D.}\ \bibnamefont {Yavuz}}, \bibinfo {author}
  {\bibfnamefont {T.~G.}\ \bibnamefont {Walker}}, \ and\ \bibinfo {author}
  {\bibfnamefont {M.}~\bibnamefont {Saffman}},\ }\href {\doibase
  10.1103/PhysRevLett.100.113003} {\bibfield  {journal} {\bibinfo  {journal}
  {Phys. Rev. Lett.}\ }\textbf {\bibinfo {volume} {100}},\ \bibinfo {pages}
  {113003} (\bibinfo {year} {2008})}\BibitemShut {NoStop}%
\bibitem [{\citenamefont {Barredo}\ \emph {et~al.}(2014)\citenamefont
  {Barredo}, \citenamefont {Ravets}, \citenamefont {Labuhn}, \citenamefont
  {B\'eguin}, \citenamefont {Vernier}, \citenamefont {Nogrette}, \citenamefont
  {Lahaye},\ and\ \citenamefont {Browaeys}}]{PRL14_Browaeys}%
  \BibitemOpen
  \bibfield  {author} {\bibinfo {author} {\bibfnamefont {D.}~\bibnamefont
  {Barredo}}, \bibinfo {author} {\bibfnamefont {S.}~\bibnamefont {Ravets}},
  \bibinfo {author} {\bibfnamefont {H.}~\bibnamefont {Labuhn}}, \bibinfo
  {author} {\bibfnamefont {L.}~\bibnamefont {B\'eguin}}, \bibinfo {author}
  {\bibfnamefont {A.}~\bibnamefont {Vernier}}, \bibinfo {author} {\bibfnamefont
  {F.}~\bibnamefont {Nogrette}}, \bibinfo {author} {\bibfnamefont
  {T.}~\bibnamefont {Lahaye}}, \ and\ \bibinfo {author} {\bibfnamefont
  {A.}~\bibnamefont {Browaeys}},\ }\href {\doibase
  10.1103/PhysRevLett.112.183002} {\bibfield  {journal} {\bibinfo  {journal}
  {Phys. Rev. Lett.}\ }\textbf {\bibinfo {volume} {112}},\ \bibinfo {pages}
  {183002} (\bibinfo {year} {2014})}\BibitemShut {NoStop}%
\bibitem [{\citenamefont {Ga\"etan}\ \emph {et~al.}(2009)\citenamefont
  {Ga\"etan}, \citenamefont {Miroshnychenko}, \citenamefont {Wilk},
  \citenamefont {Chotia}, \citenamefont {Viteau}, \citenamefont {Comparat},
  \citenamefont {Pillet}, \citenamefont {Browaeys},\ and\ \citenamefont
  {Grangier}}]{NatPhy09_Grangier}%
  \BibitemOpen
  \bibfield  {author} {\bibinfo {author} {\bibfnamefont {A.}~\bibnamefont
  {Ga\"etan}}, \bibinfo {author} {\bibfnamefont {Y.}~\bibnamefont
  {Miroshnychenko}}, \bibinfo {author} {\bibfnamefont {T.}~\bibnamefont
  {Wilk}}, \bibinfo {author} {\bibfnamefont {A.}~\bibnamefont {Chotia}},
  \bibinfo {author} {\bibfnamefont {M.}~\bibnamefont {Viteau}}, \bibinfo
  {author} {\bibfnamefont {D.}~\bibnamefont {Comparat}}, \bibinfo {author}
  {\bibfnamefont {P.}~\bibnamefont {Pillet}}, \bibinfo {author} {\bibfnamefont
  {A.}~\bibnamefont {Browaeys}}, \ and\ \bibinfo {author} {\bibfnamefont
  {P.}~\bibnamefont {Grangier}},\ }\href {\doibase 10.1038/nphys1183}
  {\bibfield  {journal} {\bibinfo  {journal} {Nat. Phys.}\ }\textbf {\bibinfo
  {volume} {5}},\ \bibinfo {pages} {115} (\bibinfo {year} {2009})}\BibitemShut
  {NoStop}%
\bibitem [{\citenamefont {Dudin}\ \emph {et~al.}(2012)\citenamefont {Dudin},
  \citenamefont {Li}, \citenamefont {Bariani},\ and\ \citenamefont
  {Kuzmich}}]{NatPhys12_Kuzmich}%
  \BibitemOpen
  \bibfield  {author} {\bibinfo {author} {\bibfnamefont {Y.~O.}\ \bibnamefont
  {Dudin}}, \bibinfo {author} {\bibfnamefont {L.}~\bibnamefont {Li}}, \bibinfo
  {author} {\bibfnamefont {F.}~\bibnamefont {Bariani}}, \ and\ \bibinfo
  {author} {\bibfnamefont {A.}~\bibnamefont {Kuzmich}},\ }\href {\doibase
  10.1038/nphys2413} {\bibfield  {journal} {\bibinfo  {journal} {Nat. Phys.}\
  }\textbf {\bibinfo {volume} {8}},\ \bibinfo {pages} {790} (\bibinfo {year}
  {2012})}\BibitemShut {NoStop}%
\bibitem [{\citenamefont {Huber}\ \emph {et~al.}(2011)\citenamefont {Huber},
  \citenamefont {Baluktsian}, \citenamefont {Schlagm\"uller}, \citenamefont
  {K\"olle}, \citenamefont {K\"ubler}, \citenamefont {L\"ow},\ and\
  \citenamefont {Pfau}}]{PRL11_Pfau}%
  \BibitemOpen
  \bibfield  {author} {\bibinfo {author} {\bibfnamefont {B.}~\bibnamefont
  {Huber}}, \bibinfo {author} {\bibfnamefont {T.}~\bibnamefont {Baluktsian}},
  \bibinfo {author} {\bibfnamefont {M.}~\bibnamefont {Schlagm\"uller}},
  \bibinfo {author} {\bibfnamefont {A.}~\bibnamefont {K\"olle}}, \bibinfo
  {author} {\bibfnamefont {H.}~\bibnamefont {K\"ubler}}, \bibinfo {author}
  {\bibfnamefont {R.}~\bibnamefont {L\"ow}}, \ and\ \bibinfo {author}
  {\bibfnamefont {T.}~\bibnamefont {Pfau}},\ }\href {\doibase
  10.1103/PhysRevLett.107.243001} {\bibfield  {journal} {\bibinfo  {journal}
  {Phys. Rev. Lett.}\ }\textbf {\bibinfo {volume} {107}},\ \bibinfo {pages}
  {243001} (\bibinfo {year} {2011})}\BibitemShut {NoStop}%
\bibitem [{\citenamefont {Kimble}(2008)}]{Nat08_Kimble}%
  \BibitemOpen
  \bibfield  {author} {\bibinfo {author} {\bibfnamefont {J.~F.}\ \bibnamefont
  {Kimble}},\ }\href {\doibase 10.1038/nature07127} {\bibfield  {journal}
  {\bibinfo  {journal} {Nature}\ }\textbf {\bibinfo {volume} {453}},\ \bibinfo
  {pages} {1023} (\bibinfo {year} {2008})}\BibitemShut {NoStop}%
\bibitem [{\citenamefont {Raimond}\ \emph {et~al.}(2001)\citenamefont
  {Raimond}, \citenamefont {Brune},\ and\ \citenamefont
  {Haroche}}]{RMP01_Haroche}%
  \BibitemOpen
  \bibfield  {author} {\bibinfo {author} {\bibfnamefont {J.~M.}\ \bibnamefont
  {Raimond}}, \bibinfo {author} {\bibfnamefont {M.}~\bibnamefont {Brune}}, \
  and\ \bibinfo {author} {\bibfnamefont {S.}~\bibnamefont {Haroche}},\ }\href
  {\doibase 10.1103/RevModPhys.73.565} {\bibfield  {journal} {\bibinfo
  {journal} {Rev. Mod. Phys.}\ }\textbf {\bibinfo {volume} {73}},\ \bibinfo
  {pages} {565} (\bibinfo {year} {2001})}\BibitemShut {NoStop}%
\bibitem [{\citenamefont {Stenholm}(1973)}]{Stenholm1973}%
  \BibitemOpen
  \bibfield  {author} {\bibinfo {author} {\bibfnamefont {S.}~\bibnamefont
  {Stenholm}},\ }\href {\doibase
  http://dx.doi.org/10.1016/0370-1573(73)90011-2} {\bibfield  {journal}
  {\bibinfo  {journal} {Phys. Rep.}\ }\textbf {\bibinfo {volume} {6}},\
  \bibinfo {pages} {1 } (\bibinfo {year} {1973})}\BibitemShut {NoStop}%
\bibitem [{\citenamefont {Zheng}\ and\ \citenamefont
  {Guo}(2000)}]{PRL00_Zheng}%
  \BibitemOpen
  \bibfield  {author} {\bibinfo {author} {\bibfnamefont {S.-B.}\ \bibnamefont
  {Zheng}}\ and\ \bibinfo {author} {\bibfnamefont {G.-C.}\ \bibnamefont
  {Guo}},\ }\href {\doibase 10.1103/PhysRevLett.85.2392} {\bibfield  {journal}
  {\bibinfo  {journal} {Phys. Rev. Lett.}\ }\textbf {\bibinfo {volume} {85}},\
  \bibinfo {pages} {2392} (\bibinfo {year} {2000})}\BibitemShut {NoStop}%
\bibitem [{\citenamefont {Lettner}\ \emph {et~al.}(2011)\citenamefont
  {Lettner}, \citenamefont {M\"ucke}, \citenamefont {Riedl}, \citenamefont
  {Vo}, \citenamefont {Hahn}, \citenamefont {Baur}, \citenamefont {Bochmann},
  \citenamefont {Ritter}, \citenamefont {D\"urr},\ and\ \citenamefont
  {Rempe}}]{PRL_11_Rempe}%
  \BibitemOpen
  \bibfield  {author} {\bibinfo {author} {\bibfnamefont {M.}~\bibnamefont
  {Lettner}}, \bibinfo {author} {\bibfnamefont {M.}~\bibnamefont {M\"ucke}},
  \bibinfo {author} {\bibfnamefont {S.}~\bibnamefont {Riedl}}, \bibinfo
  {author} {\bibfnamefont {C.}~\bibnamefont {Vo}}, \bibinfo {author}
  {\bibfnamefont {C.}~\bibnamefont {Hahn}}, \bibinfo {author} {\bibfnamefont
  {S.}~\bibnamefont {Baur}}, \bibinfo {author} {\bibfnamefont {J.}~\bibnamefont
  {Bochmann}}, \bibinfo {author} {\bibfnamefont {S.}~\bibnamefont {Ritter}},
  \bibinfo {author} {\bibfnamefont {S.}~\bibnamefont {D\"urr}}, \ and\ \bibinfo
  {author} {\bibfnamefont {G.}~\bibnamefont {Rempe}},\ }\href {\doibase
  10.1103/PhysRevLett.106.210503} {\bibfield  {journal} {\bibinfo  {journal}
  {Phys. Rev. Lett.}\ }\textbf {\bibinfo {volume} {106}},\ \bibinfo {pages}
  {210503} (\bibinfo {year} {2011})}\BibitemShut {NoStop}%
\bibitem [{\citenamefont {Reiter}\ \emph {et~al.}(2012)\citenamefont {Reiter},
  \citenamefont {Kastoryano},\ and\ \citenamefont
  {S{\o}rensen}}]{NJP_Reiter12}%
  \BibitemOpen
  \bibfield  {author} {\bibinfo {author} {\bibfnamefont {F.}~\bibnamefont
  {Reiter}}, \bibinfo {author} {\bibfnamefont {M.~J.}\ \bibnamefont
  {Kastoryano}}, \ and\ \bibinfo {author} {\bibfnamefont {A.~S.}\ \bibnamefont
  {S{\o}rensen}},\ }\href {http://stacks.iop.org/1367-2630/14/i=5/a=053022}
  {\bibfield  {journal} {\bibinfo  {journal} {New J. Phys.}\ }\textbf {\bibinfo
  {volume} {14}},\ \bibinfo {pages} {053022} (\bibinfo {year}
  {2012})}\BibitemShut {NoStop}%
\bibitem [{\citenamefont {Cabrillo}\ \emph {et~al.}(1999)\citenamefont
  {Cabrillo}, \citenamefont {Cirac}, \citenamefont {Garc\'ia-Fern\'andez},\
  and\ \citenamefont {Zoller}}]{PRA99_Zoller}%
  \BibitemOpen
  \bibfield  {author} {\bibinfo {author} {\bibfnamefont {C.}~\bibnamefont
  {Cabrillo}}, \bibinfo {author} {\bibfnamefont {J.~I.}\ \bibnamefont {Cirac}},
  \bibinfo {author} {\bibnamefont {Garc\'ia-Fern\'andez}}, \ and\ \bibinfo
  {author} {\bibfnamefont {P.}~\bibnamefont {Zoller}},\ }\href {\doibase
  10.1103/PhysRevA.59.1025} {\bibfield  {journal} {\bibinfo  {journal} {Phys.
  Rev. A}\ }\textbf {\bibinfo {volume} {59}},\ \bibinfo {pages} {1025}
  (\bibinfo {year} {1999})}\BibitemShut {NoStop}%
\bibitem [{\citenamefont {Rist}\ \emph {et~al.}(2008)\citenamefont {Rist},
  \citenamefont {Eschner}, \citenamefont {Hennrich},\ and\ \citenamefont
  {Morigi}}]{PRA08_Morigi}%
  \BibitemOpen
  \bibfield  {author} {\bibinfo {author} {\bibfnamefont {S.}~\bibnamefont
  {Rist}}, \bibinfo {author} {\bibfnamefont {J.}~\bibnamefont {Eschner}},
  \bibinfo {author} {\bibfnamefont {M.}~\bibnamefont {Hennrich}}, \ and\
  \bibinfo {author} {\bibfnamefont {G.}~\bibnamefont {Morigi}},\ }\href
  {\doibase 10.1103/PhysRevA.78.013808} {\bibfield  {journal} {\bibinfo
  {journal} {Phys. Rev. A}\ }\textbf {\bibinfo {volume} {78}},\ \bibinfo
  {pages} {013808} (\bibinfo {year} {2008})}\BibitemShut {NoStop}%
\bibitem [{\citenamefont {Kumar}\ and\ \citenamefont
  {Kumar}(2012)}]{PRA12_Kumar}%
  \BibitemOpen
  \bibfield  {author} {\bibinfo {author} {\bibfnamefont {S.}~\bibnamefont
  {Kumar}}\ and\ \bibinfo {author} {\bibfnamefont {D.}~\bibnamefont {Kumar}},\
  }\href {\doibase 10.1103/PhysRevA.85.052317} {\bibfield  {journal} {\bibinfo
  {journal} {Phys. Rev. A}\ }\textbf {\bibinfo {volume} {85}},\ \bibinfo
  {pages} {052317} (\bibinfo {year} {2012})}\BibitemShut {NoStop}%
\bibitem [{\citenamefont {Ritter}\ \emph {et~al.}(2012)\citenamefont {Ritter},
  \citenamefont {Nolleke}, \citenamefont {Hahn}, \citenamefont {Reiserer},
  \citenamefont {Neuzner}, \citenamefont {Uphoff}, \citenamefont {Mucke},
  \citenamefont {Figueroa}, \citenamefont {Bochmann},\ and\ \citenamefont
  {Rempe}}]{Nat12_Rempe}%
  \BibitemOpen
  \bibfield  {author} {\bibinfo {author} {\bibfnamefont {S.}~\bibnamefont
  {Ritter}}, \bibinfo {author} {\bibfnamefont {C.}~\bibnamefont {Nolleke}},
  \bibinfo {author} {\bibfnamefont {C.}~\bibnamefont {Hahn}}, \bibinfo {author}
  {\bibfnamefont {A.}~\bibnamefont {Reiserer}}, \bibinfo {author}
  {\bibfnamefont {A.}~\bibnamefont {Neuzner}}, \bibinfo {author} {\bibfnamefont
  {M.}~\bibnamefont {Uphoff}}, \bibinfo {author} {\bibfnamefont
  {M.}~\bibnamefont {Mucke}}, \bibinfo {author} {\bibfnamefont
  {E.}~\bibnamefont {Figueroa}}, \bibinfo {author} {\bibfnamefont
  {J.}~\bibnamefont {Bochmann}}, \ and\ \bibinfo {author} {\bibfnamefont
  {G.}~\bibnamefont {Rempe}},\ }\href {\doibase 10.1038/nature11023} {\bibfield
   {journal} {\bibinfo  {journal} {Nature}\ }\textbf {\bibinfo {volume}
  {484}},\ \bibinfo {pages} {195} (\bibinfo {year} {2012})}\BibitemShut
  {NoStop}%
\bibitem [{\citenamefont {Slodi\ifmmode~\check{c}\else \v{c}\fi{}ka}\ \emph
  {et~al.}(2013)\citenamefont {Slodi\ifmmode~\check{c}\else \v{c}\fi{}ka},
  \citenamefont {H\'etet}, \citenamefont {R\"ock}, \citenamefont {Schindler},
  \citenamefont {Hennrich},\ and\ \citenamefont {Blatt}}]{PRL_13_Blatt}%
  \BibitemOpen
  \bibfield  {author} {\bibinfo {author} {\bibfnamefont {L.}~\bibnamefont
  {Slodi\ifmmode~\check{c}\else \v{c}\fi{}ka}}, \bibinfo {author}
  {\bibfnamefont {G.}~\bibnamefont {H\'etet}}, \bibinfo {author} {\bibfnamefont
  {N.}~\bibnamefont {R\"ock}}, \bibinfo {author} {\bibfnamefont
  {P.}~\bibnamefont {Schindler}}, \bibinfo {author} {\bibfnamefont
  {M.}~\bibnamefont {Hennrich}}, \ and\ \bibinfo {author} {\bibfnamefont
  {R.}~\bibnamefont {Blatt}},\ }\href {\doibase 10.1103/PhysRevLett.110.083603}
  {\bibfield  {journal} {\bibinfo  {journal} {Phys. Rev. Lett.}\ }\textbf
  {\bibinfo {volume} {110}},\ \bibinfo {pages} {083603} (\bibinfo {year}
  {2013})}\BibitemShut {NoStop}%
\bibitem [{\citenamefont {Ye}\ \emph {et~al.}(1999)\citenamefont {Ye},
  \citenamefont {Vernooy},\ and\ \citenamefont {Kimble}}]{PRA99_Kimble}%
  \BibitemOpen
  \bibfield  {author} {\bibinfo {author} {\bibfnamefont {J.}~\bibnamefont
  {Ye}}, \bibinfo {author} {\bibfnamefont {D.~W.}\ \bibnamefont {Vernooy}}, \
  and\ \bibinfo {author} {\bibfnamefont {H.~J.}\ \bibnamefont {Kimble}},\
  }\href {\doibase 10.1103/PhysRevLett.83.4987} {\bibfield  {journal} {\bibinfo
   {journal} {Phys. Rev. Lett.}\ }\textbf {\bibinfo {volume} {83}},\ \bibinfo
  {pages} {4987} (\bibinfo {year} {1999})}\BibitemShut {NoStop}%
\bibitem [{\citenamefont {McKeever}\ \emph {et~al.}(2003)\citenamefont
  {McKeever}, \citenamefont {Buck}, \citenamefont {Boozer}, \citenamefont
  {Kuzmich}, \citenamefont {N\"agerl}, \citenamefont {Stamper-Kurn},\ and\
  \citenamefont {Kimble}}]{PRL03_Kimble}%
  \BibitemOpen
  \bibfield  {author} {\bibinfo {author} {\bibfnamefont {J.}~\bibnamefont
  {McKeever}}, \bibinfo {author} {\bibfnamefont {J.~R.}\ \bibnamefont {Buck}},
  \bibinfo {author} {\bibfnamefont {A.~D.}\ \bibnamefont {Boozer}}, \bibinfo
  {author} {\bibfnamefont {A.}~\bibnamefont {Kuzmich}}, \bibinfo {author}
  {\bibfnamefont {H.-C.}\ \bibnamefont {N\"agerl}}, \bibinfo {author}
  {\bibfnamefont {D.~M.}\ \bibnamefont {Stamper-Kurn}}, \ and\ \bibinfo
  {author} {\bibfnamefont {H.~J.}\ \bibnamefont {Kimble}},\ }\href {\doibase
  10.1103/PhysRevLett.90.133602} {\bibfield  {journal} {\bibinfo  {journal}
  {Phys. Rev. Lett.}\ }\textbf {\bibinfo {volume} {90}},\ \bibinfo {pages}
  {133602} (\bibinfo {year} {2003})}\BibitemShut {NoStop}%
\bibitem [{\citenamefont {M\"unstermann}\ \emph {et~al.}(1999)\citenamefont
  {M\"unstermann}, \citenamefont {Fischer}, \citenamefont {Maunz},
  \citenamefont {Pinkse},\ and\ \citenamefont {Rempe}}]{PRL99_Rempe}%
  \BibitemOpen
  \bibfield  {author} {\bibinfo {author} {\bibfnamefont {P.}~\bibnamefont
  {M\"unstermann}}, \bibinfo {author} {\bibfnamefont {T.}~\bibnamefont
  {Fischer}}, \bibinfo {author} {\bibfnamefont {P.}~\bibnamefont {Maunz}},
  \bibinfo {author} {\bibfnamefont {P.~W.~H.}\ \bibnamefont {Pinkse}}, \ and\
  \bibinfo {author} {\bibfnamefont {G.}~\bibnamefont {Rempe}},\ }\href
  {\doibase 10.1103/PhysRevLett.82.3791} {\bibfield  {journal} {\bibinfo
  {journal} {Phys. Rev. Lett.}\ }\textbf {\bibinfo {volume} {82}},\ \bibinfo
  {pages} {3791} (\bibinfo {year} {1999})}\BibitemShut {NoStop}%
\bibitem [{\citenamefont {Hood}\ \emph {et~al.}(2000)\citenamefont {Hood},
  \citenamefont {Lynn}, \citenamefont {Doherty}, \citenamefont {Parkins},\ and\
  \citenamefont {Kimble}}]{Science00_Kimble}%
  \BibitemOpen
  \bibfield  {author} {\bibinfo {author} {\bibfnamefont {C.~J.}\ \bibnamefont
  {Hood}}, \bibinfo {author} {\bibfnamefont {T.~W.}\ \bibnamefont {Lynn}},
  \bibinfo {author} {\bibfnamefont {A.~C.}\ \bibnamefont {Doherty}}, \bibinfo
  {author} {\bibfnamefont {A.~S.}\ \bibnamefont {Parkins}}, \ and\ \bibinfo
  {author} {\bibfnamefont {H.~J.}\ \bibnamefont {Kimble}},\ }\href {\doibase
  10.1126/science.287.5457.1447} {\bibfield  {journal} {\bibinfo  {journal}
  {Science}\ }\textbf {\bibinfo {volume} {287}},\ \bibinfo {pages} {1447}
  (\bibinfo {year} {2000})}\BibitemShut {NoStop}%
\bibitem [{\citenamefont {Fortier}\ \emph {et~al.}(2007)\citenamefont
  {Fortier}, \citenamefont {Kim}, \citenamefont {Gibbons}, \citenamefont
  {Ahmadi},\ and\ \citenamefont {Chapman}}]{PRL07_Chapman}%
  \BibitemOpen
  \bibfield  {author} {\bibinfo {author} {\bibfnamefont {K.~M.}\ \bibnamefont
  {Fortier}}, \bibinfo {author} {\bibfnamefont {S.~Y.}\ \bibnamefont {Kim}},
  \bibinfo {author} {\bibfnamefont {M.~J.}\ \bibnamefont {Gibbons}}, \bibinfo
  {author} {\bibfnamefont {P.}~\bibnamefont {Ahmadi}}, \ and\ \bibinfo {author}
  {\bibfnamefont {M.~S.}\ \bibnamefont {Chapman}},\ }\href {\doibase
  10.1103/PhysRevLett.98.233601} {\bibfield  {journal} {\bibinfo  {journal}
  {Phys. Rev. Lett.}\ }\textbf {\bibinfo {volume} {98}},\ \bibinfo {pages}
  {233601} (\bibinfo {year} {2007})}\BibitemShut {NoStop}%
\bibitem [{\citenamefont {Eto}\ \emph {et~al.}(2011)\citenamefont {Eto},
  \citenamefont {Noguchi}, \citenamefont {Zhang}, \citenamefont {Ueda},\ and\
  \citenamefont {Kozuma}}]{PRL11_Mikio}%
  \BibitemOpen
  \bibfield  {author} {\bibinfo {author} {\bibfnamefont {Y.}~\bibnamefont
  {Eto}}, \bibinfo {author} {\bibfnamefont {A.}~\bibnamefont {Noguchi}},
  \bibinfo {author} {\bibfnamefont {P.}~\bibnamefont {Zhang}}, \bibinfo
  {author} {\bibfnamefont {M.}~\bibnamefont {Ueda}}, \ and\ \bibinfo {author}
  {\bibfnamefont {M.}~\bibnamefont {Kozuma}},\ }\href {\doibase
  10.1103/PhysRevLett.106.160501} {\bibfield  {journal} {\bibinfo  {journal}
  {Phys. Rev. Lett.}\ }\textbf {\bibinfo {volume} {106}},\ \bibinfo {pages}
  {160501} (\bibinfo {year} {2011})}\BibitemShut {NoStop}%
\bibitem [{\citenamefont {Fink}\ \emph {et~al.}(2009)\citenamefont {Fink},
  \citenamefont {Bianchetti}, \citenamefont {Baur}, \citenamefont {G\"oppl},
  \citenamefont {Steffen}, \citenamefont {Filipp}, \citenamefont {Leek},
  \citenamefont {Blais},\ and\ \citenamefont {Wallraff}}]{PRL09_Wallraff}%
  \BibitemOpen
  \bibfield  {author} {\bibinfo {author} {\bibfnamefont {J.~M.}\ \bibnamefont
  {Fink}}, \bibinfo {author} {\bibfnamefont {R.}~\bibnamefont {Bianchetti}},
  \bibinfo {author} {\bibfnamefont {M.}~\bibnamefont {Baur}}, \bibinfo {author}
  {\bibfnamefont {M.}~\bibnamefont {G\"oppl}}, \bibinfo {author} {\bibfnamefont
  {L.}~\bibnamefont {Steffen}}, \bibinfo {author} {\bibfnamefont
  {S.}~\bibnamefont {Filipp}}, \bibinfo {author} {\bibfnamefont {P.~J.}\
  \bibnamefont {Leek}}, \bibinfo {author} {\bibfnamefont {A.}~\bibnamefont
  {Blais}}, \ and\ \bibinfo {author} {\bibfnamefont {A.}~\bibnamefont
  {Wallraff}},\ }\href {\doibase 10.1103/PhysRevLett.103.083601} {\bibfield
  {journal} {\bibinfo  {journal} {Phys. Rev. Lett.}\ }\textbf {\bibinfo
  {volume} {103}},\ \bibinfo {pages} {083601} (\bibinfo {year}
  {2009})}\BibitemShut {NoStop}%
\bibitem [{\citenamefont {Pritchard}\ \emph {et~al.}(2010)\citenamefont
  {Pritchard}, \citenamefont {Maxwell}, \citenamefont {Gauguet}, \citenamefont
  {Weatherill}, \citenamefont {Jones},\ and\ \citenamefont
  {Adams}}]{PRL10_Adams}%
  \BibitemOpen
  \bibfield  {author} {\bibinfo {author} {\bibfnamefont {J.~D.}\ \bibnamefont
  {Pritchard}}, \bibinfo {author} {\bibfnamefont {D.}~\bibnamefont {Maxwell}},
  \bibinfo {author} {\bibfnamefont {A.}~\bibnamefont {Gauguet}}, \bibinfo
  {author} {\bibfnamefont {K.~J.}\ \bibnamefont {Weatherill}}, \bibinfo
  {author} {\bibfnamefont {M.~P.~A.}\ \bibnamefont {Jones}}, \ and\ \bibinfo
  {author} {\bibfnamefont {C.~S.}\ \bibnamefont {Adams}},\ }\href {\doibase
  10.1103/PhysRevLett.105.193603} {\bibfield  {journal} {\bibinfo  {journal}
  {Phys. Rev. Lett.}\ }\textbf {\bibinfo {volume} {105}},\ \bibinfo {pages}
  {193603} (\bibinfo {year} {2010})}\BibitemShut {NoStop}%
\bibitem [{\citenamefont {Thompson}\ \emph {et~al.}(1998)\citenamefont
  {Thompson}, \citenamefont {Turchette}, \citenamefont {Carnal},\ and\
  \citenamefont {Kimble}}]{PRA98_Kimble}%
  \BibitemOpen
  \bibfield  {author} {\bibinfo {author} {\bibfnamefont {R.~J.}\ \bibnamefont
  {Thompson}}, \bibinfo {author} {\bibfnamefont {Q.~A.}\ \bibnamefont
  {Turchette}}, \bibinfo {author} {\bibfnamefont {O.}~\bibnamefont {Carnal}}, \
  and\ \bibinfo {author} {\bibfnamefont {H.~J.}\ \bibnamefont {Kimble}},\
  }\href {\doibase 10.1103/PhysRevA.57.3084} {\bibfield  {journal} {\bibinfo
  {journal} {Phys. Rev. A}\ }\textbf {\bibinfo {volume} {57}},\ \bibinfo
  {pages} {3084} (\bibinfo {year} {1998})}\BibitemShut {NoStop}%
\bibitem [{\citenamefont {Souza}\ \emph {et~al.}(2013)\citenamefont {Souza},
  \citenamefont {Figueroa}, \citenamefont {Chibani}, \citenamefont
  {Villas-Boas},\ and\ \citenamefont {Rempe}}]{PRL13_Rempe}%
  \BibitemOpen
  \bibfield  {author} {\bibinfo {author} {\bibfnamefont {J.~A.}\ \bibnamefont
  {Souza}}, \bibinfo {author} {\bibfnamefont {E.}~\bibnamefont {Figueroa}},
  \bibinfo {author} {\bibfnamefont {H.}~\bibnamefont {Chibani}}, \bibinfo
  {author} {\bibfnamefont {C.~J.}\ \bibnamefont {Villas-Boas}}, \ and\ \bibinfo
  {author} {\bibfnamefont {G.}~\bibnamefont {Rempe}},\ }\href {\doibase
  10.1103/PhysRevLett.111.113602} {\bibfield  {journal} {\bibinfo  {journal}
  {Phys. Rev. Lett.}\ }\textbf {\bibinfo {volume} {111}},\ \bibinfo {pages}
  {113602} (\bibinfo {year} {2013})}\BibitemShut {NoStop}%
\bibitem [{\citenamefont {Guerlin}\ \emph {et~al.}(2010)\citenamefont
  {Guerlin}, \citenamefont {Brion}, \citenamefont {Esslinger},\ and\
  \citenamefont {M\o{}lmer}}]{PRA10_Molmer}%
  \BibitemOpen
  \bibfield  {author} {\bibinfo {author} {\bibfnamefont {C.}~\bibnamefont
  {Guerlin}}, \bibinfo {author} {\bibfnamefont {E.}~\bibnamefont {Brion}},
  \bibinfo {author} {\bibfnamefont {T.}~\bibnamefont {Esslinger}}, \ and\
  \bibinfo {author} {\bibfnamefont {K.}~\bibnamefont {M\o{}lmer}},\ }\href
  {\doibase 10.1103/PhysRevA.82.053832} {\bibfield  {journal} {\bibinfo
  {journal} {Phys. Rev. A}\ }\textbf {\bibinfo {volume} {82}},\ \bibinfo
  {pages} {053832} (\bibinfo {year} {2010})}\BibitemShut {NoStop}%
\bibitem [{\citenamefont {Ritsch}\ \emph {et~al.}(1992)\citenamefont {Ritsch},
  \citenamefont {Marte},\ and\ \citenamefont {Zoller}}]{EPL92_Zoller}%
  \BibitemOpen
  \bibfield  {author} {\bibinfo {author} {\bibfnamefont {H.}~\bibnamefont
  {Ritsch}}, \bibinfo {author} {\bibfnamefont {M.~A.~M.}\ \bibnamefont
  {Marte}}, \ and\ \bibinfo {author} {\bibfnamefont {P.}~\bibnamefont
  {Zoller}},\ }\href {\doibase 10.1209/0295-5075/19/1/002} {\bibfield
  {journal} {\bibinfo  {journal} {Euro. Phys. Lett.}\ }\textbf {\bibinfo
  {volume} {19}},\ \bibinfo {pages} {7} (\bibinfo {year} {1992})}\BibitemShut
  {NoStop}%
\bibitem [{\citenamefont {Haake}\ \emph {et~al.}(1993)\citenamefont {Haake},
  \citenamefont {Kolobov}, \citenamefont {Fabre}, \citenamefont {Giacobino},\
  and\ \citenamefont {Reynaud}}]{PRL93_Haake}%
  \BibitemOpen
  \bibfield  {author} {\bibinfo {author} {\bibfnamefont {F.}~\bibnamefont
  {Haake}}, \bibinfo {author} {\bibfnamefont {M.~I.}\ \bibnamefont {Kolobov}},
  \bibinfo {author} {\bibfnamefont {C.}~\bibnamefont {Fabre}}, \bibinfo
  {author} {\bibfnamefont {E.}~\bibnamefont {Giacobino}}, \ and\ \bibinfo
  {author} {\bibfnamefont {S.}~\bibnamefont {Reynaud}},\ }\href {\doibase
  10.1103/PhysRevLett.71.995} {\bibfield  {journal} {\bibinfo  {journal} {Phys.
  Rev. Lett.}\ }\textbf {\bibinfo {volume} {71}},\ \bibinfo {pages} {995}
  (\bibinfo {year} {1993})}\BibitemShut {NoStop}%
\bibitem [{\citenamefont {Seeger}\ \emph {et~al.}(1996)\citenamefont {Seeger},
  \citenamefont {Kolobov}, \citenamefont {Ku\ifmmode~\acute{s}\else
  \'{s}\fi{}},\ and\ \citenamefont {Haake}}]{PRA96_Haake}%
  \BibitemOpen
  \bibfield  {author} {\bibinfo {author} {\bibfnamefont {C.}~\bibnamefont
  {Seeger}}, \bibinfo {author} {\bibfnamefont {M.~I.}\ \bibnamefont {Kolobov}},
  \bibinfo {author} {\bibfnamefont {M.}~\bibnamefont {Ku\ifmmode~\acute{s}\else
  \'{s}\fi{}}}, \ and\ \bibinfo {author} {\bibfnamefont {F.}~\bibnamefont
  {Haake}},\ }\href {\doibase 10.1103/PhysRevA.54.4440} {\bibfield  {journal}
  {\bibinfo  {journal} {Phys. Rev. A}\ }\textbf {\bibinfo {volume} {54}},\
  \bibinfo {pages} {4440} (\bibinfo {year} {1996})}\BibitemShut {NoStop}%
\bibitem [{\citenamefont {M\"ucke}\ \emph {et~al.}(2013)\citenamefont
  {M\"ucke}, \citenamefont {Bochmann}, \citenamefont {Hahn}, \citenamefont
  {Neuzner}, \citenamefont {N\"olleke}, \citenamefont {Reiserer}, \citenamefont
  {Rempe},\ and\ \citenamefont {Ritter}}]{PRA13_Rempe}%
  \BibitemOpen
  \bibfield  {author} {\bibinfo {author} {\bibfnamefont {M.}~\bibnamefont
  {M\"ucke}}, \bibinfo {author} {\bibfnamefont {J.}~\bibnamefont {Bochmann}},
  \bibinfo {author} {\bibfnamefont {C.}~\bibnamefont {Hahn}}, \bibinfo {author}
  {\bibfnamefont {A.}~\bibnamefont {Neuzner}}, \bibinfo {author} {\bibfnamefont
  {C.}~\bibnamefont {N\"olleke}}, \bibinfo {author} {\bibfnamefont
  {A.}~\bibnamefont {Reiserer}}, \bibinfo {author} {\bibfnamefont
  {G.}~\bibnamefont {Rempe}}, \ and\ \bibinfo {author} {\bibfnamefont
  {S.}~\bibnamefont {Ritter}},\ }\href {\doibase 10.1103/PhysRevA.87.063805}
  {\bibfield  {journal} {\bibinfo  {journal} {Phys. Rev. A}\ }\textbf {\bibinfo
  {volume} {87}},\ \bibinfo {pages} {063805} (\bibinfo {year}
  {2013})}\BibitemShut {NoStop}%
\bibitem [{\citenamefont {M\"uller}\ \emph {et~al.}(2013)\citenamefont
  {M\"uller}, \citenamefont {K\"olle}, \citenamefont {L\"ow}, \citenamefont
  {Pfau}, \citenamefont {Calarco},\ and\ \citenamefont
  {Montangero}}]{PRA13_Pfau}%
  \BibitemOpen
  \bibfield  {author} {\bibinfo {author} {\bibfnamefont {M.~M.}\ \bibnamefont
  {M\"uller}}, \bibinfo {author} {\bibfnamefont {A.}~\bibnamefont {K\"olle}},
  \bibinfo {author} {\bibfnamefont {R.}~\bibnamefont {L\"ow}}, \bibinfo
  {author} {\bibfnamefont {T.}~\bibnamefont {Pfau}}, \bibinfo {author}
  {\bibfnamefont {T.}~\bibnamefont {Calarco}}, \ and\ \bibinfo {author}
  {\bibfnamefont {S.}~\bibnamefont {Montangero}},\ }\href {\doibase
  10.1103/PhysRevA.87.053412} {\bibfield  {journal} {\bibinfo  {journal} {Phys.
  Rev. A}\ }\textbf {\bibinfo {volume} {87}},\ \bibinfo {pages} {053412}
  (\bibinfo {year} {2013})}\BibitemShut {NoStop}%
\bibitem [{\citenamefont {Schoelkopf}\ and\ \citenamefont
  {Girvin}(2008)}]{Nat08_Girvin}%
  \BibitemOpen
  \bibfield  {author} {\bibinfo {author} {\bibfnamefont {R.~J.}\ \bibnamefont
  {Schoelkopf}}\ and\ \bibinfo {author} {\bibfnamefont {S.~M.}\ \bibnamefont
  {Girvin}},\ }\href {\doibase 10.1038/451664a} {\bibfield  {journal} {\bibinfo
   {journal} {Nature}\ }\textbf {\bibinfo {volume} {451}},\ \bibinfo {pages}
  {664} (\bibinfo {year} {2008})}\BibitemShut {NoStop}%
\bibitem [{\citenamefont {M\o{}ller}\ \emph {et~al.}(2008)\citenamefont
  {M\o{}ller}, \citenamefont {Madsen},\ and\ \citenamefont
  {M\o{}lmer}}]{PRL08_Molmer}%
  \BibitemOpen
  \bibfield  {author} {\bibinfo {author} {\bibfnamefont {D.}~\bibnamefont
  {M\o{}ller}}, \bibinfo {author} {\bibfnamefont {L.~B.}\ \bibnamefont
  {Madsen}}, \ and\ \bibinfo {author} {\bibfnamefont {K.}~\bibnamefont
  {M\o{}lmer}},\ }\href {\doibase 10.1103/PhysRevLett.100.170504} {\bibfield
  {journal} {\bibinfo  {journal} {Phys. Rev. Lett.}\ }\textbf {\bibinfo
  {volume} {100}},\ \bibinfo {pages} {170504} (\bibinfo {year}
  {2008})}\BibitemShut {NoStop}%
\bibitem [{\citenamefont {Saffman}\ \emph {et~al.}(2010)\citenamefont
  {Saffman}, \citenamefont {Walker},\ and\ \citenamefont
  {M\o{}lmer}}]{RMP10_Molmer}%
  \BibitemOpen
  \bibfield  {author} {\bibinfo {author} {\bibfnamefont {M.}~\bibnamefont
  {Saffman}}, \bibinfo {author} {\bibfnamefont {T.~G.}\ \bibnamefont {Walker}},
  \ and\ \bibinfo {author} {\bibfnamefont {K.}~\bibnamefont {M\o{}lmer}},\
  }\href {\doibase 10.1103/RevModPhys.82.2313} {\bibfield  {journal} {\bibinfo
  {journal} {Rev. Mod. Phys.}\ }\textbf {\bibinfo {volume} {82}},\ \bibinfo
  {pages} {2313} (\bibinfo {year} {2010})}\BibitemShut {NoStop}%
\bibitem [{\citenamefont {Thompson}\ \emph {et~al.}(2006)\citenamefont
  {Thompson}, \citenamefont {Simon}, \citenamefont {Loh},\ and\ \citenamefont
  {Vuleti\'{c}}}]{Thompson06}%
  \BibitemOpen
  \bibfield  {author} {\bibinfo {author} {\bibfnamefont {J.~K.}\ \bibnamefont
  {Thompson}}, \bibinfo {author} {\bibfnamefont {J.}~\bibnamefont {Simon}},
  \bibinfo {author} {\bibfnamefont {H.}~\bibnamefont {Loh}}, \ and\ \bibinfo
  {author} {\bibfnamefont {V.}~\bibnamefont {Vuleti\'{c}}},\ }\href {\doibase
  10.1126/science.1127676} {\bibfield  {journal} {\bibinfo  {journal}
  {Science}\ }\textbf {\bibinfo {volume} {313}},\ \bibinfo {pages} {74}
  (\bibinfo {year} {2006})}\BibitemShut {NoStop}%
\bibitem [{\citenamefont {Massar}\ and\ \citenamefont
  {Polzik}(2003)}]{PRL03_Polzik}%
  \BibitemOpen
  \bibfield  {author} {\bibinfo {author} {\bibfnamefont {S.}~\bibnamefont
  {Massar}}\ and\ \bibinfo {author} {\bibfnamefont {E.~S.}\ \bibnamefont
  {Polzik}},\ }\href {\doibase 10.1103/PhysRevLett.91.060401} {\bibfield
  {journal} {\bibinfo  {journal} {Phys. Rev. Lett.}\ }\textbf {\bibinfo
  {volume} {91}},\ \bibinfo {pages} {060401} (\bibinfo {year}
  {2003})}\BibitemShut {NoStop}%
\bibitem [{\citenamefont {Hammerer}\ \emph {et~al.}(2010)\citenamefont
  {Hammerer}, \citenamefont {S\o{}rensen},\ and\ \citenamefont
  {Polzik}}]{RMP10_Polzik}%
  \BibitemOpen
  \bibfield  {author} {\bibinfo {author} {\bibfnamefont {K.}~\bibnamefont
  {Hammerer}}, \bibinfo {author} {\bibfnamefont {A.~S.}\ \bibnamefont
  {S\o{}rensen}}, \ and\ \bibinfo {author} {\bibfnamefont {E.~S.}\ \bibnamefont
  {Polzik}},\ }\href {\doibase 10.1103/RevModPhys.82.1041} {\bibfield
  {journal} {\bibinfo  {journal} {Rev. Mod. Phys.}\ }\textbf {\bibinfo {volume}
  {82}},\ \bibinfo {pages} {1041} (\bibinfo {year} {2010})}\BibitemShut
  {NoStop}%
\bibitem [{\citenamefont {Brekke}\ \emph {et~al.}(2008)\citenamefont {Brekke},
  \citenamefont {Day},\ and\ \citenamefont {Walker}}]{PRA08_Walker}%
  \BibitemOpen
  \bibfield  {author} {\bibinfo {author} {\bibfnamefont {E.}~\bibnamefont
  {Brekke}}, \bibinfo {author} {\bibfnamefont {J.~O.}\ \bibnamefont {Day}}, \
  and\ \bibinfo {author} {\bibfnamefont {T.~G.}\ \bibnamefont {Walker}},\
  }\href {\doibase 10.1103/PhysRevA.78.063830} {\bibfield  {journal} {\bibinfo
  {journal} {Phys. Rev. A}\ }\textbf {\bibinfo {volume} {78}},\ \bibinfo
  {pages} {063830} (\bibinfo {year} {2008})}\BibitemShut {NoStop}%
\bibitem [{\citenamefont {K\"olle}\ \emph {et~al.}(2012)\citenamefont
  {K\"olle}, \citenamefont {Epple}, \citenamefont {K\"ubler}, \citenamefont
  {L\"ow},\ and\ \citenamefont {Pfau}}]{PRA_12_Pfau}%
  \BibitemOpen
  \bibfield  {author} {\bibinfo {author} {\bibfnamefont {A.}~\bibnamefont
  {K\"olle}}, \bibinfo {author} {\bibfnamefont {G.}~\bibnamefont {Epple}},
  \bibinfo {author} {\bibfnamefont {H.}~\bibnamefont {K\"ubler}}, \bibinfo
  {author} {\bibfnamefont {R.}~\bibnamefont {L\"ow}}, \ and\ \bibinfo {author}
  {\bibfnamefont {T.}~\bibnamefont {Pfau}},\ }\href {\doibase
  10.1103/PhysRevA.85.063821} {\bibfield  {journal} {\bibinfo  {journal} {Phys.
  Rev. A}\ }\textbf {\bibinfo {volume} {85}},\ \bibinfo {pages} {063821}
  (\bibinfo {year} {2012})}\BibitemShut {NoStop}%
\bibitem [{\citenamefont {Ates}\ \emph {et~al.}(2007)\citenamefont {Ates},
  \citenamefont {Pohl}, \citenamefont {Pattard},\ and\ \citenamefont
  {Rost}}]{PRA07_Rost}%
  \BibitemOpen
  \bibfield  {author} {\bibinfo {author} {\bibfnamefont {C.}~\bibnamefont
  {Ates}}, \bibinfo {author} {\bibfnamefont {T.}~\bibnamefont {Pohl}}, \bibinfo
  {author} {\bibfnamefont {T.}~\bibnamefont {Pattard}}, \ and\ \bibinfo
  {author} {\bibfnamefont {J.~M.}\ \bibnamefont {Rost}},\ }\href {\doibase
  10.1103/PhysRevA.76.013413} {\bibfield  {journal} {\bibinfo  {journal} {Phys.
  Rev. A}\ }\textbf {\bibinfo {volume} {76}},\ \bibinfo {pages} {013413}
  (\bibinfo {year} {2007})}\BibitemShut {NoStop}%
\bibitem [{\citenamefont {Singer}\ \emph {et~al.}(2004)\citenamefont {Singer},
  \citenamefont {Reetz-Lamour}, \citenamefont {Amthor}, \citenamefont
  {Marcassa},\ and\ \citenamefont {Weidem\"uller}}]{PRL04_Weidemuller}%
  \BibitemOpen
  \bibfield  {author} {\bibinfo {author} {\bibfnamefont {K.}~\bibnamefont
  {Singer}}, \bibinfo {author} {\bibfnamefont {M.}~\bibnamefont
  {Reetz-Lamour}}, \bibinfo {author} {\bibfnamefont {T.}~\bibnamefont
  {Amthor}}, \bibinfo {author} {\bibfnamefont {L.~G.}\ \bibnamefont
  {Marcassa}}, \ and\ \bibinfo {author} {\bibfnamefont {M.}~\bibnamefont
  {Weidem\"uller}},\ }\href {\doibase 10.1103/PhysRevLett.93.163001} {\bibfield
   {journal} {\bibinfo  {journal} {Phys. Rev. Lett.}\ }\textbf {\bibinfo
  {volume} {93}},\ \bibinfo {pages} {163001} (\bibinfo {year}
  {2004})}\BibitemShut {NoStop}%
\bibitem [{\citenamefont {Stanojevic}\ and\ \citenamefont
  {C\^ot\'e}(2009)}]{PRA09_Cote}%
  \BibitemOpen
  \bibfield  {author} {\bibinfo {author} {\bibfnamefont {J.}~\bibnamefont
  {Stanojevic}}\ and\ \bibinfo {author} {\bibfnamefont {R.}~\bibnamefont
  {C\^ot\'e}},\ }\href {\doibase 10.1103/PhysRevA.80.033418} {\bibfield
  {journal} {\bibinfo  {journal} {Phys. Rev. A}\ }\textbf {\bibinfo {volume}
  {80}},\ \bibinfo {pages} {033418} (\bibinfo {year} {2009})}\BibitemShut
  {NoStop}%
\bibitem [{\citenamefont {Robicheaux}\ and\ \citenamefont
  {Hern\'andez}(2005)}]{PRA05_Robicheaux}%
  \BibitemOpen
  \bibfield  {author} {\bibinfo {author} {\bibfnamefont {F.}~\bibnamefont
  {Robicheaux}}\ and\ \bibinfo {author} {\bibfnamefont {J.~V.}\ \bibnamefont
  {Hern\'andez}},\ }\href {\doibase 10.1103/PhysRevA.72.063403} {\bibfield
  {journal} {\bibinfo  {journal} {Phys. Rev. A}\ }\textbf {\bibinfo {volume}
  {72}},\ \bibinfo {pages} {063403} (\bibinfo {year} {2005})}\BibitemShut
  {NoStop}%
\bibitem [{\citenamefont {Saffman}\ and\ \citenamefont
  {Walker}(2002)}]{PRA02_Saffman}%
  \BibitemOpen
  \bibfield  {author} {\bibinfo {author} {\bibfnamefont {M.}~\bibnamefont
  {Saffman}}\ and\ \bibinfo {author} {\bibfnamefont {T.~G.}\ \bibnamefont
  {Walker}},\ }\href {\doibase 10.1103/PhysRevA.66.065403} {\bibfield
  {journal} {\bibinfo  {journal} {Phys. Rev. A}\ }\textbf {\bibinfo {volume}
  {66}},\ \bibinfo {pages} {065403} (\bibinfo {year} {2002})}\BibitemShut
  {NoStop}%
\bibitem [{\citenamefont {Fleischhauer}\ \emph {et~al.}(2000)\citenamefont
  {Fleischhauer}, \citenamefont {Yelin},\ and\ \citenamefont
  {Lukin}}]{OptCom00_Fleischhauer}%
  \BibitemOpen
  \bibfield  {author} {\bibinfo {author} {\bibfnamefont {M.}~\bibnamefont
  {Fleischhauer}}, \bibinfo {author} {\bibfnamefont {S.}~\bibnamefont {Yelin}},
  \ and\ \bibinfo {author} {\bibfnamefont {M.}~\bibnamefont {Lukin}},\ }\href
  {\doibase http://dx.doi.org/10.1016/S0030-4018(99)00679-3} {\bibfield
  {journal} {\bibinfo  {journal} {Opt. Commun.}\ }\textbf {\bibinfo {volume}
  {179}},\ \bibinfo {pages} {395 } (\bibinfo {year} {2000})}\BibitemShut
  {NoStop}%
\bibitem [{\citenamefont {Overstreet}\ \emph {et~al.}(2009)\citenamefont
  {Overstreet}, \citenamefont {Schwettmann}, \citenamefont {Tallant},
  \citenamefont {Booth},\ and\ \citenamefont {Shaffer}}]{NatPhy09_Richard}%
  \BibitemOpen
  \bibfield  {author} {\bibinfo {author} {\bibfnamefont {K.~R.}\ \bibnamefont
  {Overstreet}}, \bibinfo {author} {\bibfnamefont {A.}~\bibnamefont
  {Schwettmann}}, \bibinfo {author} {\bibfnamefont {J.}~\bibnamefont
  {Tallant}}, \bibinfo {author} {\bibfnamefont {D.}~\bibnamefont {Booth}}, \
  and\ \bibinfo {author} {\bibfnamefont {J.}~\bibnamefont {Shaffer}},\ }\href
  {\doibase 10.1038/nphys1307} {\bibfield  {journal} {\bibinfo  {journal} {Nat.
  Phys.}\ }\textbf {\bibinfo {volume} {5}},\ \bibinfo {pages} {581} (\bibinfo
  {year} {2009})}\BibitemShut {NoStop}%
\bibitem [{\citenamefont {Overstreet}\ \emph {et~al.}(2007)\citenamefont
  {Overstreet}, \citenamefont {Schwettmann}, \citenamefont {Tallant},\ and\
  \citenamefont {Shaffer}}]{PRA07_Richard}%
  \BibitemOpen
  \bibfield  {author} {\bibinfo {author} {\bibfnamefont {K.~R.}\ \bibnamefont
  {Overstreet}}, \bibinfo {author} {\bibfnamefont {A.}~\bibnamefont
  {Schwettmann}}, \bibinfo {author} {\bibfnamefont {J.}~\bibnamefont
  {Tallant}}, \ and\ \bibinfo {author} {\bibfnamefont {J.~P.}\ \bibnamefont
  {Shaffer}},\ }\href {\doibase 10.1103/PhysRevA.76.011403} {\bibfield
  {journal} {\bibinfo  {journal} {Phys. Rev. A}\ }\textbf {\bibinfo {volume}
  {76}},\ \bibinfo {pages} {011403} (\bibinfo {year} {2007})}\BibitemShut
  {NoStop}%
\bibitem [{\citenamefont {Beterov}\ \emph {et~al.}(2009)\citenamefont
  {Beterov}, \citenamefont {Ryabtsev}, \citenamefont {Tretyakov},\ and\
  \citenamefont {Entin}}]{PRA09_Beterov}%
  \BibitemOpen
  \bibfield  {author} {\bibinfo {author} {\bibfnamefont {I.~I.}\ \bibnamefont
  {Beterov}}, \bibinfo {author} {\bibfnamefont {I.~I.}\ \bibnamefont
  {Ryabtsev}}, \bibinfo {author} {\bibfnamefont {D.~B.}\ \bibnamefont
  {Tretyakov}}, \ and\ \bibinfo {author} {\bibfnamefont {V.~M.}\ \bibnamefont
  {Entin}},\ }\href {\doibase 10.1103/PhysRevA.79.052504} {\bibfield  {journal}
  {\bibinfo  {journal} {Phys. Rev. A}\ }\textbf {\bibinfo {volume} {79}},\
  \bibinfo {pages} {052504} (\bibinfo {year} {2009})}\BibitemShut {NoStop}%
\bibitem [{\citenamefont {Grynberg}\ \emph {et~al.}(2010)\citenamefont
  {Grynberg}, \citenamefont {Aspect},\ and\ \citenamefont
  {Fabre}}]{QO_Fabre10}%
  \BibitemOpen
  \bibfield  {author} {\bibinfo {author} {\bibfnamefont {G.}~\bibnamefont
  {Grynberg}}, \bibinfo {author} {\bibfnamefont {A.}~\bibnamefont {Aspect}}, \
  and\ \bibinfo {author} {\bibfnamefont {C.}~\bibnamefont {Fabre}},\ }\href
  {http://dx.doi.org/} {\emph {\bibinfo {title} {Quantum Optics}}}\ (\bibinfo
  {publisher} {Cambridge University Press, New York},\ \bibinfo {year}
  {2010})\BibitemShut {NoStop}%
\bibitem [{\citenamefont {Beterov}\ \emph {et~al.}(2011)\citenamefont
  {Beterov}, \citenamefont {Tretyakov}, \citenamefont {Entin}, \citenamefont
  {Yakshina}, \citenamefont {Ryabtsev}, \citenamefont {MacCormick},\ and\
  \citenamefont {Bergamini}}]{PRA11_Bergamini}%
  \BibitemOpen
  \bibfield  {author} {\bibinfo {author} {\bibfnamefont {I.~I.}\ \bibnamefont
  {Beterov}}, \bibinfo {author} {\bibfnamefont {D.~B.}\ \bibnamefont
  {Tretyakov}}, \bibinfo {author} {\bibfnamefont {V.~M.}\ \bibnamefont
  {Entin}}, \bibinfo {author} {\bibfnamefont {E.~A.}\ \bibnamefont {Yakshina}},
  \bibinfo {author} {\bibfnamefont {I.~I.}\ \bibnamefont {Ryabtsev}}, \bibinfo
  {author} {\bibfnamefont {C.}~\bibnamefont {MacCormick}}, \ and\ \bibinfo
  {author} {\bibfnamefont {S.}~\bibnamefont {Bergamini}},\ }\href {\doibase
  10.1103/PhysRevA.84.023413} {\bibfield  {journal} {\bibinfo  {journal} {Phys.
  Rev. A}\ }\textbf {\bibinfo {volume} {84}},\ \bibinfo {pages} {023413}
  (\bibinfo {year} {2011})}\BibitemShut {NoStop}%
\bibitem [{\citenamefont {Petrosyan}\ and\ \citenamefont
  {M\o{}lmer}(2013)}]{PRA13_Molmer}%
  \BibitemOpen
  \bibfield  {author} {\bibinfo {author} {\bibfnamefont {D.}~\bibnamefont
  {Petrosyan}}\ and\ \bibinfo {author} {\bibfnamefont {K.}~\bibnamefont
  {M\o{}lmer}},\ }\href {\doibase 10.1103/PhysRevA.87.033416} {\bibfield
  {journal} {\bibinfo  {journal} {Phys. Rev. A}\ }\textbf {\bibinfo {volume}
  {87}},\ \bibinfo {pages} {033416} (\bibinfo {year} {2013})}\BibitemShut
  {NoStop}%
\bibitem [{\citenamefont {Grynberg}\ and\ \citenamefont
  {Stora}(1982)}]{Grynberg82}%
  \BibitemOpen
  \bibfield  {author} {\bibinfo {author} {\bibfnamefont {G.}~\bibnamefont
  {Grynberg}}\ and\ \bibinfo {author} {\bibfnamefont {R.}~\bibnamefont
  {Stora}},\ }\href@noop {} {\emph {\bibinfo {title} {New Trends in Atomic
  Physics, Les Houches, Session XXXVIII}}}\ (\bibinfo  {publisher} {Elsevier
  science publishers},\ \bibinfo {year} {1982})\BibitemShut {NoStop}%
\bibitem [{\citenamefont {Lambropoulos}\ and\ \citenamefont
  {Petrosyan}(2006)}]{Petrosyan06}%
  \BibitemOpen
  \bibfield  {author} {\bibinfo {author} {\bibfnamefont {P.}~\bibnamefont
  {Lambropoulos}}\ and\ \bibinfo {author} {\bibfnamefont {D.}~\bibnamefont
  {Petrosyan}},\ }\href {http://dx.doi.org/10.1017/CBO9780511524530} {\emph
  {\bibinfo {title} {Fundamentals of Quantum Optics and Quantum Information}}}\
  (\bibinfo  {publisher} {Springer Berlin Heidelberg, New York},\ \bibinfo
  {year} {2006})\BibitemShut {NoStop}%
\bibitem [{\citenamefont {Agarwal}\ and\ \citenamefont
  {Puri}(1986)}]{PRA86_Agarwal}%
  \BibitemOpen
  \bibfield  {author} {\bibinfo {author} {\bibfnamefont {G.~S.}\ \bibnamefont
  {Agarwal}}\ and\ \bibinfo {author} {\bibfnamefont {R.~R.}\ \bibnamefont
  {Puri}},\ }\href {\doibase 10.1103/PhysRevA.33.1757} {\bibfield  {journal}
  {\bibinfo  {journal} {Phys. Rev. A}\ }\textbf {\bibinfo {volume} {33}},\
  \bibinfo {pages} {1757} (\bibinfo {year} {1986})}\BibitemShut {NoStop}%
\bibitem [{\citenamefont {Law}\ and\ \citenamefont {Kimble}(1997)}]{Law97}%
  \BibitemOpen
  \bibfield  {author} {\bibinfo {author} {\bibfnamefont {C.~K.}\ \bibnamefont
  {Law}}\ and\ \bibinfo {author} {\bibfnamefont {H.~J.}\ \bibnamefont
  {Kimble}},\ }\href
  {hhttp://www.tandfonline.com/doi/abs/10.1080/09500349708231869} {\bibfield
  {journal} {\bibinfo  {journal} {J. Mod. Opt.}\ }\textbf {\bibinfo {volume}
  {44}},\ \bibinfo {pages} {2067} (\bibinfo {year} {1997})}\BibitemShut
  {NoStop}%
\bibitem [{\citenamefont {Kuhn}\ and\ \citenamefont
  {Ljunggren}(2010)}]{Khun10}%
  \BibitemOpen
  \bibfield  {author} {\bibinfo {author} {\bibfnamefont {A.}~\bibnamefont
  {Kuhn}}\ and\ \bibinfo {author} {\bibfnamefont {D.}~\bibnamefont
  {Ljunggren}},\ }\href
  {http://www.tandfonline.com/doi/abs/10.1080/00107511003602990} {\bibfield
  {journal} {\bibinfo  {journal} {Contemporary Phys.}\ }\textbf {\bibinfo
  {volume} {51}},\ \bibinfo {pages} {289} (\bibinfo {year} {2010})}\BibitemShut
  {NoStop}%
\bibitem [{\citenamefont {Lee}\ \emph {et~al.}(2012)\citenamefont {Lee},
  \citenamefont {H\"affner},\ and\ \citenamefont {Cross}}]{PRL12_Cross}%
  \BibitemOpen
  \bibfield  {author} {\bibinfo {author} {\bibfnamefont {T.~E.}\ \bibnamefont
  {Lee}}, \bibinfo {author} {\bibfnamefont {H.}~\bibnamefont {H\"affner}}, \
  and\ \bibinfo {author} {\bibfnamefont {M.~C.}\ \bibnamefont {Cross}},\ }\href
  {\doibase 10.1103/PhysRevLett.108.023602} {\bibfield  {journal} {\bibinfo
  {journal} {Phys. Rev. Lett.}\ }\textbf {\bibinfo {volume} {108}},\ \bibinfo
  {pages} {023602} (\bibinfo {year} {2012})}\BibitemShut {NoStop}%
\bibitem [{\citenamefont {Johnson}\ and\ \citenamefont
  {Rolston}(2010)}]{PRA_Rolston10}%
  \BibitemOpen
  \bibfield  {author} {\bibinfo {author} {\bibfnamefont {J.~E.}\ \bibnamefont
  {Johnson}}\ and\ \bibinfo {author} {\bibfnamefont {S.~L.}\ \bibnamefont
  {Rolston}},\ }\href {\doibase 10.1103/PhysRevA.82.033412} {\bibfield
  {journal} {\bibinfo  {journal} {Phys. Rev. A}\ }\textbf {\bibinfo {volume}
  {82}},\ \bibinfo {pages} {033412} (\bibinfo {year} {2010})}\BibitemShut
  {NoStop}%
\bibitem [{\citenamefont {Schwettmann}\ \emph {et~al.}(2007)\citenamefont
  {Schwettmann}, \citenamefont {Overstreet}, \citenamefont {Tallant},\ and\
  \citenamefont {Shaffer}}]{SchwettJMODOPT}%
  \BibitemOpen
  \bibfield  {author} {\bibinfo {author} {\bibfnamefont {A.}~\bibnamefont
  {Schwettmann}}, \bibinfo {author} {\bibfnamefont {K.~R.}\ \bibnamefont
  {Overstreet}}, \bibinfo {author} {\bibfnamefont {J.}~\bibnamefont {Tallant}},
  \ and\ \bibinfo {author} {\bibfnamefont {J.~P.}\ \bibnamefont {Shaffer}},\
  }\href {\doibase 10.1080/09500340701584076} {\bibfield  {journal} {\bibinfo
  {journal} {J. Mod. Opt.}\ }\textbf {\bibinfo {volume} {54}},\ \bibinfo
  {pages} {2551} (\bibinfo {year} {2007})}\BibitemShut {NoStop}%
\bibitem [{\citenamefont {Phoenix}\ and\ \citenamefont
  {Knight}(1988)}]{Phoenix1988}%
  \BibitemOpen
  \bibfield  {author} {\bibinfo {author} {\bibfnamefont {S.~J.~D.}\
  \bibnamefont {Phoenix}}\ and\ \bibinfo {author} {\bibfnamefont {P.~L.}\
  \bibnamefont {Knight}},\ }\href {\doibase
  http://dx.doi.org/10.1016/0003-4916(88)90006-1} {\bibfield  {journal}
  {\bibinfo  {journal} {Ann. Phys.}\ }\textbf {\bibinfo {volume} {186}},\
  \bibinfo {pages} {381 } (\bibinfo {year} {1988})}\BibitemShut {NoStop}%
\bibitem [{\citenamefont {Phoenix}\ and\ \citenamefont
  {Knight}(1991)}]{PRA91_Knight}%
  \BibitemOpen
  \bibfield  {author} {\bibinfo {author} {\bibfnamefont {S.~J.~D.}\
  \bibnamefont {Phoenix}}\ and\ \bibinfo {author} {\bibfnamefont {P.~L.}\
  \bibnamefont {Knight}},\ }\href {\doibase 10.1103/PhysRevA.44.6023}
  {\bibfield  {journal} {\bibinfo  {journal} {Phys. Rev. A}\ }\textbf {\bibinfo
  {volume} {44}},\ \bibinfo {pages} {6023} (\bibinfo {year}
  {1991})}\BibitemShut {NoStop}%
\bibitem [{\citenamefont {Olaya-Castro1}\ \emph {et~al.}(2004)\citenamefont
  {Olaya-Castro1}, \citenamefont {Johnson},\ and\ \citenamefont
  {Quiroga}}]{JPB04_Johnson}%
  \BibitemOpen
  \bibfield  {author} {\bibinfo {author} {\bibfnamefont {A.}~\bibnamefont
  {Olaya-Castro1}}, \bibinfo {author} {\bibfnamefont {N.~F.}\ \bibnamefont
  {Johnson}}, \ and\ \bibinfo {author} {\bibfnamefont {L.}~\bibnamefont
  {Quiroga}},\ }\href {http://iopscience.iop.org/1464-4266/6/8/016/cites}
  {\bibfield  {journal} {\bibinfo  {journal} {J. Opt. B: Quantum Semiclass.
  Opt.}\ }\textbf {\bibinfo {volume} {6}},\ \bibinfo {pages} {S730} (\bibinfo
  {year} {2004})}\BibitemShut {NoStop}%
\bibitem [{\citenamefont {Araki}\ and\ \citenamefont {Lieb}(1970)}]{Araki1970}%
  \BibitemOpen
  \bibfield  {author} {\bibinfo {author} {\bibfnamefont {H.}~\bibnamefont
  {Araki}}\ and\ \bibinfo {author} {\bibfnamefont {E.~H.}\ \bibnamefont
  {Lieb}},\ }\href
  {http://link.springer.com/article/10.1007%2FBF01646092#page-1} {\bibfield
  {journal} {\bibinfo  {journal} {Commun. Math. Phys.}\ }\textbf {\bibinfo
  {volume} {18}},\ \bibinfo {pages} {160 } (\bibinfo {year}
  {1970})}\BibitemShut {NoStop}%
\bibitem [{\citenamefont {Gea-Banacloche}(1990)}]{PRL_90_Julio}%
  \BibitemOpen
  \bibfield  {author} {\bibinfo {author} {\bibfnamefont {J.}~\bibnamefont
  {Gea-Banacloche}},\ }\href {\doibase 10.1103/PhysRevLett.65.3385} {\bibfield
  {journal} {\bibinfo  {journal} {Phys. Rev. Lett.}\ }\textbf {\bibinfo
  {volume} {65}},\ \bibinfo {pages} {3385} (\bibinfo {year}
  {1990})}\BibitemShut {NoStop}%
\bibitem [{\citenamefont {Plenio}\ \emph {et~al.}(1999)\citenamefont {Plenio},
  \citenamefont {Huelga}, \citenamefont {Beige},\ and\ \citenamefont
  {Knight}}]{PRA99_Knight}%
  \BibitemOpen
  \bibfield  {author} {\bibinfo {author} {\bibfnamefont {M.~B.}\ \bibnamefont
  {Plenio}}, \bibinfo {author} {\bibfnamefont {S.~F.}\ \bibnamefont {Huelga}},
  \bibinfo {author} {\bibfnamefont {A.}~\bibnamefont {Beige}}, \ and\ \bibinfo
  {author} {\bibfnamefont {P.~L.}\ \bibnamefont {Knight}},\ }\href {\doibase
  10.1103/PhysRevA.59.2468} {\bibfield  {journal} {\bibinfo  {journal} {Phys.
  Rev. A}\ }\textbf {\bibinfo {volume} {59}},\ \bibinfo {pages} {2468}
  (\bibinfo {year} {1999})}\BibitemShut {NoStop}%
\bibitem [{\citenamefont {Lounis}\ and\ \citenamefont
  {Moerner}(2000)}]{Nat00_Moerner}%
  \BibitemOpen
  \bibfield  {author} {\bibinfo {author} {\bibfnamefont {B.}~\bibnamefont
  {Lounis}}\ and\ \bibinfo {author} {\bibfnamefont {W.~E.}\ \bibnamefont
  {Moerner}},\ }\href
  {http://www.nature.com/nature/journal/v407/n6803/abs/407491a0.html}
  {\bibfield  {journal} {\bibinfo  {journal} {Nature}\ }\textbf {\bibinfo
  {volume} {407}},\ \bibinfo {pages} {491 } (\bibinfo {year}
  {2000})}\BibitemShut {NoStop}%
\bibitem [{\citenamefont {Eisaman}\ \emph {et~al.}(2011)\citenamefont
  {Eisaman}, \citenamefont {Fan}, \citenamefont {Migdall},\ and\ \citenamefont
  {Polyakov}}]{AIP11_Migdall}%
  \BibitemOpen
  \bibfield  {author} {\bibinfo {author} {\bibfnamefont {M.~D.}\ \bibnamefont
  {Eisaman}}, \bibinfo {author} {\bibfnamefont {J.}~\bibnamefont {Fan}},
  \bibinfo {author} {\bibfnamefont {A.}~\bibnamefont {Migdall}}, \ and\
  \bibinfo {author} {\bibfnamefont {S.~V.}\ \bibnamefont {Polyakov}},\ }\href
  {http://dx.doi.org/10.1063/1.3610677} {\bibfield  {journal} {\bibinfo
  {journal} {Rev. Sci. Instrum.}\ }\textbf {\bibinfo {volume} {82}},\ \bibinfo
  {pages} {071101} (\bibinfo {year} {2011})}\BibitemShut {NoStop}%
\bibitem [{\citenamefont {Castelletto}\ \emph {et~al.}(2014)\citenamefont
  {Castelletto}, \citenamefont {Johnson}, \citenamefont {Iv\'ady},
  \citenamefont {Stavrias}, \citenamefont {Umeda}, \citenamefont {Gali},\ and\
  \citenamefont {Ohshima}}]{Nat14_Ohshima}%
  \BibitemOpen
  \bibfield  {author} {\bibinfo {author} {\bibfnamefont {S.}~\bibnamefont
  {Castelletto}}, \bibinfo {author} {\bibfnamefont {B.~C.}\ \bibnamefont
  {Johnson}}, \bibinfo {author} {\bibfnamefont {V.}~\bibnamefont {Iv\'ady}},
  \bibinfo {author} {\bibfnamefont {N.}~\bibnamefont {Stavrias}}, \bibinfo
  {author} {\bibfnamefont {T.}~\bibnamefont {Umeda}}, \bibinfo {author}
  {\bibfnamefont {A.}~\bibnamefont {Gali}}, \ and\ \bibinfo {author}
  {\bibfnamefont {T.}~\bibnamefont {Ohshima}},\ }\href {\doibase
  10.1038/nature11361} {\bibfield  {journal} {\bibinfo  {journal} {Nat. Mat.}\
  }\textbf {\bibinfo {volume} {13}},\ \bibinfo {pages} {151} (\bibinfo {year}
  {2014})}\BibitemShut {NoStop}%
\bibitem [{\citenamefont {Schwarzkopf}\ \emph {et~al.}(2011)\citenamefont
  {Schwarzkopf}, \citenamefont {Sapiro},\ and\ \citenamefont
  {Raithel}}]{PRL_11_Raithel}%
  \BibitemOpen
  \bibfield  {author} {\bibinfo {author} {\bibfnamefont {A.}~\bibnamefont
  {Schwarzkopf}}, \bibinfo {author} {\bibfnamefont {R.~E.}\ \bibnamefont
  {Sapiro}}, \ and\ \bibinfo {author} {\bibfnamefont {G.}~\bibnamefont
  {Raithel}},\ }\href {\doibase 10.1103/PhysRevLett.107.103001} {\bibfield
  {journal} {\bibinfo  {journal} {Phys. Rev. Lett.}\ }\textbf {\bibinfo
  {volume} {107}},\ \bibinfo {pages} {103001} (\bibinfo {year}
  {2011})}\BibitemShut {NoStop}%
\bibitem [{\citenamefont {Schau\ss}\ \emph {et~al.}(2012)\citenamefont
  {Schau\ss}, \citenamefont {Cheneau}, \citenamefont {Endres}, \citenamefont
  {Fukuhara}, \citenamefont {Hild}, \citenamefont {Omran}, \citenamefont
  {Pohl}, \citenamefont {Gross}, \citenamefont {Kuhr},\ and\ \citenamefont
  {Bloch}}]{Nat12_Schausz}%
  \BibitemOpen
  \bibfield  {author} {\bibinfo {author} {\bibfnamefont {P.}~\bibnamefont
  {Schau\ss}}, \bibinfo {author} {\bibfnamefont {M.}~\bibnamefont {Cheneau}},
  \bibinfo {author} {\bibfnamefont {M.}~\bibnamefont {Endres}}, \bibinfo
  {author} {\bibfnamefont {T.}~\bibnamefont {Fukuhara}}, \bibinfo {author}
  {\bibfnamefont {S.}~\bibnamefont {Hild}}, \bibinfo {author} {\bibfnamefont
  {A.}~\bibnamefont {Omran}}, \bibinfo {author} {\bibfnamefont
  {T.}~\bibnamefont {Pohl}}, \bibinfo {author} {\bibfnamefont {C.}~\bibnamefont
  {Gross}}, \bibinfo {author} {\bibfnamefont {S.}~\bibnamefont {Kuhr}}, \ and\
  \bibinfo {author} {\bibfnamefont {I.}~\bibnamefont {Bloch}},\ }\href
  {http://dx.doi.org/10.1038/nature11596} {\bibfield  {journal} {\bibinfo
  {journal} {Nature}\ }\textbf {\bibinfo {volume} {491}},\ \bibinfo {pages}
  {87} (\bibinfo {year} {2012})}\BibitemShut {NoStop}%
\bibitem [{\citenamefont {Malossi}\ \emph {et~al.}(2014)\citenamefont
  {Malossi}, \citenamefont {Valado}, \citenamefont {Scotto}, \citenamefont
  {Huillery}, \citenamefont {Pillet}, \citenamefont {Ciampini}, \citenamefont
  {Arimondo},\ and\ \citenamefont {Morsch}}]{PRL14_Arimondo}%
  \BibitemOpen
  \bibfield  {author} {\bibinfo {author} {\bibfnamefont {N.}~\bibnamefont
  {Malossi}}, \bibinfo {author} {\bibfnamefont {M.~M.}\ \bibnamefont {Valado}},
  \bibinfo {author} {\bibfnamefont {S.}~\bibnamefont {Scotto}}, \bibinfo
  {author} {\bibfnamefont {P.}~\bibnamefont {Huillery}}, \bibinfo {author}
  {\bibfnamefont {P.}~\bibnamefont {Pillet}}, \bibinfo {author} {\bibfnamefont
  {D.}~\bibnamefont {Ciampini}}, \bibinfo {author} {\bibfnamefont
  {E.}~\bibnamefont {Arimondo}}, \ and\ \bibinfo {author} {\bibfnamefont
  {O.}~\bibnamefont {Morsch}},\ }\href {\doibase
  10.1103/PhysRevLett.113.023006} {\bibfield  {journal} {\bibinfo  {journal}
  {Phys. Rev. Lett.}\ }\textbf {\bibinfo {volume} {113}},\ \bibinfo {pages}
  {023006} (\bibinfo {year} {2014})}\BibitemShut {NoStop}%
\bibitem [{\citenamefont {Urvoy}\ \emph {et~al.}(2015)\citenamefont {Urvoy},
  \citenamefont {Ripka}, \citenamefont {Lesanovsky}, \citenamefont {Booth},
  \citenamefont {Shaffer}, \citenamefont {Pfau},\ and\ \citenamefont
  {L\"ow}}]{Urvoy15}%
  \BibitemOpen
  \bibfield  {author} {\bibinfo {author} {\bibfnamefont {A.}~\bibnamefont
  {Urvoy}}, \bibinfo {author} {\bibfnamefont {F.}~\bibnamefont {Ripka}},
  \bibinfo {author} {\bibfnamefont {I.}~\bibnamefont {Lesanovsky}}, \bibinfo
  {author} {\bibfnamefont {D.}~\bibnamefont {Booth}}, \bibinfo {author}
  {\bibfnamefont {J.~P.}\ \bibnamefont {Shaffer}}, \bibinfo {author}
  {\bibfnamefont {T.}~\bibnamefont {Pfau}}, \ and\ \bibinfo {author}
  {\bibfnamefont {R.}~\bibnamefont {L\"ow}},\ }\href {\doibase
  10.1103/PhysRevLett.114.203002} {\bibfield  {journal} {\bibinfo  {journal}
  {Phys. Rev. Lett.}\ }\textbf {\bibinfo {volume} {114}},\ \bibinfo {pages}
  {203002} (\bibinfo {year} {2015})}\BibitemShut {NoStop}%
\bibitem [{\citenamefont {Schempp}\ \emph {et~al.}(2014)\citenamefont
  {Schempp}, \citenamefont {G\"unter}, \citenamefont {Robert-de Saint-Vincent},
  \citenamefont {Hofmann}, \citenamefont {Breyel}, \citenamefont {Komnik},
  \citenamefont {Sch\"onleber}, \citenamefont {G\"arttner}, \citenamefont
  {Evers}, \citenamefont {Whitlock},\ and\ \citenamefont
  {Weidem\"uller}}]{PRL_12_Schempp}%
  \BibitemOpen
  \bibfield  {author} {\bibinfo {author} {\bibfnamefont {H.}~\bibnamefont
  {Schempp}}, \bibinfo {author} {\bibfnamefont {G.}~\bibnamefont {G\"unter}},
  \bibinfo {author} {\bibfnamefont {M.}~\bibnamefont {Robert-de
  Saint-Vincent}}, \bibinfo {author} {\bibfnamefont {C.~S.}\ \bibnamefont
  {Hofmann}}, \bibinfo {author} {\bibfnamefont {D.}~\bibnamefont {Breyel}},
  \bibinfo {author} {\bibfnamefont {A.}~\bibnamefont {Komnik}}, \bibinfo
  {author} {\bibfnamefont {D.~W.}\ \bibnamefont {Sch\"onleber}}, \bibinfo
  {author} {\bibfnamefont {M.}~\bibnamefont {G\"arttner}}, \bibinfo {author}
  {\bibfnamefont {J.}~\bibnamefont {Evers}}, \bibinfo {author} {\bibfnamefont
  {S.}~\bibnamefont {Whitlock}}, \ and\ \bibinfo {author} {\bibfnamefont
  {M.}~\bibnamefont {Weidem\"uller}},\ }\href {\doibase
  10.1103/PhysRevLett.112.013002} {\bibfield  {journal} {\bibinfo  {journal}
  {Phys. Rev. Lett.}\ }\textbf {\bibinfo {volume} {112}},\ \bibinfo {pages}
  {013002} (\bibinfo {year} {2014})}\BibitemShut {NoStop}%
\bibitem [{\citenamefont {Pohl}\ \emph {et~al.}(2010)\citenamefont {Pohl},
  \citenamefont {Demler},\ and\ \citenamefont {Lukin}}]{PRL10_Phol1}%
  \BibitemOpen
  \bibfield  {author} {\bibinfo {author} {\bibfnamefont {T.}~\bibnamefont
  {Pohl}}, \bibinfo {author} {\bibfnamefont {E.}~\bibnamefont {Demler}}, \ and\
  \bibinfo {author} {\bibfnamefont {M.~D.}\ \bibnamefont {Lukin}},\ }\href
  {\doibase 10.1103/PhysRevLett.104.043002} {\bibfield  {journal} {\bibinfo
  {journal} {Phys. Rev. Lett.}\ }\textbf {\bibinfo {volume} {104}},\ \bibinfo
  {pages} {043002} (\bibinfo {year} {2010})}\BibitemShut {NoStop}%
\bibitem [{\citenamefont {Petrosyan}\ and\ \citenamefont
  {M\o{}lmer}(2014)}]{Petrosyan_PRL1132014}%
  \BibitemOpen
  \bibfield  {author} {\bibinfo {author} {\bibfnamefont {D.}~\bibnamefont
  {Petrosyan}}\ and\ \bibinfo {author} {\bibfnamefont {K.}~\bibnamefont
  {M\o{}lmer}},\ }\href {\doibase 10.1103/PhysRevLett.113.123003} {\bibfield
  {journal} {\bibinfo  {journal} {Phys. Rev. Lett.}\ }\textbf {\bibinfo
  {volume} {113}},\ \bibinfo {pages} {123003} (\bibinfo {year}
  {2014})}\BibitemShut {NoStop}%
\bibitem [{\citenamefont {Bendkowsky}\ \emph {et~al.}(2009)\citenamefont
  {Bendkowsky}, \citenamefont {Butscher}, \citenamefont {Nipper}, \citenamefont
  {Shaffer}, \citenamefont {L\"ow},\ and\ \citenamefont
  {Pfau}}]{Nat09_Bendkowsky}%
  \BibitemOpen
  \bibfield  {author} {\bibinfo {author} {\bibfnamefont {V.}~\bibnamefont
  {Bendkowsky}}, \bibinfo {author} {\bibfnamefont {B.}~\bibnamefont
  {Butscher}}, \bibinfo {author} {\bibfnamefont {J.}~\bibnamefont {Nipper}},
  \bibinfo {author} {\bibfnamefont {J.~P.}\ \bibnamefont {Shaffer}}, \bibinfo
  {author} {\bibfnamefont {R.}~\bibnamefont {L\"ow}}, \ and\ \bibinfo {author}
  {\bibfnamefont {T.}~\bibnamefont {Pfau}},\ }\href {\doibase
  10.1038/nature07945} {\bibfield  {journal} {\bibinfo  {journal} {Nature}\
  }\textbf {\bibinfo {volume} {458}},\ \bibinfo {pages} {1005} (\bibinfo {year}
  {2009})}\BibitemShut {NoStop}%
\bibitem [{\citenamefont {Booth}\ \emph {et~al.}(2015)\citenamefont {Booth},
  \citenamefont {Rittenhouse}, \citenamefont {Yang}, \citenamefont
  {Sadeghpour},\ and\ \citenamefont {Shaffer}}]{Booth2015}%
  \BibitemOpen
  \bibfield  {author} {\bibinfo {author} {\bibfnamefont {D.}~\bibnamefont
  {Booth}}, \bibinfo {author} {\bibfnamefont {S.~T.}\ \bibnamefont
  {Rittenhouse}}, \bibinfo {author} {\bibfnamefont {J.}~\bibnamefont {Yang}},
  \bibinfo {author} {\bibfnamefont {H.~R.}\ \bibnamefont {Sadeghpour}}, \ and\
  \bibinfo {author} {\bibfnamefont {J.~P.}\ \bibnamefont {Shaffer}},\ }\href
  {\doibase 10.1126/science.1260722} {\bibfield  {journal} {\bibinfo  {journal}
  {Science}\ }\textbf {\bibinfo {volume} {348}},\ \bibinfo {pages} {99}
  (\bibinfo {year} {2015})}\BibitemShut {NoStop}%
\bibitem [{\citenamefont {Heinzen}\ \emph {et~al.}(1987)\citenamefont
  {Heinzen}, \citenamefont {Childs}, \citenamefont {Thomas},\ and\
  \citenamefont {Feld}}]{Heinzen87}%
  \BibitemOpen
  \bibfield  {author} {\bibinfo {author} {\bibfnamefont {D.~J.}\ \bibnamefont
  {Heinzen}}, \bibinfo {author} {\bibfnamefont {J.~J.}\ \bibnamefont {Childs}},
  \bibinfo {author} {\bibfnamefont {J.~E.}\ \bibnamefont {Thomas}}, \ and\
  \bibinfo {author} {\bibfnamefont {M.~S.}\ \bibnamefont {Feld}},\ }\href
  {\doibase 10.1103/PhysRevLett.58.1320} {\bibfield  {journal} {\bibinfo
  {journal} {Phys. Rev. Lett.}\ }\textbf {\bibinfo {volume} {58}},\ \bibinfo
  {pages} {1320} (\bibinfo {year} {1987})}\BibitemShut {NoStop}%
\bibitem [{\citenamefont {Hunger}\ \emph {et~al.}(2010)\citenamefont {Hunger},
  \citenamefont {Steinmetz}, \citenamefont {Colombe}, \citenamefont {Deutsch},
  \citenamefont {H\"ansch},\ and\ \citenamefont {Reichel}}]{Reichel10}%
  \BibitemOpen
  \bibfield  {author} {\bibinfo {author} {\bibfnamefont {D.}~\bibnamefont
  {Hunger}}, \bibinfo {author} {\bibfnamefont {T.}~\bibnamefont {Steinmetz}},
  \bibinfo {author} {\bibfnamefont {Y.}~\bibnamefont {Colombe}}, \bibinfo
  {author} {\bibfnamefont {C.}~\bibnamefont {Deutsch}}, \bibinfo {author}
  {\bibfnamefont {T.~W.}\ \bibnamefont {H\"ansch}}, \ and\ \bibinfo {author}
  {\bibfnamefont {J.}~\bibnamefont {Reichel}},\ }\href@noop {} {\bibfield
  {journal} {\bibinfo  {journal} {New J. Phys.}\ }\textbf {\bibinfo {volume}
  {12}},\ \bibinfo {pages} {065038} (\bibinfo {year} {2010})}\BibitemShut
  {NoStop}%
\bibitem [{\citenamefont {Yalla}\ \emph {et~al.}(2014)\citenamefont {Yalla},
  \citenamefont {Sadgrove}, \citenamefont {Nayak},\ and\ \citenamefont
  {Hakuta}}]{Yalla14}%
  \BibitemOpen
  \bibfield  {author} {\bibinfo {author} {\bibfnamefont {R.}~\bibnamefont
  {Yalla}}, \bibinfo {author} {\bibfnamefont {M.}~\bibnamefont {Sadgrove}},
  \bibinfo {author} {\bibfnamefont {K.~P.}\ \bibnamefont {Nayak}}, \ and\
  \bibinfo {author} {\bibfnamefont {K.}~\bibnamefont {Hakuta}},\ }\href
  {\doibase 10.1103/PhysRevLett.113.143601} {\bibfield  {journal} {\bibinfo
  {journal} {Phys. Rev. Lett.}\ }\textbf {\bibinfo {volume} {113}},\ \bibinfo
  {pages} {143601} (\bibinfo {year} {2014})}\BibitemShut {NoStop}%
\bibitem [{\citenamefont {Gross}\ and\ \citenamefont
  {Haroche}(1982)}]{PhyRep93_Gross}%
  \BibitemOpen
  \bibfield  {author} {\bibinfo {author} {\bibfnamefont {M.}~\bibnamefont
  {Gross}}\ and\ \bibinfo {author} {\bibfnamefont {S.}~\bibnamefont
  {Haroche}},\ }\href {\doibase
  "http://dx.doi.org/10.1016/0370-1573(82)90102-8"} {\bibfield  {journal}
  {\bibinfo  {journal} {Phys. Rep.}\ }\textbf {\bibinfo {volume} {93}},\
  \bibinfo {pages} {301 } (\bibinfo {year} {1982})}\BibitemShut {NoStop}%
\end{thebibliography}
%

\end{document}